\documentclass[namedreferences]{SolarPhysics}
\usepackage[optionalrh]{spr-sola-addons} 
\usepackage{graphicx}                    
\usepackage{courier}                    
\usepackage{amssymb}                     

\newcommand{\etal}{{\it et al.}}

\newcommand{\Bv}{\mathbf{v}}
\newcommand{\BB}{\mathbf{B}}
\newcommand{\Bj}{\mathbf{j}}
\newcommand{\BA}{\mathbf{A}}
\newcommand{\BE}{\mathbf{E}}
\newcommand{\Br}{\mathbf{r}}

\usepackage{color}                       
\usepackage{url}                         

\begin{document}
\begin{article}
\begin{opening}

\title{Solar Magnetic Carpet III: Coronal Modelling of Synthetic Magnetograms}
\author{K.A. \surname{Meyer}$^{1}$\sep D.H. \surname{Mackay}$^{1}$\sep A.A.~\surname{van~Ballegooijen}$^{2}$ \sep C.E. \surname{Parnell}$^{1}$}

\runningauthor{Meyer, K.A. \etal}
\runningtitle{Coronal Modelling of Synthetic Magnetograms}

\institute{ $^{1}$ School of Mathematics and Statistics, University of 
                  St Andrews, North Haugh, St Andrews, Fife, KY16 9SS, Scotland, U.K. \\
                  \url{karen@mcs.st-and.ac.uk}\\
                  $^{2}$ Harvard-Smithsonian Center for Astrophysics, 60 Garden Street, Cambridge, MA 02138}

\date{Received 21 September 2012 / accepted 5 March 2013}
\begin{abstract}\\
This paper is the third in a series of papers working towards the construction
of a realistic, evolving, non-linear force-free coronal field model for the solar magnetic
carpet. Here, we present preliminary results of 3D time-dependent simulations of the
small-scale coronal field of the magnetic carpet. Four simulations are
considered, each with the same evolving photospheric boundary condition: a 48
hr time series of synthetic magnetograms produced from the model of
\inlinecite{meyer2011}. Three simulations include a uniform, overlying coronal
magnetic field of differing strength, the fourth simulation includes no overlying field. The build-up,
storage and dissipation of magnetic energy within the simulations is studied. In particular, we study their dependence upon the evolution of the photospheric magnetic field and the strength of the
overlying coronal field. We also consider where energy is stored and dissipated
within the coronal field. The free magnetic energy built up is found to be more than
sufficient to power small-scale, transient phenomena such as nanoflares and X-ray
bright points, with the bulk of the free energy found to be stored low down,
between $0.5-0.8$ Mm. The energy dissipated is presently found to be too small to account for the heating
of the entire quiet Sun corona. However, the form and location of
energy dissipation regions are in qualitative agreement with what is observed
on small scales on the Sun. Future MHD modelling using the same synthetic magnetograms may lead to a higher energy release.
\end{abstract}
\end{opening}

\section{Introduction}\label{intro}

The small-scale photospheric magnetic field of the quiet Sun, the \emph{magnetic
carpet}, is complex and constantly evolving. This evolution is due to underlying
photospheric flows produced by convection on many scales ({\it e.g.} granulation and supergranulation,
\inlinecite{leighton1962}, \inlinecite{wang1989}, \inlinecite{schrijver1997}, \inlinecite{rieutord2010}) and the flux evolution
processes of emergence, cancellation, coalescence and fragmentation. Small-scale
magnetic features classified as ephemeral regions
\cite{harvey1973,harvey1993,hagenaar2008,schrijver2010}, internetwork features
\cite{livingston1975,wang1996,dewijn2008,zhou2010} and network features
\cite{simon1964,zirin1985,martin1984,martin1988} continually interact with one
another, resulting in a photospheric recycle time of just $1-2$ hr (hours)
\cite{hagenaar2008}. This is the time taken for all flux within the quiet Sun
photosphere to be replaced. Since magnetic fields from the magnetic carpet
extend up into the solar chromosphere and lower corona, it is expected that the
quiet Sun corona is also highly dynamic. Complex interactions of magnetic
features on the photosphere may result in significant heating of the corona,
for example through braiding and reconnection of magnetic field lines ({\it e.g.}
\inlinecite{galsgaard1996}, \inlinecite{parnell2004}, \inlinecite{haynes2007} \inlinecite{rappazzo2008}, \inlinecite{berger2009}, \inlinecite{wilmot2009}, \inlinecite{pontin2011}). Therefore, it is of interest to simulate the small-scale
coronal field resulting from the evolution of the solar magnetic carpet.

Previous magnetic carpet coronal field models using extrapolation methods have studied, for example, flux
topology and connectivity \cite{schrijver2002, close2003}; the number density
and locations of coronal null points \cite{schrijver2002,regnier2008,longcope2009}; coronal
remap times \cite{close2004,cranmer2010}; and whether the solar wind can be
driven by reconnection in the magnetic carpet \cite{cranmer2010}. However, each
of these studies considered only potential field extrapolations of the
small-scale coronal field.

Within this paper, we present preliminary results of 3D simulations of the
network-scale coronal field of the magnetic carpet. In contrast to the models
described above, which produce independent potential field extrapolations, we
model a continuous evolution of a non-linear force-free coronal field. A non-linear
force-free field satisfies the conditions $\nabla\cdot\BB=0$ and
$\nabla\times\BB=\alpha\BB$, where $\alpha=\alpha(\Br)$ is a scalar function of position (but constant
along a given field line) describing the twist of the field.
This approximation is a step up in complexity from a potential field as it allows
for the existence of electric currents and free magnetic energy. In
\inlinecite{meyer2011} (hereafter Paper I) we presented a realistic model for
the photospheric evolution of the magnetic carpet, that reproduced many
observational properties. Synthetic magnetograms produced from this model will
provide the photospheric boundary condition to drive the evolution of the full
3D coronal field. The coronal field is evolved through a series of quasi-static,
non-linear force-free states in response to the evolution of the photospheric
magnetic field, using a magnetofrictional technique that is described in
\inlinecite{meyer2012} (hereafter Paper II). There are several advantages to
using synthetic magnetograms as opposed to actual observed data: (i) The total
magnetic flux within the synthetic magnetograms is always known $-$ there is no
noise or instrumental limitations. (ii) We have complete control over the
photospheric evolution $-$ magnetic features do not drift into or out of the
field of view, which is unavoidable in real magnetograms. (iii) The synthetic
magnetograms are always in flux balance (total positive flux = total negative
flux). This is a requirement for the coronal model due to our set-up. (iv) We
know exactly where and when each of the processes of emergence, cancellation,
coalescence and fragmentation occur within the synthetic magnetograms, as well
as the exact flux involved in each event. Many feature tracking techniques exist
to follow the evolution of magnetic features in real magnetogram data, {\it e.g.}
\inlinecite{deforest2007}. However, these are limited by factors such as spatial
resolution and time cadence. In a future study, we will consider in detail how
various photospheric events in our synthetic magnetograms affect the evolution of the coronal field.

In this paper we consider four simulations in total, each with the same
photospheric boundary condition: a $48$ hr set of synthetic
magnetograms. Three of the simulations have a uniform, overlying coronal magnetic field (of varying strength) which points in the positive
$x-$direction; this simulates the influence of larger scale magnetic features,
such as exist on the Sun, especially during solar maximum. The fourth simulation has no overlying field and thus
represents a very quiet region of the Sun, such as may be found during solar minimum. We consider the magnetic energy both
stored and dissipated within the coronal volume, along with the square of the current
density, $j^2$. Some of the properties of these quantities that we consider are:
how they evolve in time, where they are located spatially with respect to the underlying photospheric magnetic field, and the effect of
varying the strength of the overlying coronal magnetic field.

Section~\ref{sec:model} describes the lower boundary condition and the set-up of
the 3D simulations. Section~\ref{sec:result} provides the results, while
Section~\ref{sec:conc} gives a discussion and conclusions.

\section{Lower Boundary Condition and Set-Up}\label{sec:model}

  \begin{figure}

   \centerline{\hspace*{0.015\textwidth}
              \includegraphics[width=0.515\textwidth,clip=]{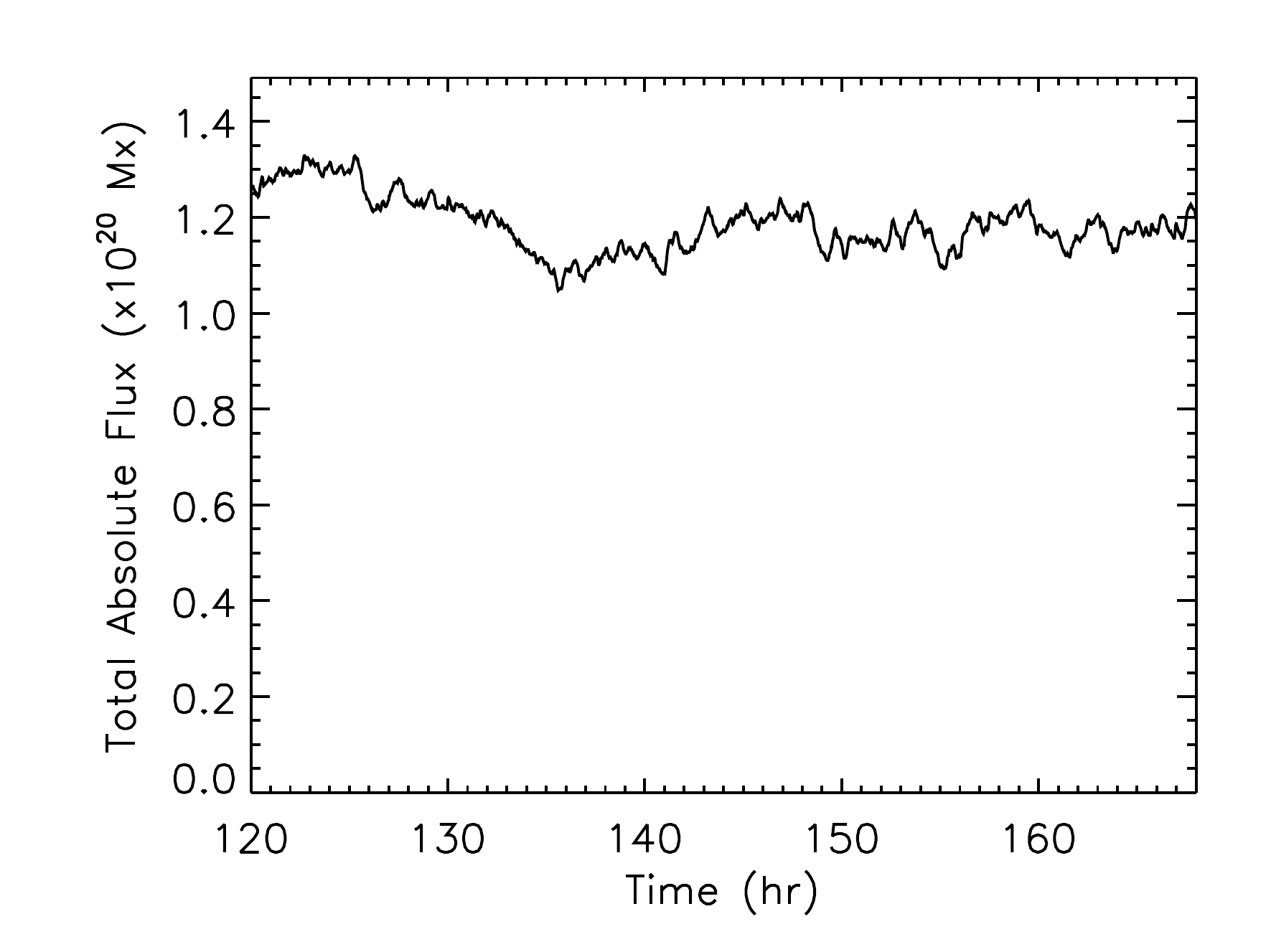}
               \hspace*{-0.03\textwidth}
              \includegraphics[width=0.515\textwidth,clip=]{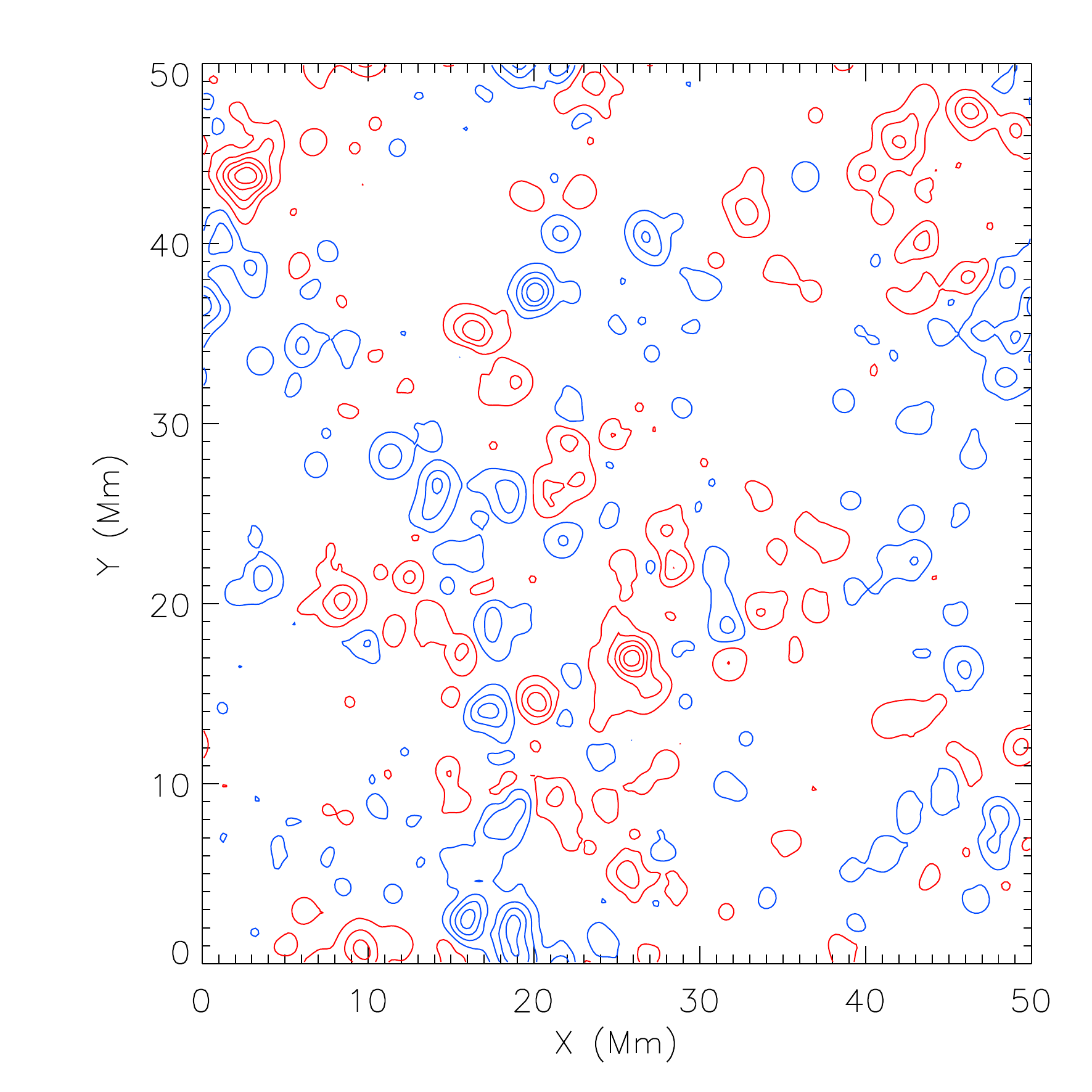}
              }
     \vspace{-0.4\textwidth}   
     \centerline{ \bf     
      \hspace{-0.03 \textwidth}  \color{black}{(a)}
      \hspace{0.45\textwidth}  \color{black}{(b)}
         \hfill}
     \vspace{0.37\textwidth}    

\caption{(a) Total absolute flux through the photosphere as a function
of time for the 48 hr series of
magnetograms used in our 3D models. (b) Synthetic magnetogram at $t=120$ hr, composed of 1499 individual magnetic elements. Ten contour levels of positive (red) and negative (blue) magnetic field are shown, with
absolute values spaced evenly between 7.5 G and 142.5 G.}\label{fig:lower}
   \end{figure}

We choose the most realistic simulation from Paper I to provide the lower
boundary condition for 3D modelling. This simulation, which covers an area of $50 \times
50$ Mm$^2$, is of length $250$ hr with a cadence of 1 min, and
has a flux emergence range of $\phi_\textrm{\tiny bp}=4\times 10^{16}-10^{19}$
Mx. Flux emergence is determined by the probability distribution of \inlinecite{thornton2011}, giving an emergence rate of $1.2\times 10^{-3}$ Mx cm$^{-2}$ s$^{-1}$. The smallest magnetic features that occur within this simulation have a flux of $10^{16}$ Mx. The magnetograms are all in flux balance and are
periodic in the $x-$ and $y-$directions. For full details of the simulation, see Paper I. For
the 3D simulations, we select a $48$ hr window of the synthetic magnetograms from
the full $250$ hr set, as explained below, providing a series of $2881$
synthetic magnetograms.

Figure~\ref{fig:lower}(a) shows the total absolute flux for the $48$ hr section of the 2D
simulation that is used for 3D modelling. The curve oscillates about a mean value of $1.19\times 10^{20}$ Mx, with a standard deviation of $5.79\times 10^{18}$ Mx, or $4.9\%$, indicating that the 2D simulation is
in a steady state at this time.

Figures~\ref{fig:lower}(b) shows an example of a synthetic magnetogram,
taken at $t=120$ hr, where red and blue contours represent positive and negative magnetic field, respectively. A movie
showing the photospheric evolution of $B_z$ for the first $2$ hr is available (\textcolor{blue}{magnet48\_bz\_2hr.mpg}). The magnetic
elements are mainly located around the boundaries of the supergranules, forming
the magnetic network (an image of the simulated supergranules can be seen in
Figure 3(b) of Paper I). The supergranular flow profile is not time-evolving, so
the general shape of the magnetic network does not vary much. Despite this limitation, the exact distribution of magnetic elements does change significantly
throughout the $48$ hr period. In Paper I it was determined that the
photospheric recycle time for this series of synthetic magnetograms was $1.48$
hr, in excellent agreement with \inlinecite{hagenaar2008}'s recycle time of
$1-2$ hr. Therefore the synthetic magnetogram series realistically simulates the
dynamic nature of the magnetic carpet, making it a suitable lower boundary
condition for 3D modelling.

\subsection{Coronal Magnetic Field Evolution}

We now discuss the set-up of the 3D model. We choose a numerical box of size
$50\times 50\times 25$ Mm$^3$, composed of $512\times 512\times 256$ grid cells.
The box is periodic in the $x-$ and $y-$directions and closed at the top. Since
magnetic flux may only enter and exit the box through the lower boundary, the
synthetic magnetograms are required to be in complete flux balance. The periodicity of the side
boundaries gives the effect of the region being surrounded by similar regions of
small-scale magnetic carpet features. The initial condition for each simulation is a potential field extrapolated from the first magnetogram. We subsequently evolve
the coronal magnetic field in
response to photospheric boundary motions using a magnetofrictional relaxation
technique \cite{yang1986,mackay2006,yeates2008,mackay2009,mackay2011,meyer2012}. The
magnetofrictional technique produces a continuous evolution of the coronal
magnetic field, so that a `memory' of flux connectivity and electric current systems is maintained from one
step to the next. The coronal field induction equation is given by
\begin{equation}\label{eqn:ind}
 \frac{\partial \BA}{\partial t}=\Bv\times\BB + \textrm{\boldmath$\epsilon$}
\end{equation}
where $\BB=\nabla\times\BA$ is the magnetic field and $\BA$ its the associated vector potential.  The plasma velocity, $\Bv$, is given by
\begin{equation}\label{eqn:v}
\Bv = \frac{1}{\nu}\frac{\Bj\times\BB}{B^2},
\end{equation}
where $\nu$ is the coefficient of friction, $\frac{1}{\nu}=8\times 10^4$ km$^2$ s$^{-1}$, and $\Bj=\nabla\times\BB$. This
velocity describes the relaxation of the coronal magnetic field towards a
non-linear force-free state in response to perturbations, and takes advantage of
the fact that the Lorentz force, $\Bj\times\BB$, is the dominant force within
the corona. The term $\textrm{\boldmath$\epsilon$}$ within
Equation~\ref{eqn:ind} represents \emph{hyperdiffusion} ({\it e.g.}
\inlinecite{boozer1986}, \inlinecite{vanballegooijen2008}) and is chosen to be
of the form
\begin{equation}\label{eqn:hd}
 \textrm{\boldmath$\epsilon$} = \frac{\BB}{B^2}\nabla\cdot(\eta_4 B^2
\nabla\alpha),
\end{equation}
where, $\eta_4=7.6\times 10^5$ km$^4$ s$^{-1}$ and $\alpha$ is the scalar coefficient from the definition of a non-linear force-free field such that $\nabla\times\BB=\alpha\BB$. It is calculated as
\begin{equation}
 \alpha = \frac{\Bj\cdot\BB}{B^2}.
\end{equation}
Hyperdiffusion aids the stability of the code
by smoothing gradients in $\alpha$ and allows reconnection to occur. It acts to reduce the magnetic field to a
linear force-free state, however the time scales of the present simulations are
too short for such a state to be reached.

The evolution of the photospheric boundary perturbs the coronal field, which
responds through Equations~\ref{eqn:ind} and \ref{eqn:v}. This continual stressing and relaxing of the
coronal field, in response to the lower boundary motions, evolves the field through a
series of quasi-static, non-linear force-free equilibria.

\subsection{Lower Boundary Treatment}\label{sec:lower}

The lower boundary treatment is the same as is described in Paper II, where a full description may be found. Since
Equation~\ref{eqn:ind} is specified in terms of the vector potential $\BA$, the
lower boundary is also required in terms of $\BA$. A linear interpolation of the
$A_x$ and $A_y$ corresponding to $B_z$ at $z=0$ Mm is carried out between each
synthetic magnetogram. $B_z$ at $z=0$ Mm is analytically specified at each time
step, rather than advected numerically (see Paper I). This avoids certain
undesirable numerical effects such as numerical overshoot or pile-up at
cancellation sites. Between each analytical specification of
the photospheric magnetic field, $500$ interpolation steps are taken,
corresponding to $0.12$ s each. The magnetofrictional technique is applied
during each of these interpolation steps so that the relaxation is gradual.

In total, four simulations are run, each with the same photospheric boundary
evolution. Three of the simulations have an overlying, uniform magnetic field of
strength $B_0=1$ G, 3 G or 10 G, which points in the positive $x-$direction. The
fourth simulation has no overlying field. For the cases with a uniform overlying field, we add this field to the initial potential field as follows:
The potential field is computed in terms of the vector potential $\BA$. We add the term $B_0(z_\textrm{\small max} - z)$ to $A_y$ throughout the volume, where $B_0$ is the strength of the overlying field and $z$ is the height above the photosphere. Then, when we compute $\BB=\nabla\times\BA$, since the $x-$component is
\begin{displaymath}
B_x=\frac{\partial A_z}{\partial y}-\frac{\partial A_y}{\partial z},                                                           
\end{displaymath}
this has the effect of adding a constant $B_0$ to $B_x$ throughout the volume.
We also add a contribution $B_0 z_\textrm{\small max}$ to the $A_y$ component of all of the synthetic magnetograms that provide the lower boundary condition.

\inlinecite{west2011} investigated an EIT wave during solar minimum and estimated the strength of the quiet Sun coronal magnetic field to be $0.7\pm 0.7$ G. A variety of authors have attempted to estimate the strength of active region coronal loops using coronal seismology, with values obtained anywhere in the range $3-90$ G \cite{nakariakov2001,aschwanden2002,verwichte2004,vandoorsselaere2008,wang2007}, although \inlinecite{demoortel2009} stress that coronal seismology does not necessarily reliably determine the strength of the magnetic field. Indeed, they find that when they apply the method of \inlinecite{nakariakov2001} to their 3D model of a coronal loop, it overestimates the strength of the magnetic field by $50$ \%. However, our no overlying field and 1 G overlying field cases can be taken to represent the quiet Sun during solar minimum, while the 3 G and 10 G cases represent a quiet region of the Sun influenced by nearby active regions during solar maximum. The following sections give
preliminary results of these 3D non-potential simulations. In Section~\ref{sec:result}, we briefly discuss
field line connectivity between magnetic elements. Section~\ref{sec:free}
focuses on the free magnetic energy, Section~\ref{sec:j} the square of the
current density ($j^2$) and Section~\ref{sec:q} on the energy dissipated. For
both energies, we consider the time evolution of the quantity integrated over
the whole volume, and also how it is affected by varying the overlying field strength.
We then consider where the energy is located spatially.

\section{Results}\label{sec:result}

  \begin{figure}

   \centerline{\hspace*{0.015\textwidth}
              \includegraphics[width=0.49\textwidth,clip=]{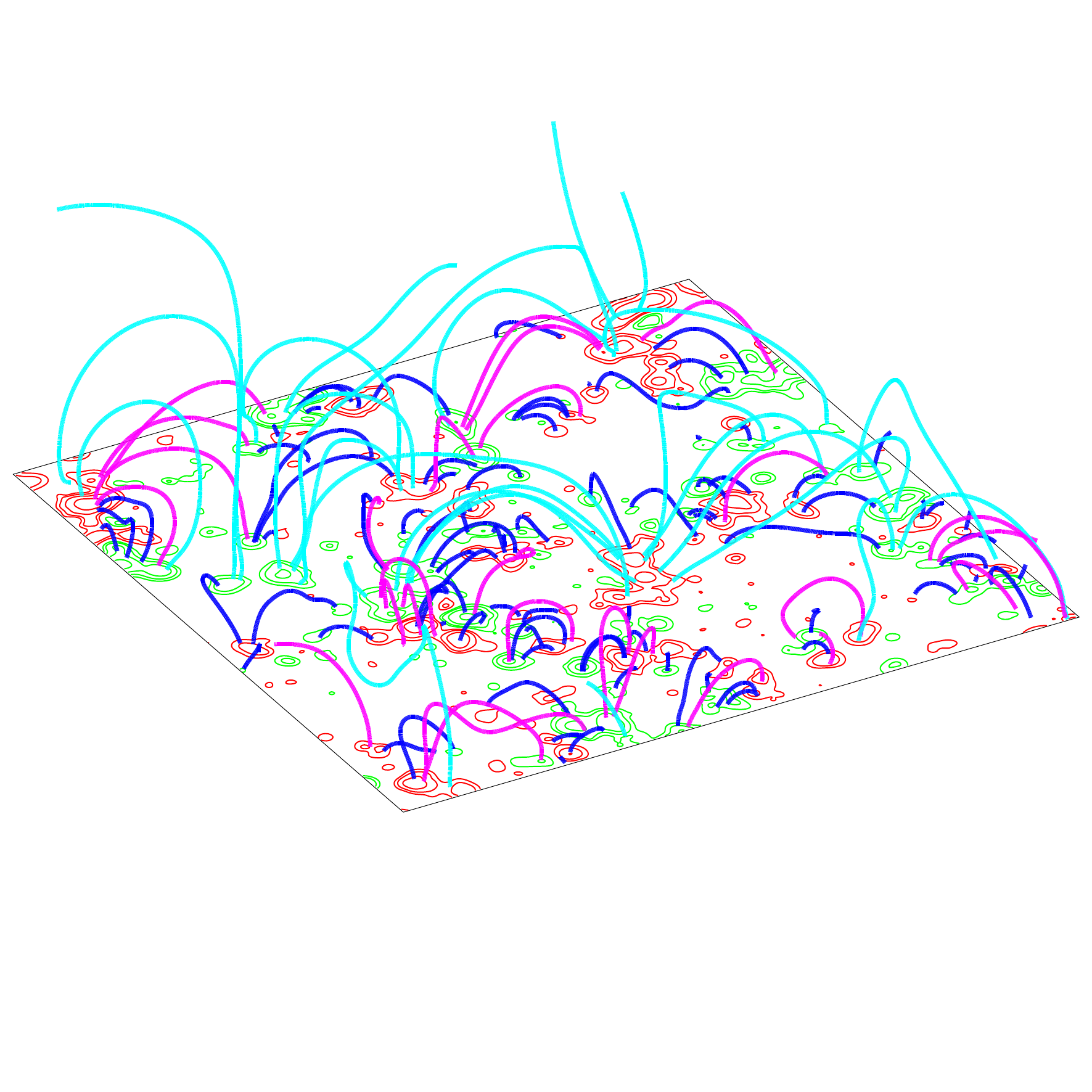}
               \hspace*{-0.01\textwidth}
              \includegraphics[width=0.49\textwidth,clip=]{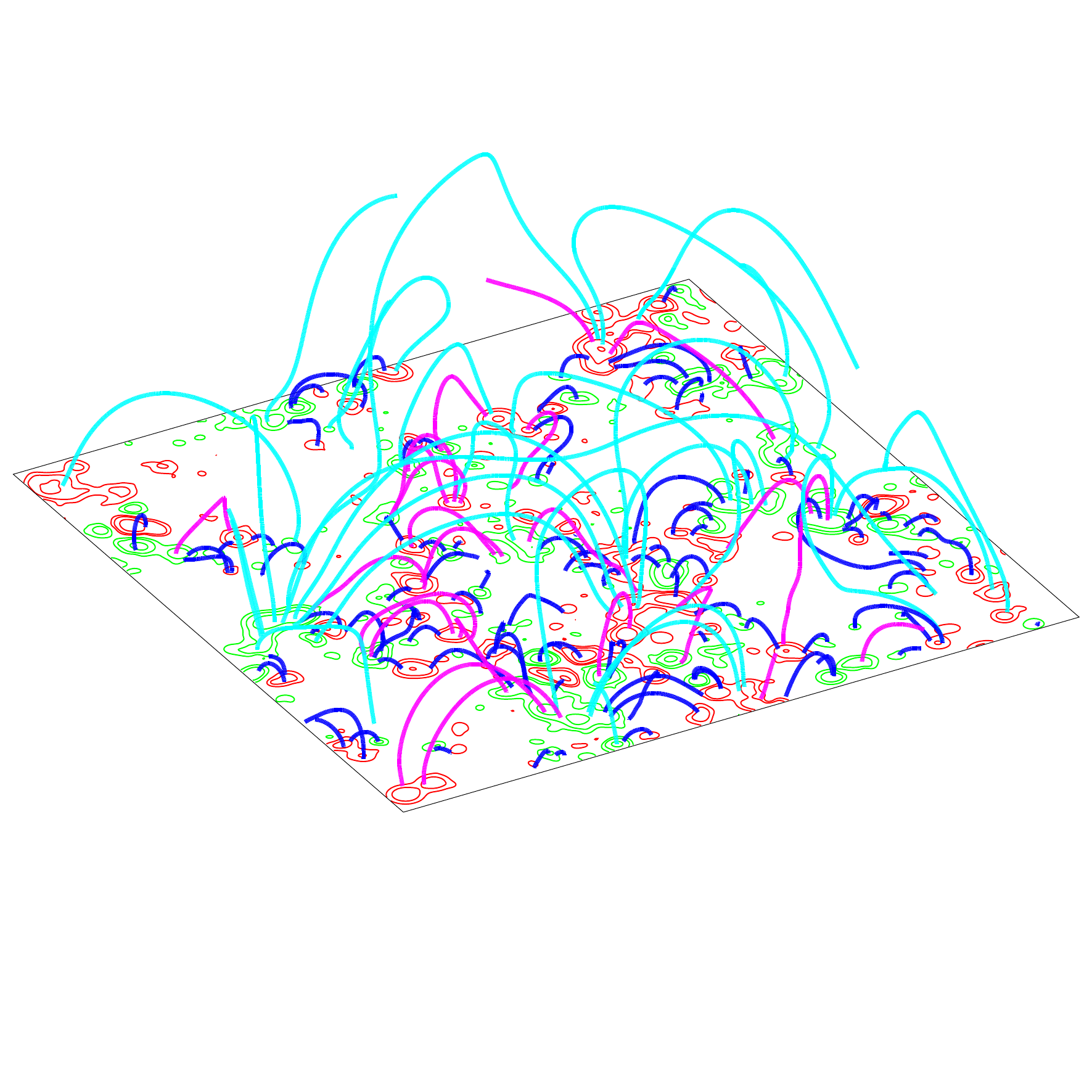}
              }
     \vspace{-0.4\textwidth}   
     \centerline{ \bf     
      \hspace{-0.03 \textwidth}  \color{black}{(a)}
      \hspace{0.46\textwidth}  \color{black}{(b)}
         \hfill}
     \vspace{0.22\textwidth}    

   \centerline{\hspace*{0.015\textwidth}
              \includegraphics[width=0.49\textwidth,clip=]{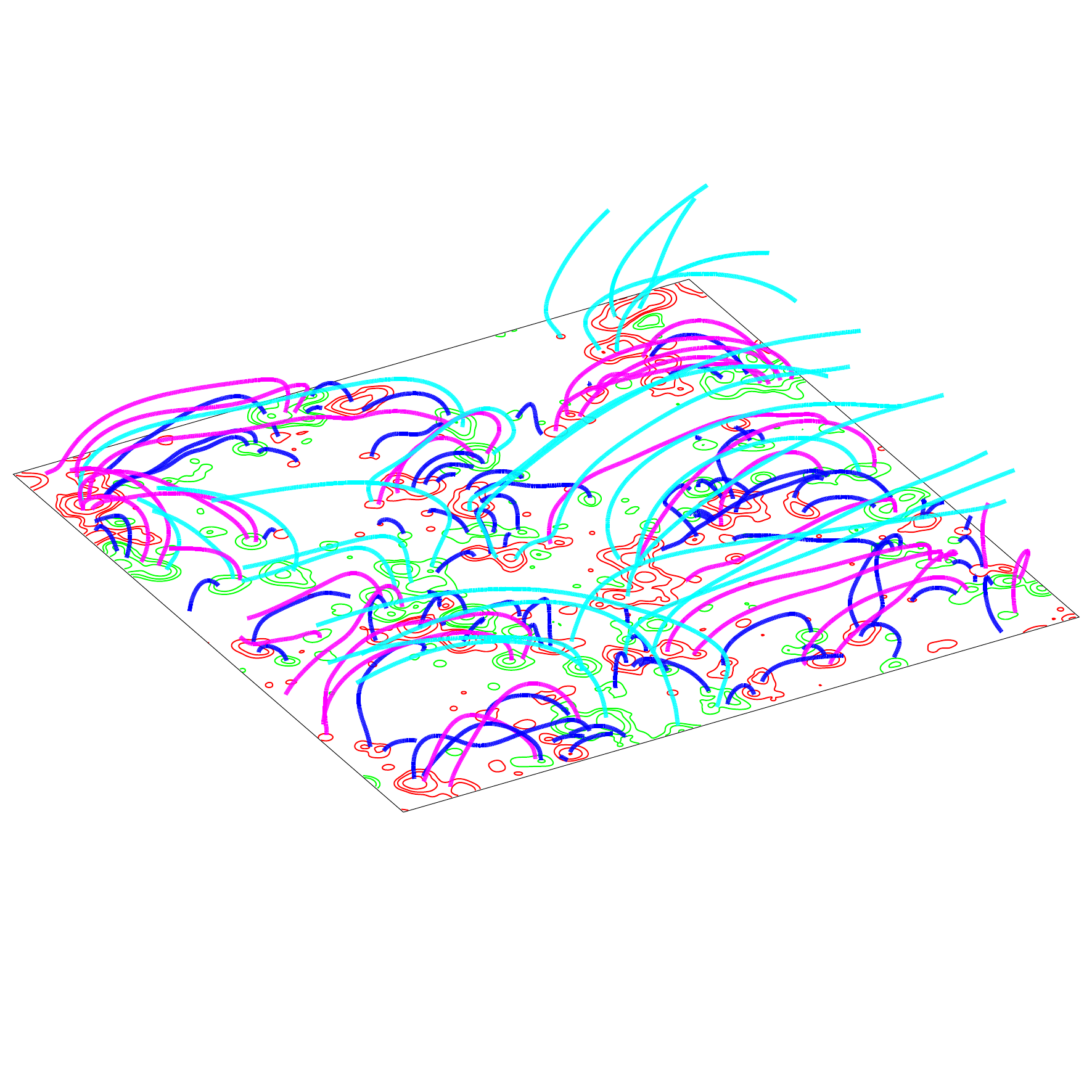}
               \hspace*{-0.01\textwidth}
              \includegraphics[width=0.49\textwidth,clip=]{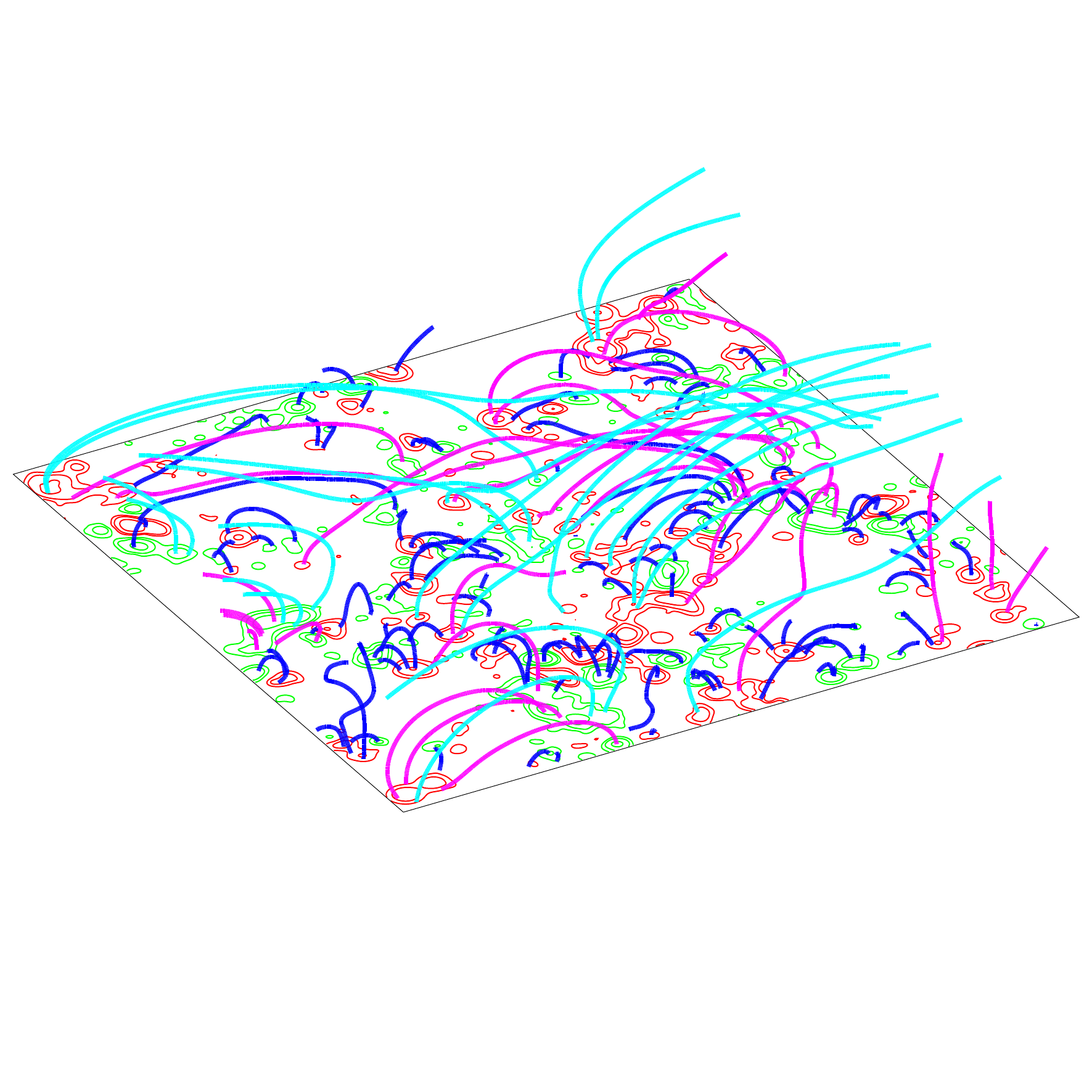}
              }
     \vspace{-0.4\textwidth}   
     \centerline{ \bf     
      \hspace{-0.03 \textwidth}  \color{black}{(c)}
      \hspace{0.46\textwidth}  \color{black}{(d)}
         \hfill}
     \vspace{0.3\textwidth}    
\caption{Coronal field images for the 3D simulation with (a) and (b) no
overlying field, (c) and (d) a 3 G overlying field. On the base, contours of
positive (red) and negative (green) magnetic field are plotted. A selection of
coronal magnetic field lines are shown in each case reaching maximum heights of $<2.5$ Mm (dark blue), $2.5-5$ Mm (magenta) and $>5$ Mm (pale blue). The images are taken at (a)
and (c) $t=128$ hr, (b) and (d) $t=168$ hr.}\label{fig:fl}
   \end{figure}

Figures~\ref{fig:fl}(a) and (b) show images from the simulation with no overlying field, (c) and (d) the 3 G overlying field simulation. The images on the
left are both taken at $t=128$ hr and the images on the right are shown at
$t=168$ hr. In all four images, contours of $B_z$ at $z=0$ Mm are plotted on the base (red=positive, green=negative). A selection of magnetic field lines are also over-plotted in each case,
where dark blue, magenta and pale blue field lines reach a maximum height of $<2.5$ Mm, $2.5-5$ Mm and $>5$ Mm respectively. For comparison, field lines in Figures~\ref{fig:fl}(c) and (d) are
plotted from the same photospheric starting points as field lines in (a) and (b) respectively. As expected, we find that connections between magnetic elements
within the simulation with no overlying field may reach much greater heights than those in the 3 G
simulation, as the overlying field suppresses the extension of the magnetic
elements' field into the corona. Also, the connectivity is quite different between the images at $t=128$ hr and $t=168$ hr, showing that the coronal field has changed significantly during this interval, as expected. The images here are intended only to give an
indication of what the connectivity is like between the magnetic elements. A more in
depth analysis will be carried out in future.

\subsection{Free Magnetic Energy}\label{sec:free}

  \begin{figure}

   \centerline{\hspace*{0.015\textwidth}
              \includegraphics[width=0.515\textwidth,clip=]{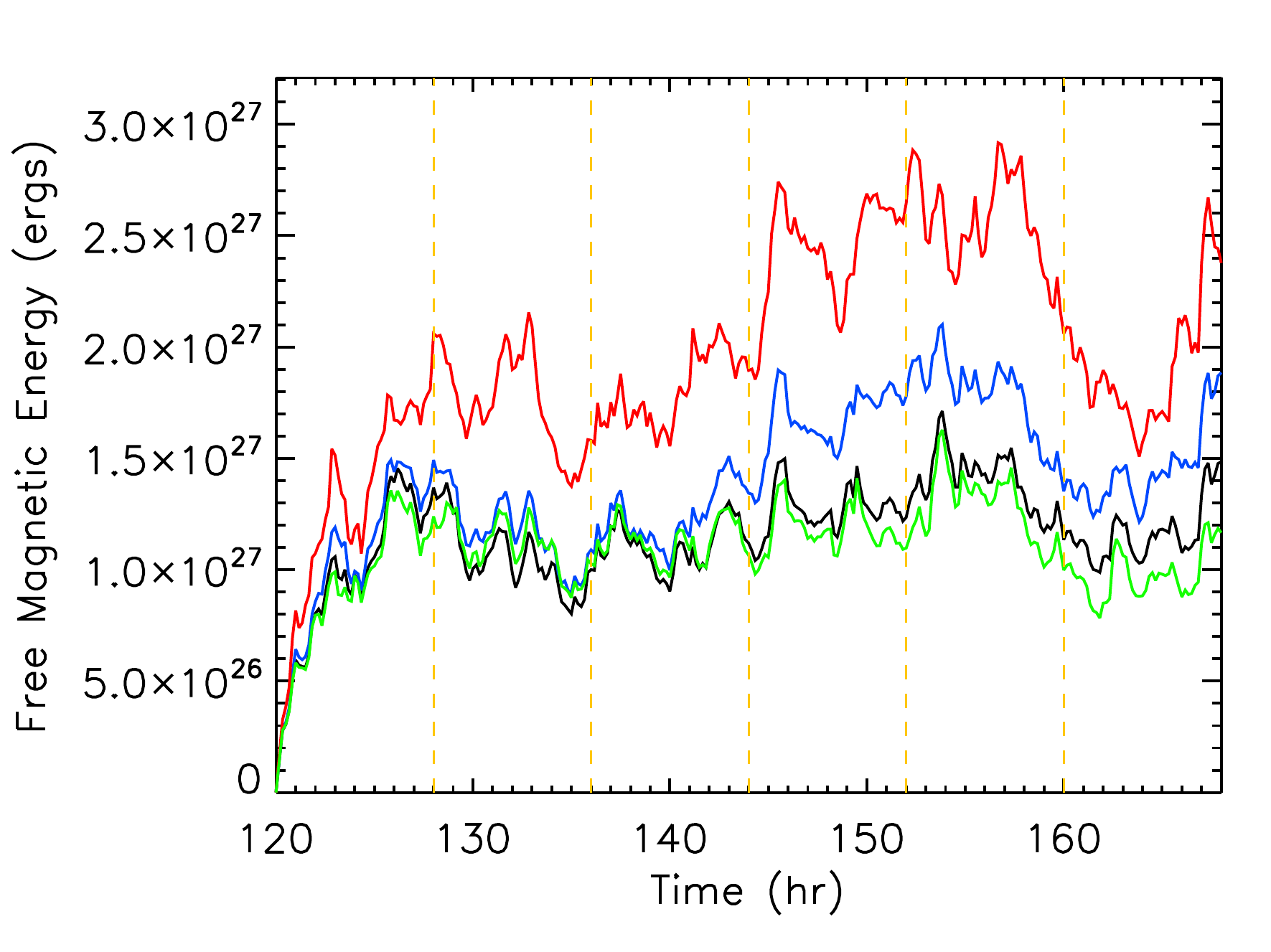}
               \hspace*{-0.01\textwidth}
              \includegraphics[width=0.515\textwidth,clip=]{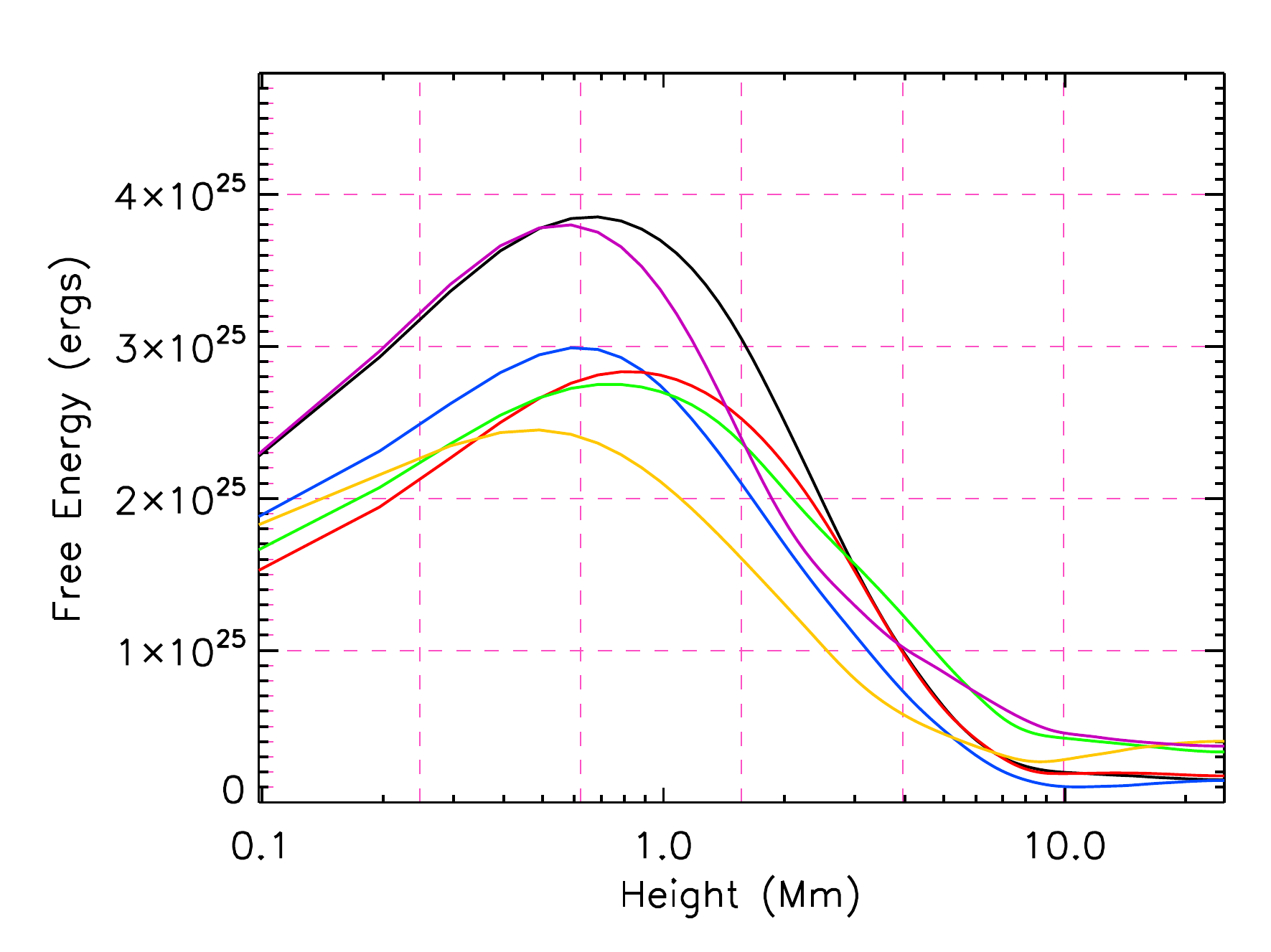}
              }
     \vspace{-0.4\textwidth}   
     \centerline{ \bf     
      \hspace{-0.03 \textwidth}  \color{black}{(a)}
      \hspace{0.46\textwidth}  \color{black}{(b)}
         \hfill}
     \vspace{0.37\textwidth}    

   \centerline{\hspace*{0.07\textwidth}
              \includegraphics[width=0.49\textwidth,clip=]{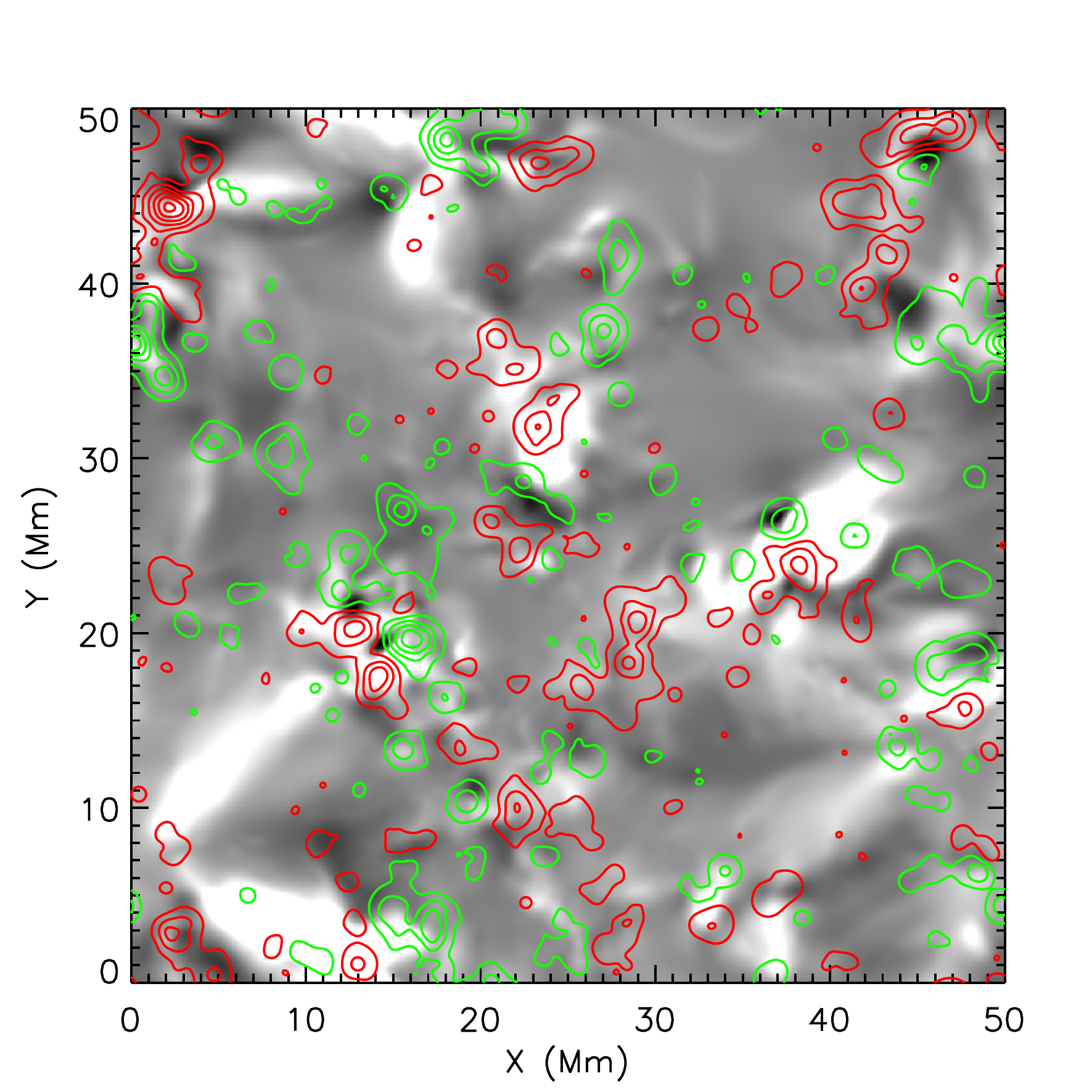}
               \hspace*{0.03\textwidth}
              \includegraphics[width=0.49\textwidth,clip=]{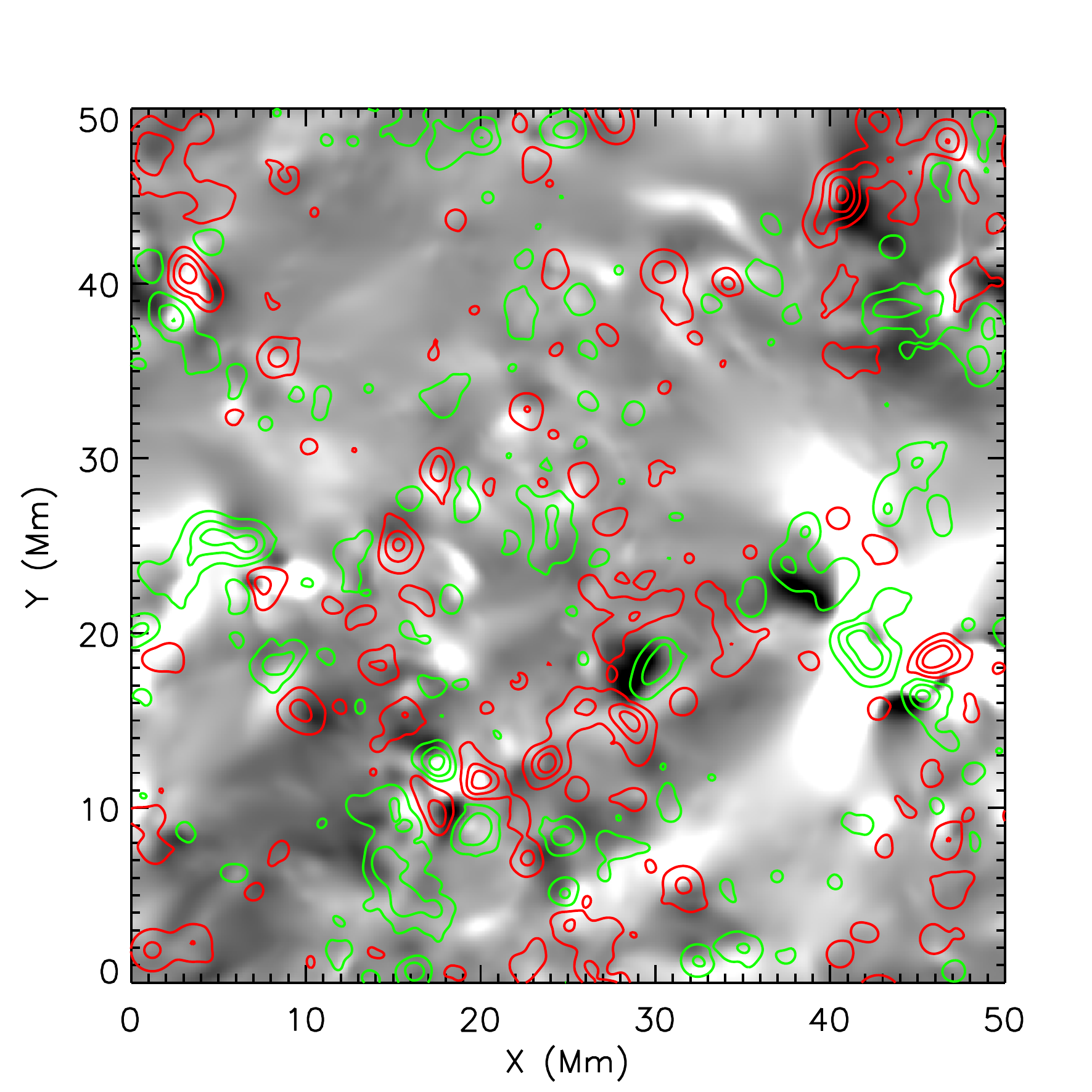}
              }
     \vspace{-0.45\textwidth}   
     \centerline{ \bf     
      \hspace{-0.03 \textwidth}  \color{black}{(c)}
      \hspace{0.46\textwidth}  \color{black}{(d)}
         \hfill}
     \vspace{0.42\textwidth}    
   \centerline{\hspace*{0.03\textwidth}
              \includegraphics[width=0.49\textwidth,clip=]{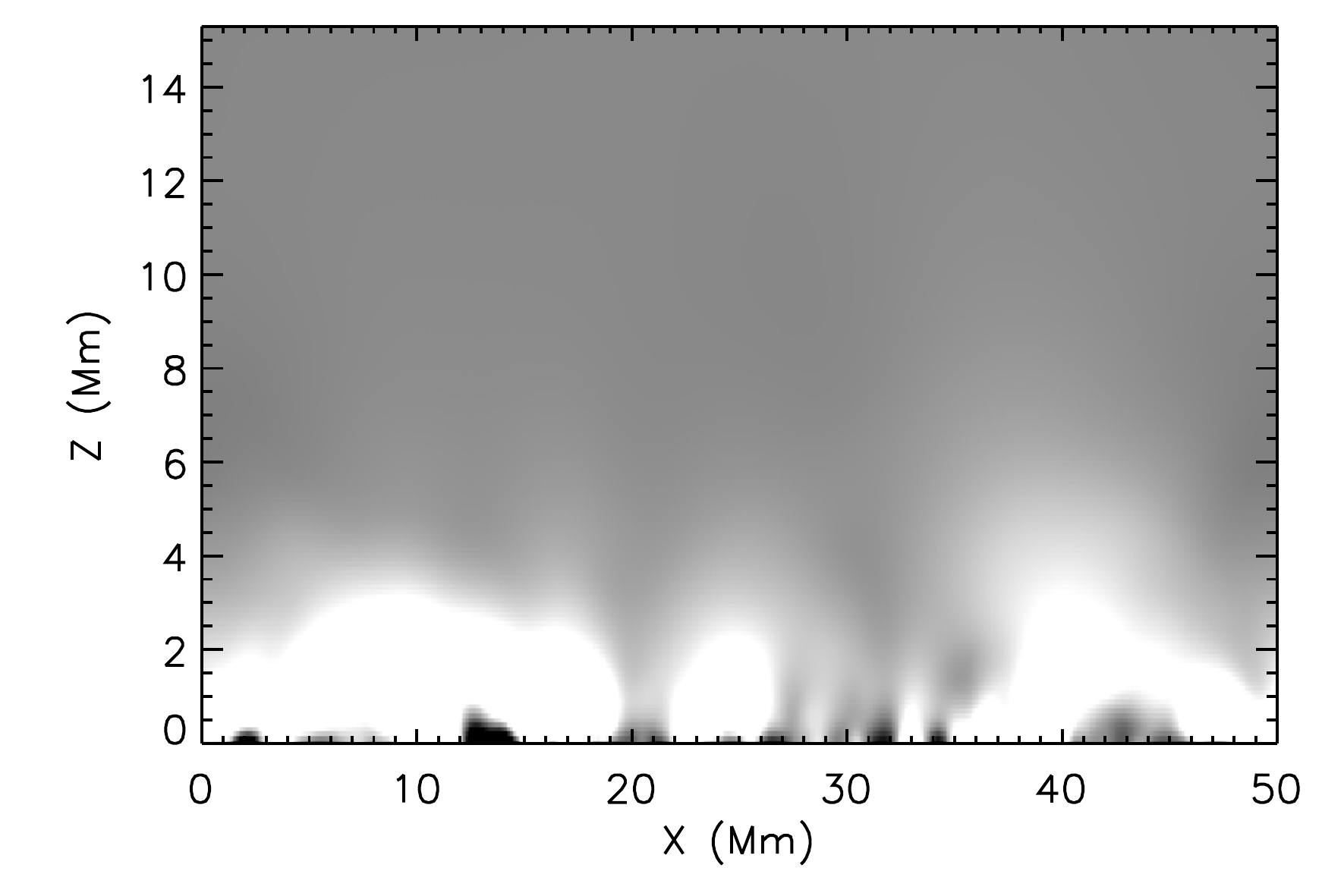}
               \hspace*{0.03\textwidth}
              \includegraphics[width=0.49\textwidth,clip=]{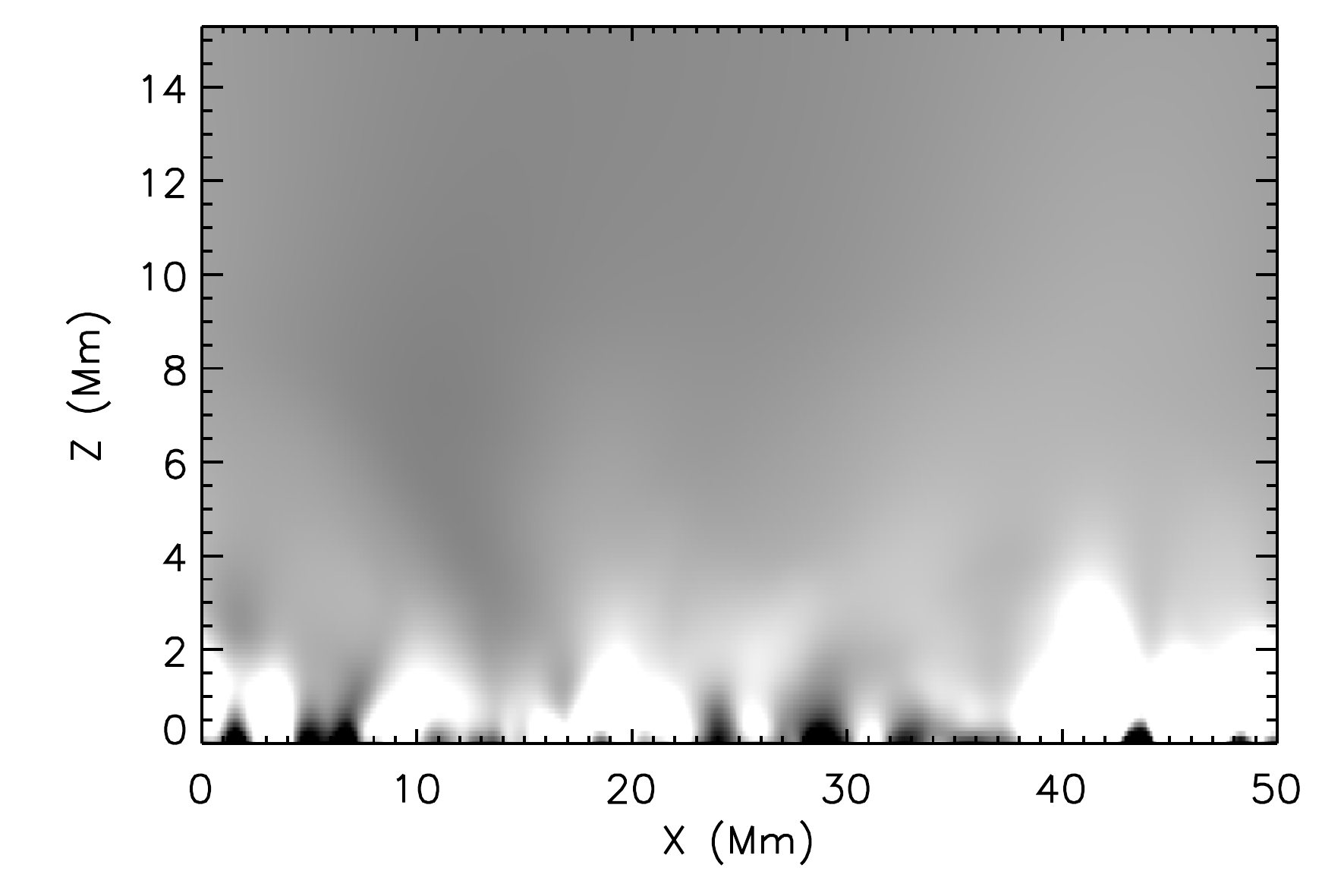}
              }
     \vspace{-0.34\textwidth}   
     \centerline{ \bf     
      \hspace{-0.03 \textwidth}  \color{black}{(e)}
      \hspace{0.46\textwidth}  \color{black}{(f)}
         \hfill}
     \vspace{0.31\textwidth}    

\caption{(a) Free magnetic energy as a function of time for the 3D simulations
with no overlying field (green), 1 G (black), 3 G (blue) and 10 G (red) overlying field. (b)
Total free magnetic energy as a function of height for the simulation with a 3 G
overlying field. The curves show the free energy at $t=128$ hr (black), $t=136$
hr (blue), $t=144$ hr (red), $t=152$ hr (green), $t=160$ hr (yellow) and $t=168$
hr (purple). (c) and (d) Free magnetic energy density integrated along the
LOS from above ($x-y$ plane), for the 3 G overlying field simulation, at (c) $t=128$ hr and (d) $t=168$ hr. White denotes regions where the free energy density is positive and black where it's negative,
saturated at $\pm 1.9\times 10^{22}$ ergs.
Positive (red) and negative (green) contours of $B_z$ at $z=0$ Mm are over-plotted at levels of
$\pm[7, 13, 27, 53, 106]$ G. (e) and (f) Free energy density integrated along the LOS, viewed from the side ($x-z$ plane)
saturated at $\pm 4.8\times 10^{22}$ ergs, at (e) $t=128$ hr and (f) $t=168$
hr.}\label{fig:free1}
   \end{figure}

The presence of free magnetic energy within our 3D model is a significant
difference to the potential field (minimum energy) models of the magnetic carpet coronal field discussed in the introduction.
Figure~\ref{fig:free1}(a) shows a plot of the free magnetic energy (ergs) as a
function of time, defined as
\begin{equation}
 E_\textrm{\small f}(t)=
\int_V \frac{|\BB_{\textrm{\small nl}}|^2-|\BB_\textrm{\small p}|^2}{8\pi} dV,
\end{equation}
where $\BB_{\textrm{\small nl}}$ is the non-linear
force-free magnetic field and $\BB_{\textrm{\small p}}$ is the magnetic field of the
corresponding potential field.
Results are shown for the no overlying field (green), 1 G (black), 3 G (blue) and 10 G (red)
simulations. For each simulation, the free energy initially increases rapidly as the
coronal field diverges from a potential state due to surface motions. For the no overlying field and 1 G cases, the free energy then levels off and oscillates around a mean
value of $1.13\times 10^{27}$ ergs and  $1.20\times 10^{27}$ ergs, respectively, with standard deviations of $1.52\times 10^{26}$ ergs and $1.74\times 10^{26}$ ergs.
For the 3 G and 10 G cases, the free energy also levels off to a lesser extent,
but is less steady, with mean values of $1.46\times 10^{27}$ ergs and  $2.10\times 10^{27}$ ergs and standard deviations of $2.84\times 10^{26}$ ergs and $4.06\times 10^{26}$ ergs respectively. The mean and maximum values of free energy per unit area for each simulation are given in Table~\ref{tab:free}. The trend is that a stronger
overlying field leads to a greater build-up of free energy. In each case, the individual peaks in free energy differ between the simulations, however the general shape of the curve is the same for all four overlying field strengths. Therefore the overall behaviour of the free magnetic energy
is largely dependent on the evolution of the photospheric magnetic field, rather
than the strength of the overlying coronal field.

\begin{table}
\begin{center}
\begin{tabular}{ccc}
\hline
Simulation & Mean Free Magnetic & Maximum Free Magnetic  \\
($B_0$) & Energy ($\times 10^{7}$ ergs cm$^{-2}$) & Energy ($\times 10^{7}$ ergs cm$^{-2}$) \\
\hline
0 G  & 4.53 & 6.51 \\
1 G  & 4.79 & 6.85 \\
3 G  & 5.85 & 8.40 \\
10 G & 8.38 & 11.67 \\
\hline
\end{tabular}
\caption{Mean and maximum values of free magnetic energy for each
simulation.}\label{tab:free}
\end{center}
\end{table}

The variation of free energy with height is similar for all simulations, so here we just discuss the results from the 3 G overlying field simulation.
Figure~\ref{fig:free1}(b) shows plots of the free magnetic energy
integrated over $x$ and $y$, as a function of height, for the simulation with a 3
G overlying field.
This is computed as follows:
\begin{displaymath}
 E_\textrm{\small f}(z)=L_z\int_{y_{\textrm{\tiny min}}}^{y_{\textrm{\tiny
max}}}\int_{x_{\textrm{\tiny min}}}^{x_{\textrm{\tiny max}}}
\frac{|\BB(x,y,z)_{\textrm{\small nl}}|^2-|\BB(x,y,z)_\textrm{\small p}|^2}{8\pi} dx dy,
\end{displaymath}
where $x_{\textrm{\small min}}=y_{\textrm{\small min}}=0$ Mm, $x_{\textrm{\small
max}}=y_{\textrm{\small max}}=50$ Mm and $L_z=0.098$ Mm is the length of a cell
in $z$.
The total free magnetic energy as a function of height is shown at times
spaced evenly throughout the simulations, at $t=128$ hr (black), $t=136$ hr
(blue), $t=144$ hr (red), $t=152$ hr (green), $t=160$ hr (yellow) and $t=168$ hr
(purple). In Figure~\ref{fig:free1}(a), which shows the total free magnetic
energy as a function of time, the vertical yellow dashed lines are over-plotted on
the graph at intervals of 8 hr, indicating the times at which the lines in
Figure~\ref{fig:free1}(b) are taken. In Figure~\ref{fig:free1}(b), it can be seen that the ordering of the
curves is not time dependent.
Once the coronal field has evolved away from its initial potential state, the
total amount of free magnetic energy within the volume depends upon how much is
both built up and stored due to surface motions. For each of the curves in
Figure~\ref{fig:free1}(b), there is a peak between roughly $z=0.5$ Mm and
$z=0.8$ Mm, indicating that this is where the majority of the free magnetic
energy is stored. The free energy then drops off rapidly after $z=1$ Mm.
As in Paper II, we find that the field departs most from a potential state low
down in the corona, as this is close to where we are driving the evolution of
the field by photospheric motions. In addition, most closed connections between
magnetic elements (as opposed to connections from the magnetic elements to the overlying field) are found to be low lying ({\it e.g.} \inlinecite{priest2002},
\inlinecite{close2003}), and it is along these connections that free energy is
stored.
A movie showing the time evolution of the free magnetic energy as a function of
height for the 3 G simulation is available
(\textcolor{blue}{magnet48b\_free\_ht.mpg}). The movie shows that the free energy
is highly dynamic and rapidly evolving. The height of the curve is continually
changing, however it can be seen that the peak in the curve tends to remain
between $z=0.5$ Mm and $z=0.8$ Mm.

Figures~\ref{fig:free1}(c) and (d) show the free magnetic energy density,
$\frac{B_{\textrm{\tiny nl}}^2-B_\textrm{\tiny p}^2}{8\pi}$, integrated in $z$,
in the $x-y$ plane for the 3 G simulation at $t=128$ hr and $t=168$ hr, respectively. This is computed as follows:
\begin{displaymath}
 E_\textrm{\small f}(x,y)=A \int_{z_{\textrm{\tiny min}}}^{z_{\textrm{\tiny
max}}} \frac{|\BB(x,y,z)_{\textrm{\small nl}}|^2-|\BB(x,y,z)_\textrm{\small p}|^2}{8\pi}
dz,
\end{displaymath}
where $A=L_x\; L_y$, $L_x=0.098$ Mm is the length of a cell in $x$ and $L_y=0.098$
Mm is the length of a cell in $y$. White patches indicate where the free
energy density is positive, {\it i.e.} where $B^2_{\textrm{\small nl}} >
B^2_{\textrm{\small p}}$, black where the free energy
density is negative, $B^2_{\textrm{\small nl}} < B^2_{\textrm{\small p}}$. Note
that the total free magnetic energy integrated over the volume is always
positive (see Figure~\ref{fig:free1}(a)). Positive (red) and negative (green) contours of $B_z$ at $z=0$ Mm are
over-plotted. We define free magnetic energy to be `stored' at
locations where the line-of-sight (LOS) integrated free magnetic energy density is
positive. From these images, it can be seen that free magnetic energy may be
stored both at the boundaries between supergranular cells and within the cells
themselves.

\begin{figure}

   \centerline{\hspace*{0.01\textwidth}
              \includegraphics[width=0.5\textwidth,clip=]{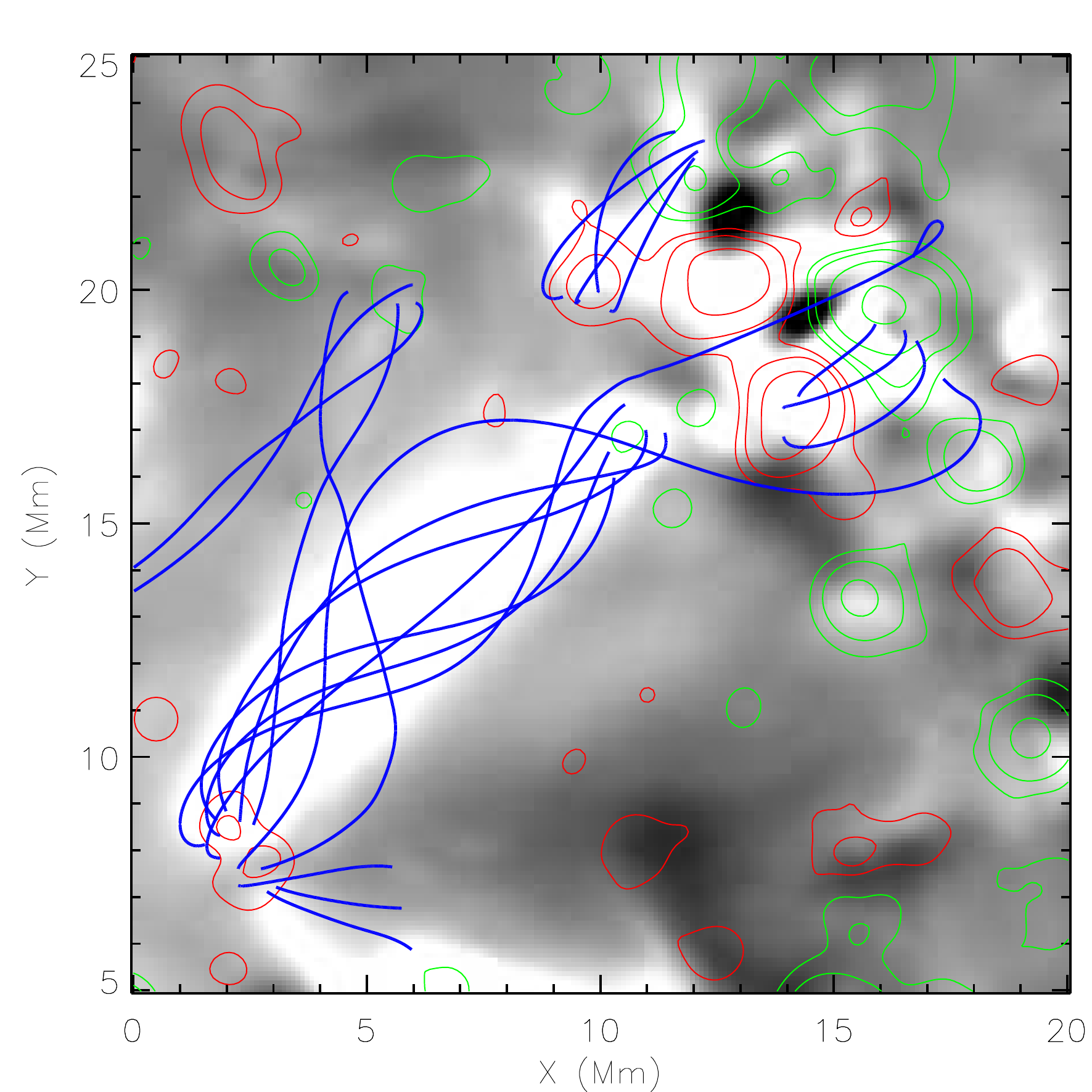}
               \hspace*{0.0\textwidth}
              \includegraphics[width=0.5\textwidth,clip=]{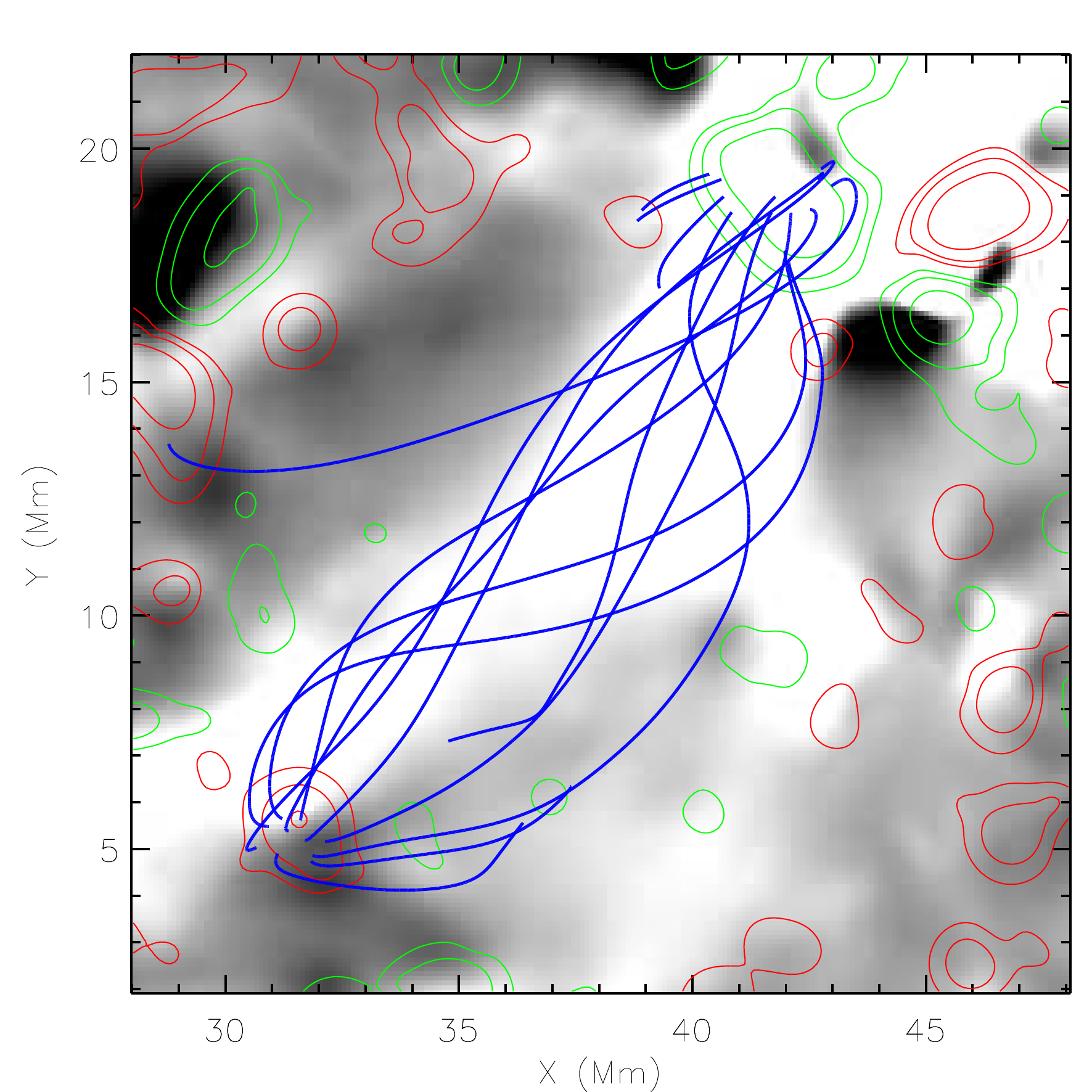}
              }
     \vspace{-0.48\textwidth}   
     \centerline{ \bf     
      \hspace{-0.03 \textwidth}  \color{black}{(a)}
      \hspace{0.46\textwidth}  \color{black}{(b)}
         \hfill}
     \vspace{0.38\textwidth}    

   \centerline{\hspace*{0.02\textwidth}
              \includegraphics[width=0.5\textwidth,clip=]{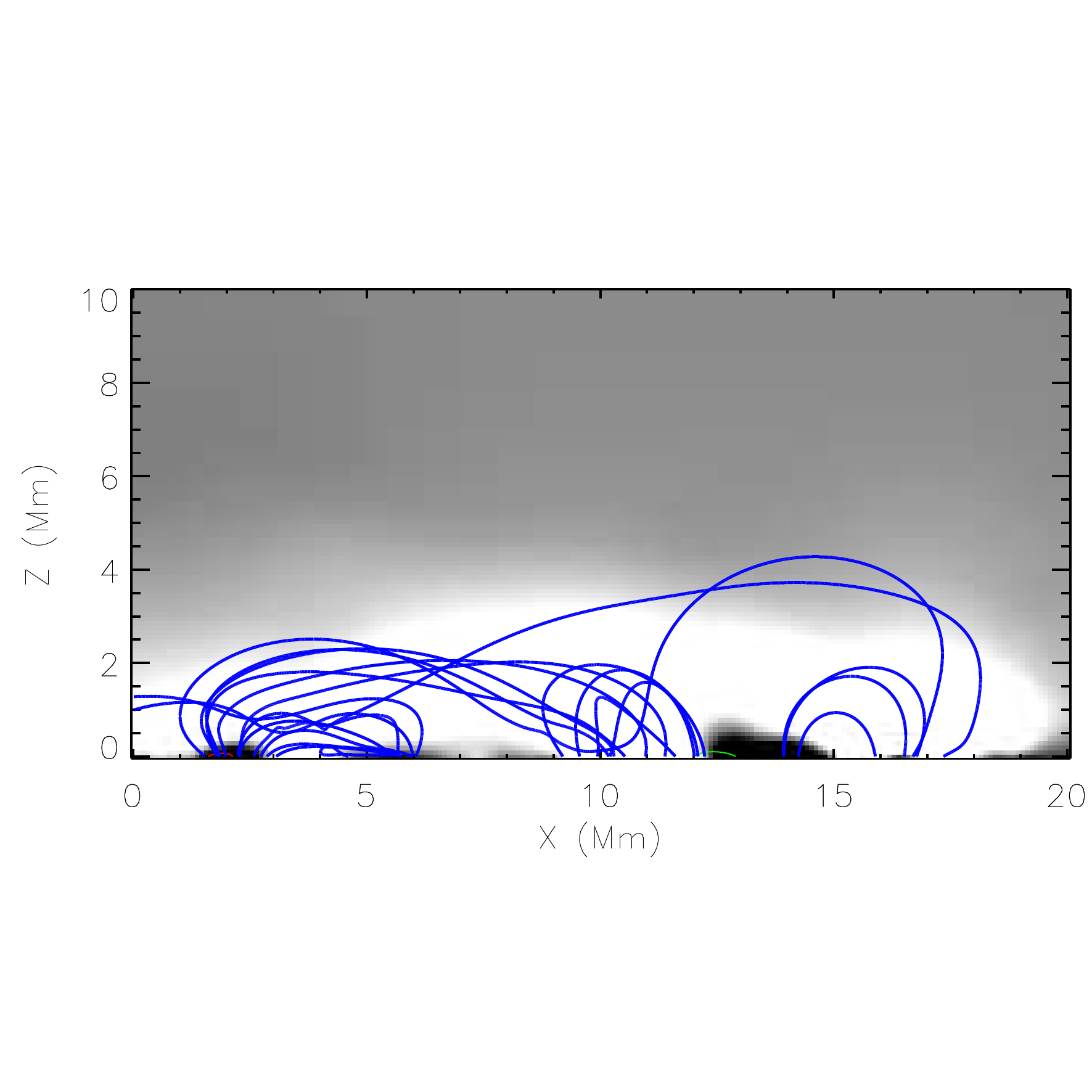}
               \hspace*{0.0\textwidth}
              \includegraphics[width=0.5\textwidth,clip=]{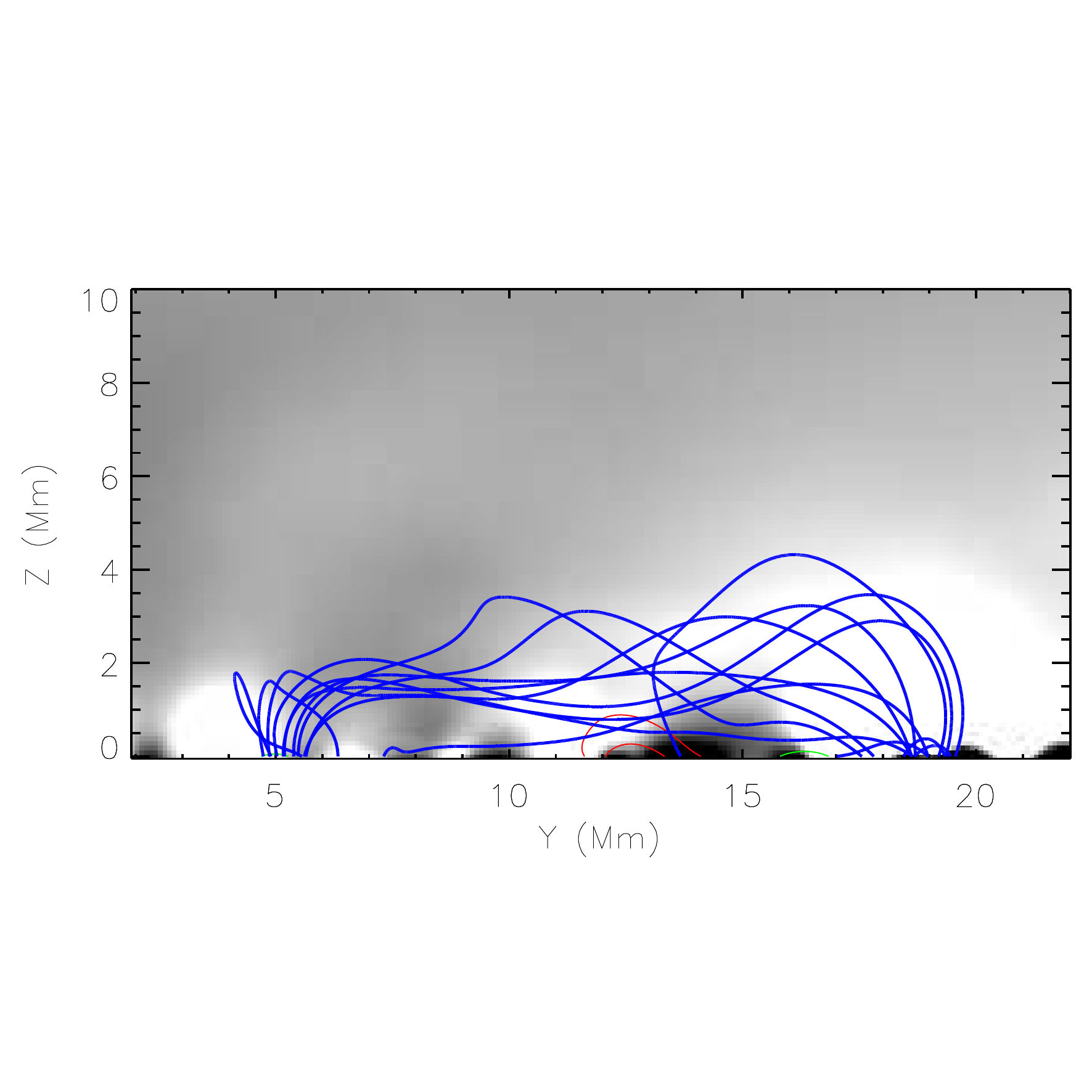}
              }
     \vspace{-0.4\textwidth}   
     \centerline{ \bf     
      \hspace{-0.03 \textwidth}  \color{black}{(c)}
      \hspace{0.46\textwidth}  \color{black}{(d)}
         \hfill}
     \vspace{0.26\textwidth}    

\caption{All images are for the 3 G simulation. (a) and (b) Free magnetic energy
density integrated in $z$, shown in the $x-y$ plane. The images are white in
regions where the free energy density is positive, black where the free energy
density is negative. Positive (red) and negative (green) contours of $B_z$ at $z=0$ Mm are over-plotted at levels of $\pm[7, 13, 27, 53, 106]$ G. The images are shown at (a)
$t=128$ hr and (b) $t=168$ hr. (c) Free magnetic energy integrated in $y$, shown
in the $x-z$ plane at $t=128$ hr. (d) Free magnetic energy density integrated in
$x$, shown in the $y-z$ plane at $t=168$ hr. A selection of field lines are
over-plotted in blue on each image.}\label{fig:free2}
   \end{figure}

In Paper II, it was found that when the evolution of the magnetic elements disturbed a larger
volume of the overlying field, a greater amount of free energy was built up. It
was also found that closed connections between the magnetic elements are
required, along which the free energy may be stored. In agreement with Paper II,
we find that free energy is stored mainly in two locations. Firstly, we see that
white patches are located around supergranule cell boundaries where the magnetic
network is formed. Large numbers of magnetic elements are swept to these
locations by supergranular flows, and continually interact with one another,
cancelling, coalescing and fragmenting. This continual evolution of the magnetic
elements results in a large build up of free energy, which may then be stored
along the multiple connections that form between nearby magnetic elements that
lie in the network. The second location where we see free energy stored is along
long-lived, far-reaching, twisted connections between magnetic elements. Such
connections may stretch across supergranule cells, between magnetic elements
located at opposite boundaries (examples of this will be given later, in
Figure~\ref{fig:free2}). Longer connections will clearly disturb a larger volume
of the surrounding coronal magnetic field, hence building up free energy. A
movie showing the free magnetic energy density, integrated in $z$, in the $x-y$
plane is available
(\textcolor{blue}{magnet48b\_free\_xy\_bz.mpg}), with contours of $B_z$ at $z=0$ Mm over-plotted. Within the
movie, one can see that regions of positive free energy density are continually
evolving in response to photospheric motions. In particular, occasional large patches of
positive free energy density develop around the magnetic network. One can also
see long-lived bands of positive free magnetic energy density stretching across supergranules,
between magnetic elements. Many of the regions of free magnetic energy density last for several hours. For readers unable to view the movie, six still images spaced 1 hr apart are included in Appendix~\ref{sec:app1} (Figure~\ref{fig:app1}), to give an impression of the time-scale of the evolution.

Figures~\ref{fig:free1}(e) and (f) show $x-z$ plane images of the LOS
integrated free magnetic energy density, saturated at $\pm 4.8\times 10^{22}$
ergs, for the 3 G simulation at $t=128$ hr and
$t=168$ hr, respectively. A movie of the free magnetic energy density in the $x-z$ plane,
integrated in $y$, is available
(\textcolor{blue}{magnet48b\_free\_xz.mpg}). In both the movie and the still images, it can be
seen that the free energy is mainly located low down, with the bulk of it being
below $z=3$ Mm. This is where many closed connections exist between the magnetic
elements and a larger departure of the magnetic field from a potential state is
found. One can see that the locations of positive free energy density are highly
dynamic and there exist long-lived `bulbs' of positive free energy density,
where it is stored within the corona along closed connections between magnetic
elements. Similar evolution of the LOS integrated free magnetic energy
density is seen in the $y-z$ plane and for different strengths of overlying
field.

Figures~\ref{fig:free2}(a) and (b) show two zoomed in sections of the $x-y$
plane images of the LOS integrated free energy density taken at
$t=128$ hr and $t=168$ hr, respectively, from Figures~\ref{fig:free1}(c) and (d). A selection of closed field lines
has been over-plotted in blue in each case.
Figure~\ref{fig:free2}(a) shows the band of positive free energy density that
can be seen lying across the lower left supergranule in
Figure~\ref{fig:free1}(c), while Figure~\ref{fig:free2}(b) shows the band of
positive free energy density across the lower right supergranule in
Figure~\ref{fig:free1}(d). In both zoomed images, twisted magnetic field
lines connect between various magnetic elements on
either side of the supergranule. The free energy is stored in these regions of complex
connections. Figure~\ref{fig:free2}(c) shows an $x-z$ plane image of free energy
density integrated in $y$ at $t=128$ hr, and is a side view of the band of
positive free energy density in Figure~\ref{fig:free2}(a). Similarly,
Figure~\ref{fig:free2}(d) is a side view of the band of positive free energy
density in Figure~\ref{fig:free2}(b) and shows a $y-z$ plane image of free
energy density integrated in $x$ at $t=168$ hr. For each of the cases, a complex
structure of the magnetic field can be seen.

  \begin{figure}[!h]

   \centerline{\hspace*{0.07\textwidth}
              \includegraphics[width=0.49\textwidth,clip=]{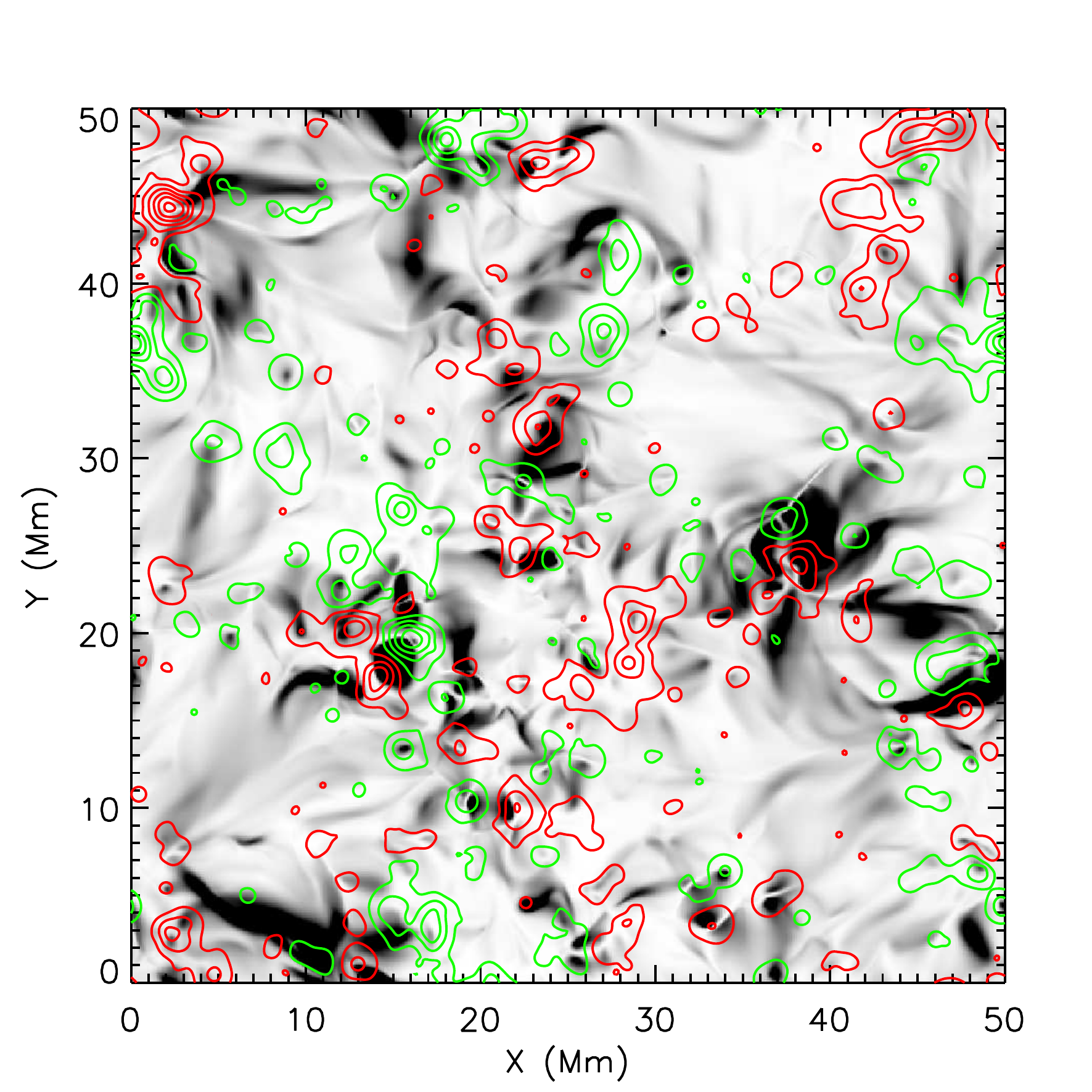}
               \hspace*{0.03\textwidth}
              \includegraphics[width=0.49\textwidth,clip=]{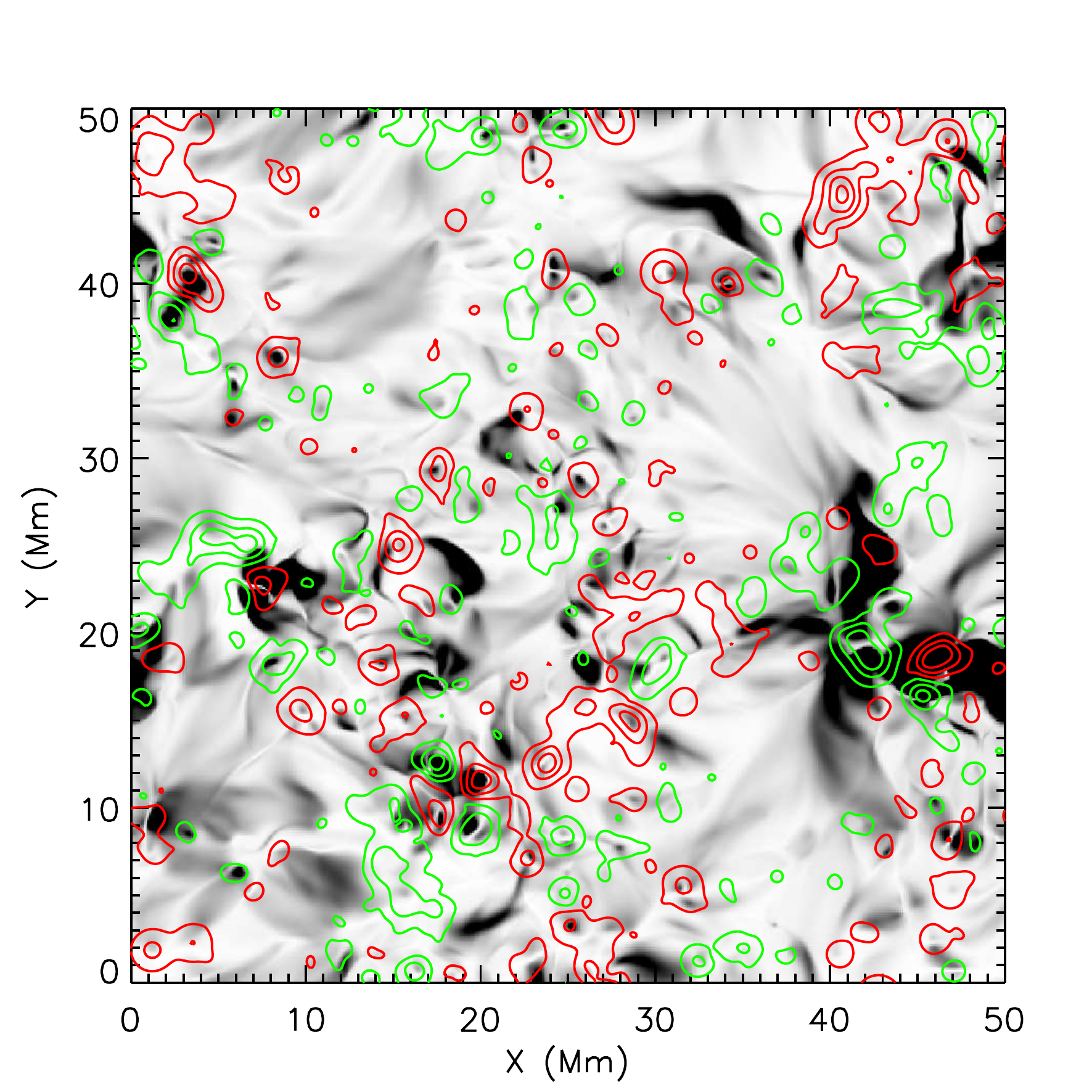}
              }
     \vspace{-0.45\textwidth}   
     \centerline{ \bf     
      \hspace{-0.03 \textwidth}  \color{black}{(a)}
      \hspace{0.46\textwidth}  \color{black}{(b)}
         \hfill}
     \vspace{0.45\textwidth}    

   \centerline{\hspace*{0.03\textwidth}
              \includegraphics[width=0.49\textwidth,clip=]{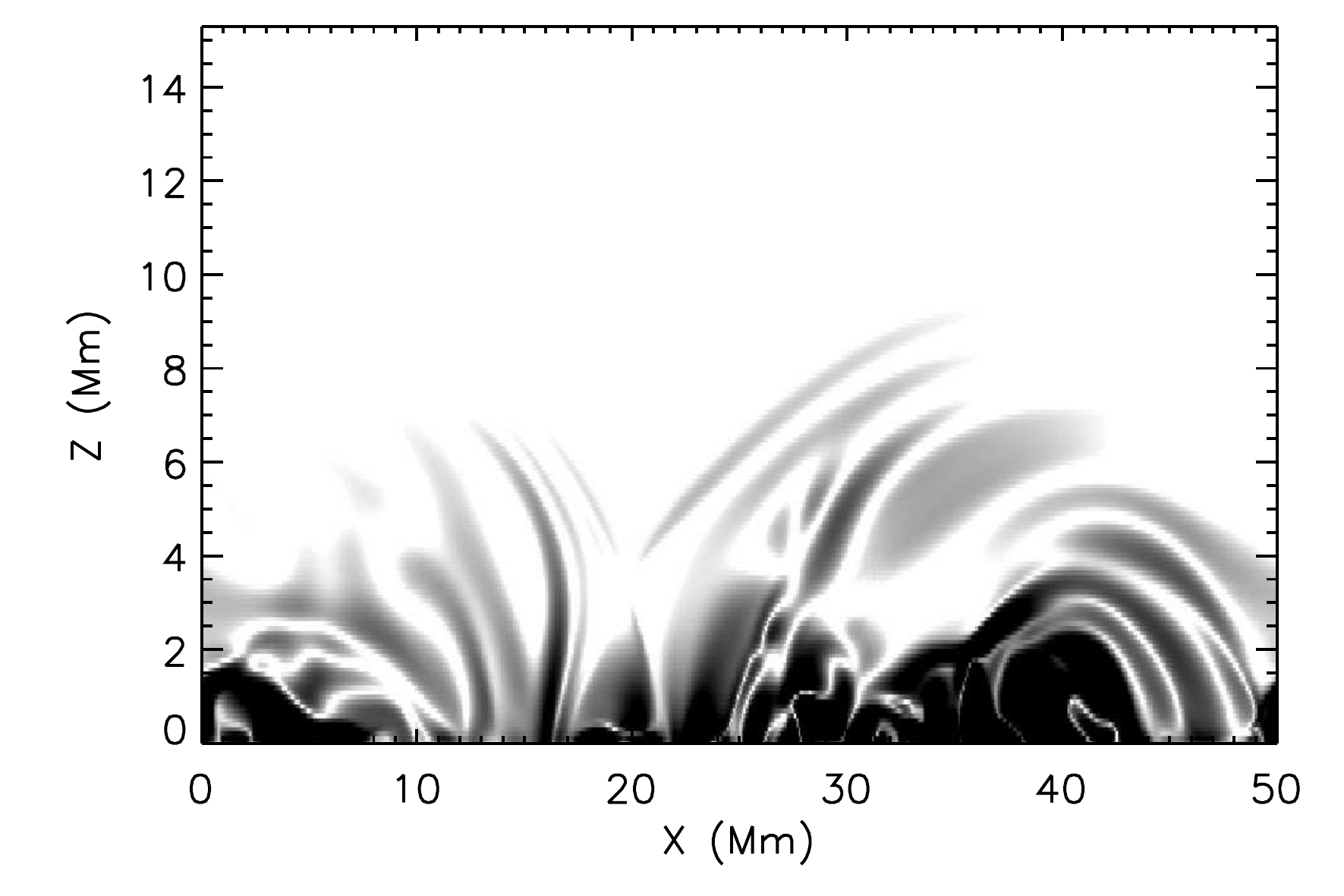}
               \hspace*{0.03\textwidth}
              \includegraphics[width=0.49\textwidth,clip=]{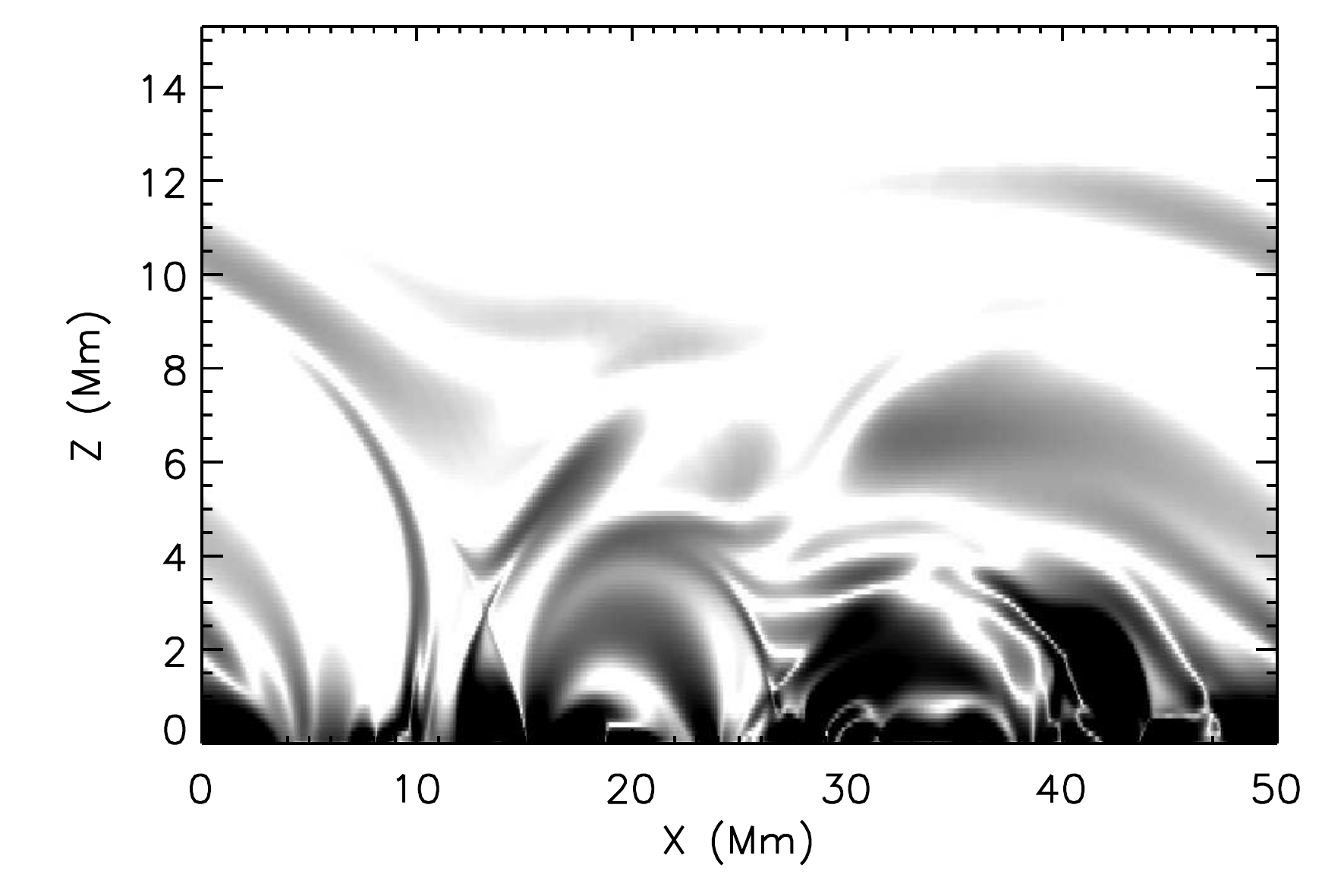}
              }
     \vspace{-0.34\textwidth}   
     \centerline{ \bf     
      \hspace{-0.03 \textwidth}  \color{black}{(c)}
      \hspace{0.46\textwidth}  \color{black}{(d)}
         \hfill}
     \vspace{0.34\textwidth}    
   \centerline{\hspace*{0.03\textwidth}
              \includegraphics[width=0.49\textwidth,clip=]{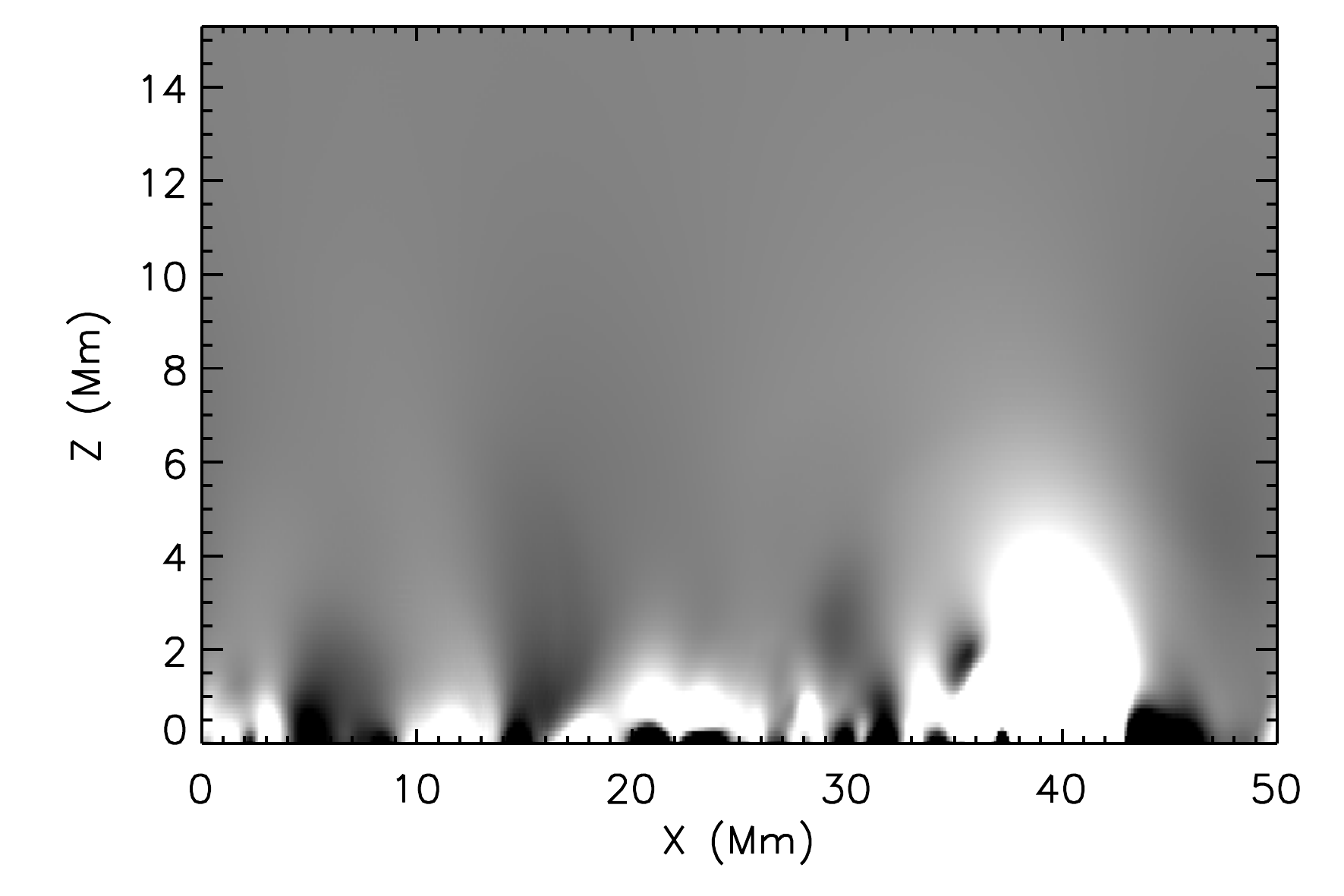}
               \hspace*{0.03\textwidth}
              \includegraphics[width=0.49\textwidth,clip=]{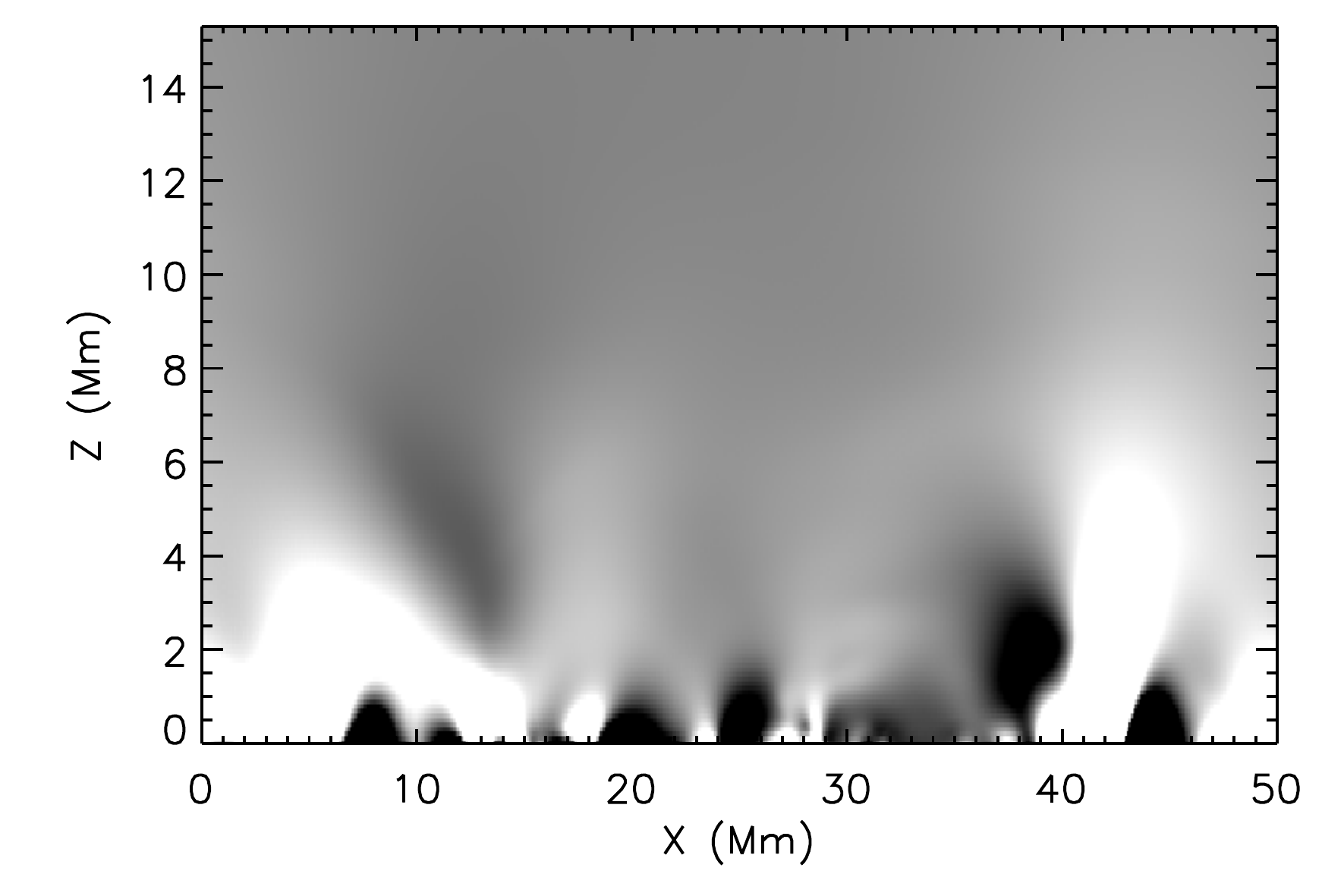}
              }
     \vspace{-0.34\textwidth}   
     \centerline{ \bf     
      \hspace{-0.03 \textwidth}  \color{black}{(e)}
      \hspace{0.46\textwidth}  \color{black}{(f)}
         \hfill}
     \vspace{0.31\textwidth}    
\caption{All images are for the 3 G simulation. For (a)-(d), darker regions correspond to higher values. (a) and (b) Normalised $j^2$
integrated in $z$, shown in the $x-y$ plane. Contours of positive (red) and negative (green) magnetic field at $z=0$ Mm are over-plotted, at levels of $\pm[7, 13, 27, 53, 106]$ G. (c)
and (d) logarithm of normalised $j^2$ in the $x-z$ plane at $y=25$ Mm. (e) and
(f) free magnetic energy density shown in the $x-z$ plane at $y=25$ Mm,
saturated at $\pm 1.9\times 10^{20}$ ergs. Images in the left-hand column are
shown at $t=128$ hr and in the right-hand column at $t=168$ hr.}\label{fig:j}
   \end{figure}

\subsection{Current Density}\label{sec:j}

The square of the current density, $j^2$, is of interest because it indicates
locations of possible Ohmic heating, $\frac{j^2}{\sigma}$. However, it should be noted that in the present simulations, due to the simplified form used, we do not have an energy equation. Figures~\ref{fig:j}(a) and (b) show images in the $x-y$ plane of $j^2$ integrated in $z$, at $t=128$ hr and $t=168$ hr, respectively. This is computed as
follows:
\begin{displaymath}
 E_j(x,y)=\int_{z_{\textrm{\tiny min}}}^{z_{\textrm{\tiny max}}}j(x,y,z)^2 dV.
\end{displaymath}
Note that the colour table has been reversed, so that darker regions correspond to higher values of $E_j(x,y)$.
Contours of $B_z$ at $z=0$ Mm
are over-plotted at the same levels as in Figures~\ref{fig:free1}(c) and (d). On comparison
with the $x-y$ plane images of free magnetic energy in
Figures~\ref{fig:free1}(c) and (d), the locations of high $j^2$ and of positive
free magnetic energy density seem to match very well. The regions of high $j^2$
appear to be strongest in the magnetic network.
Within a force-free field, by definition, $\Bj\times\BB=0$, hence $\Bj$ is
parallel to $\BB$. This means that we can express $\Bj$ as a scalar multiple of $\BB$, $\Bj=\alpha\BB$,
where $\alpha$ is a scalar representing the twist of the magnetic field with respect to
the corresponding potential field. Therefore, it makes sense that $j^2$ is at its
largest near the magnetic sources, where $\BB$ is largest.
We also see fainter bands of $j^2$ stretching across supergranules, often in the
same places as bands of positive free magnetic energy density. Again, it makes
sense for $j^2$ to be high in such locations as free magnetic energy is built up
in regions of high non-potentiality (large $|\alpha|$) which arise due to
non-zero $\Bj$ ($=\alpha\BB$). The evolution of $j^2$, integrated in $z$, in the $x-y$
plane can be seen in the movie,
\textcolor{blue}{magnet48b\_j\_xy\_bz.mpg}, for the 3 G simulation.
Six still images from this movie, spaced 1 hr apart, are included in Appendix~\ref{sec:app1} (Figure~\ref{fig:app2}), to give an impression of the evolution for those who cannot view the movie. Although regions of $j^2$ are often co-located with regions of positive free energy density, the regions of $j^2$ appear to be more rapidly evolving than those of free energy. The spatial distribution of $j^2$ is quite different from one hour to the next.

Figures~\ref{fig:j}(c) and (d) show images of the logarithm of $j^2$ in the
$x-z$ plane at $y=25$ Mm, at $t=128$ hr and $t=168$ hr, respectively. The regions of $j^2$ appear to be well structured, and obviously follow
the shape of the magnetic field (as $\Bj$ is parallel to $\BB$).
Figures~\ref{fig:j}(e) and (f) show images of the free magnetic energy density
at $y=25$ Mm, shown at the same times as (c) and (d) respectively. It can be
seen that regions of positive free energy density tend to be co-located
with regions of high $j^2$.

\subsection{Energy Dissipated}\label{sec:q}

  \begin{figure}[!h]

   \centerline{\hspace*{0.015\textwidth}
              \includegraphics[width=0.515\textwidth,clip=]{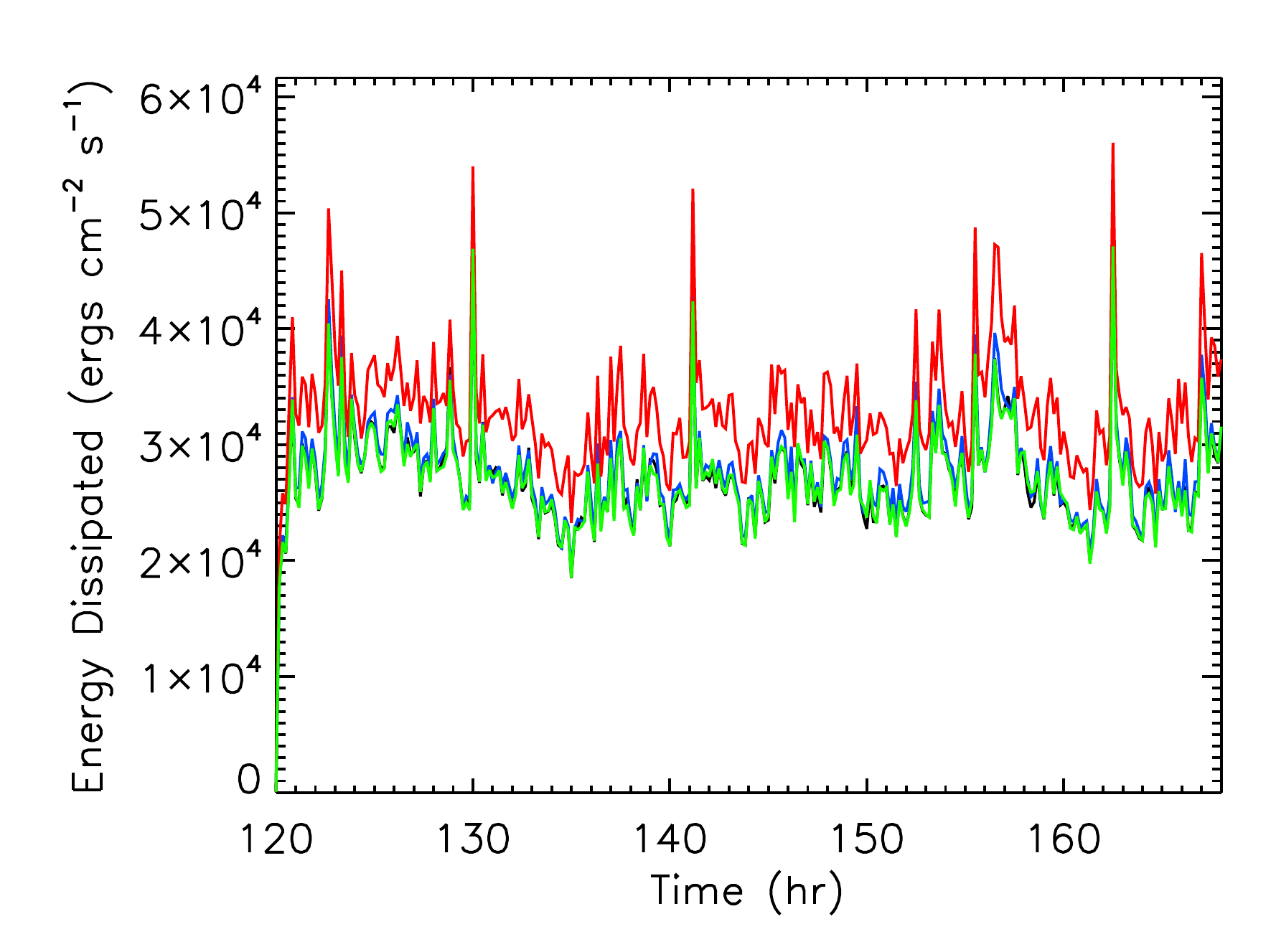}
               \hspace*{0.03\textwidth}
               \includegraphics[width=0.515\textwidth,clip=]{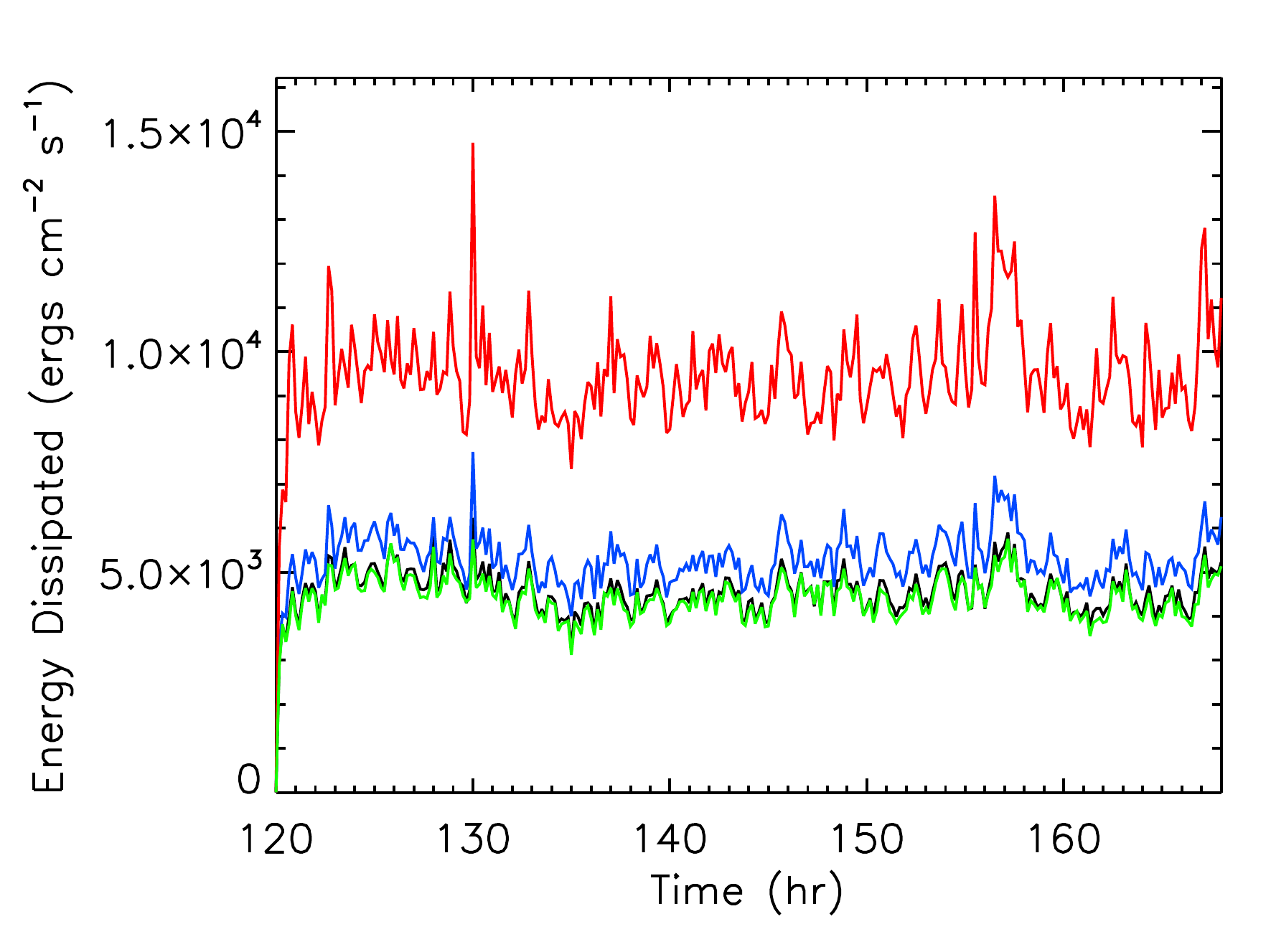}
              }
     \vspace{-0.39\textwidth}   
     \centerline{ \bf     
      \hspace{-0.08 \textwidth}  \color{black}{(a)}
      \hspace{0.49\textwidth}  \color{black}{(b)}
         \hfill}
     \vspace{0.36\textwidth}    
   \centerline{\hspace*{0.015\textwidth}
               \includegraphics[width=0.515\textwidth,clip=]{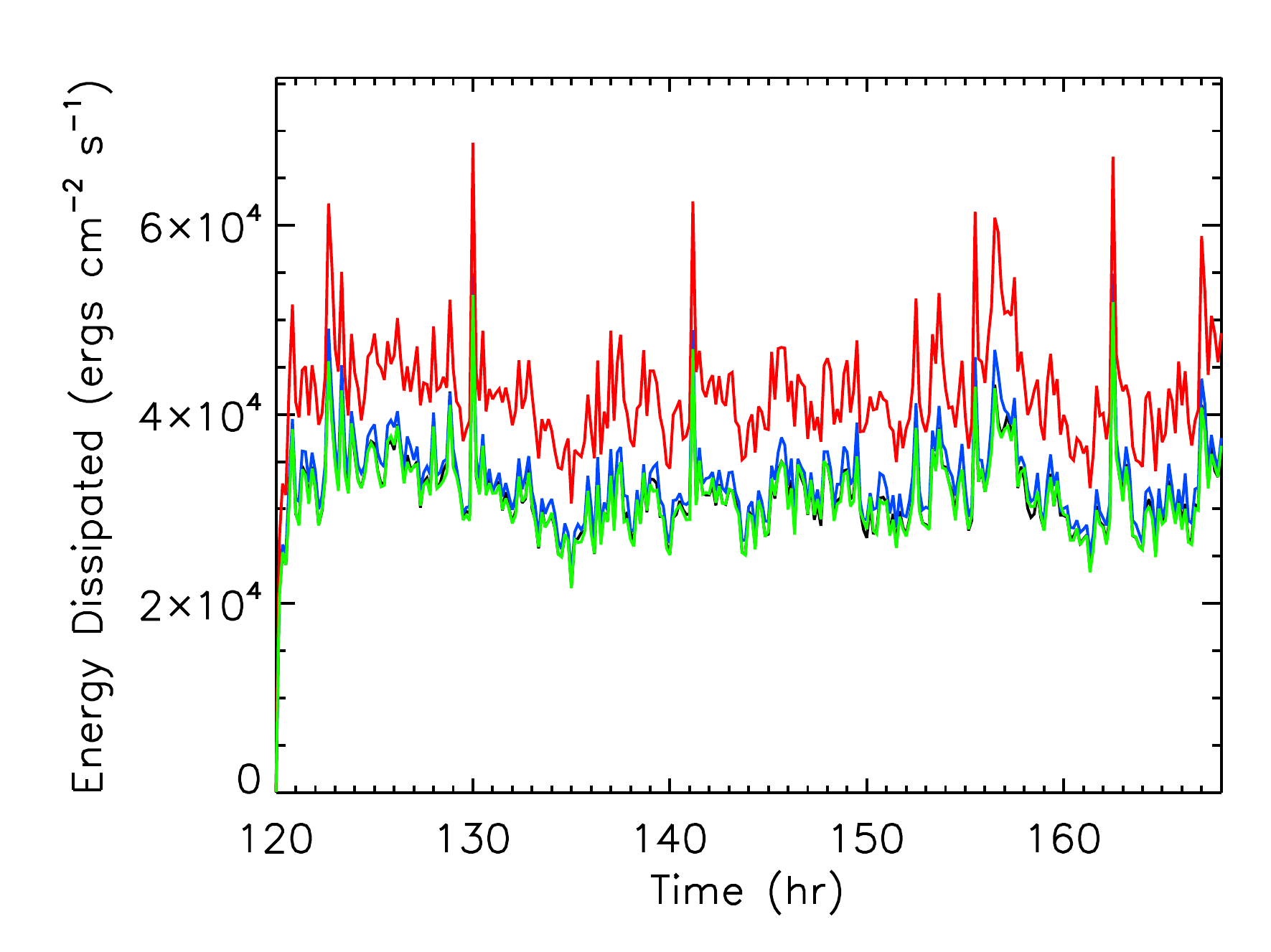}
               \hspace*{0.03\textwidth}
              \includegraphics[width=0.515\textwidth,clip=]{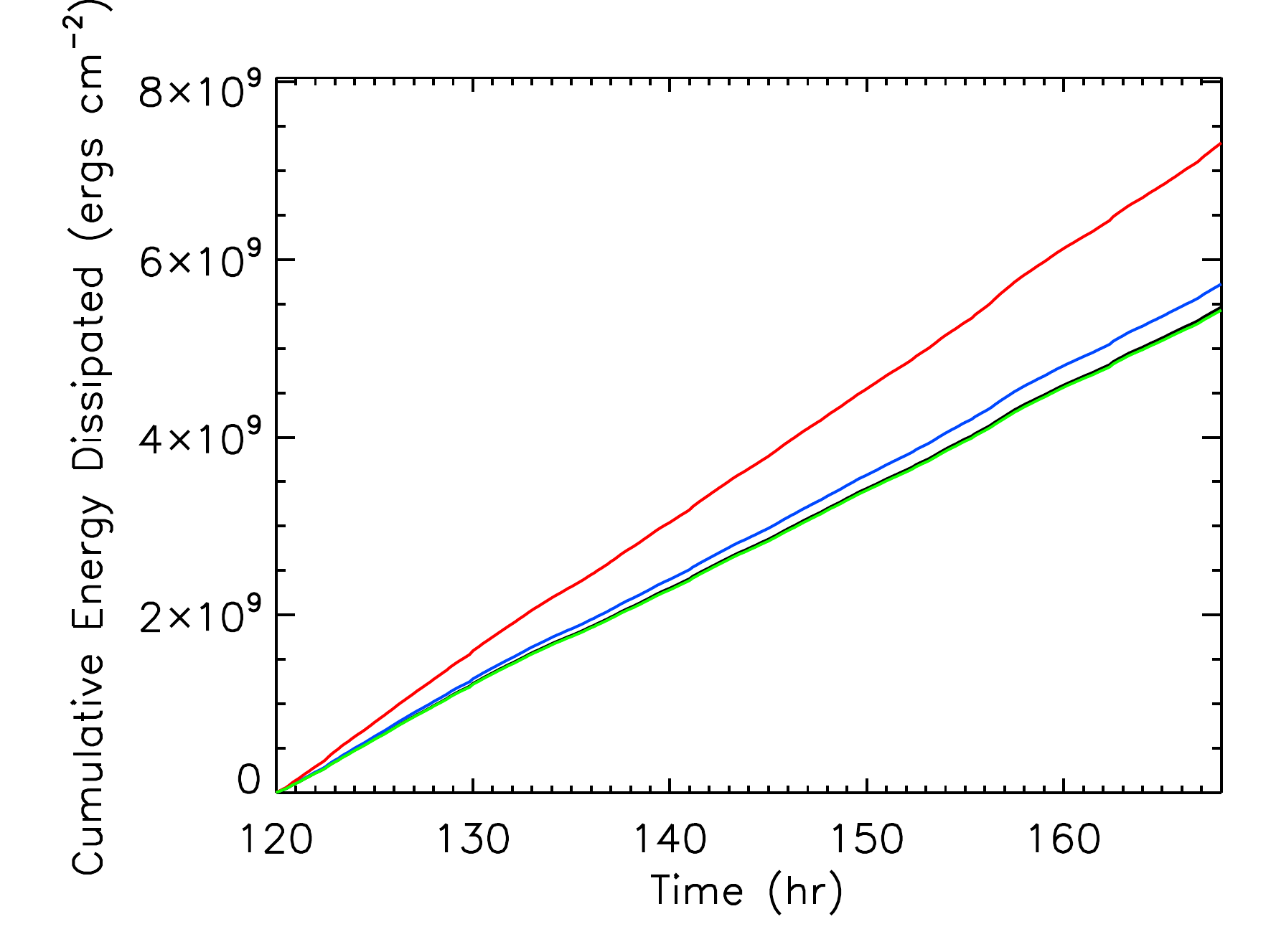}
              }
     \vspace{-0.39\textwidth}   
     \centerline{ \bf     
      \hspace{-0.08 \textwidth}  \color{black}{(c)}
      \hspace{0.49\textwidth}  \color{black}{(d)}
         \hfill}
     \vspace{0.36\textwidth}    
   \centerline{\hspace*{0.015\textwidth}
              \includegraphics[width=0.515\textwidth,clip=]{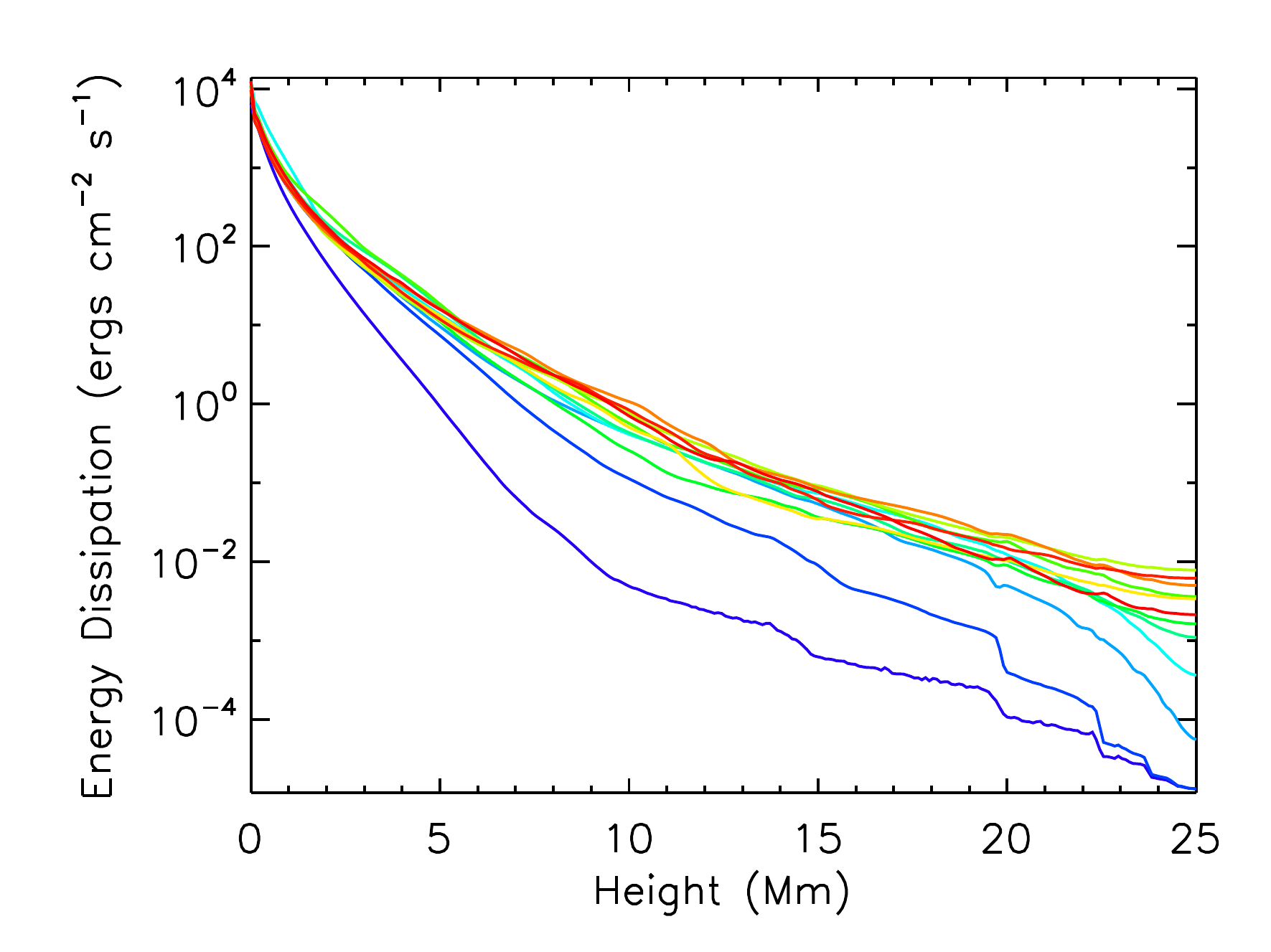}
               \hspace*{0.03\textwidth}
              \includegraphics[width=0.515\textwidth,clip=]{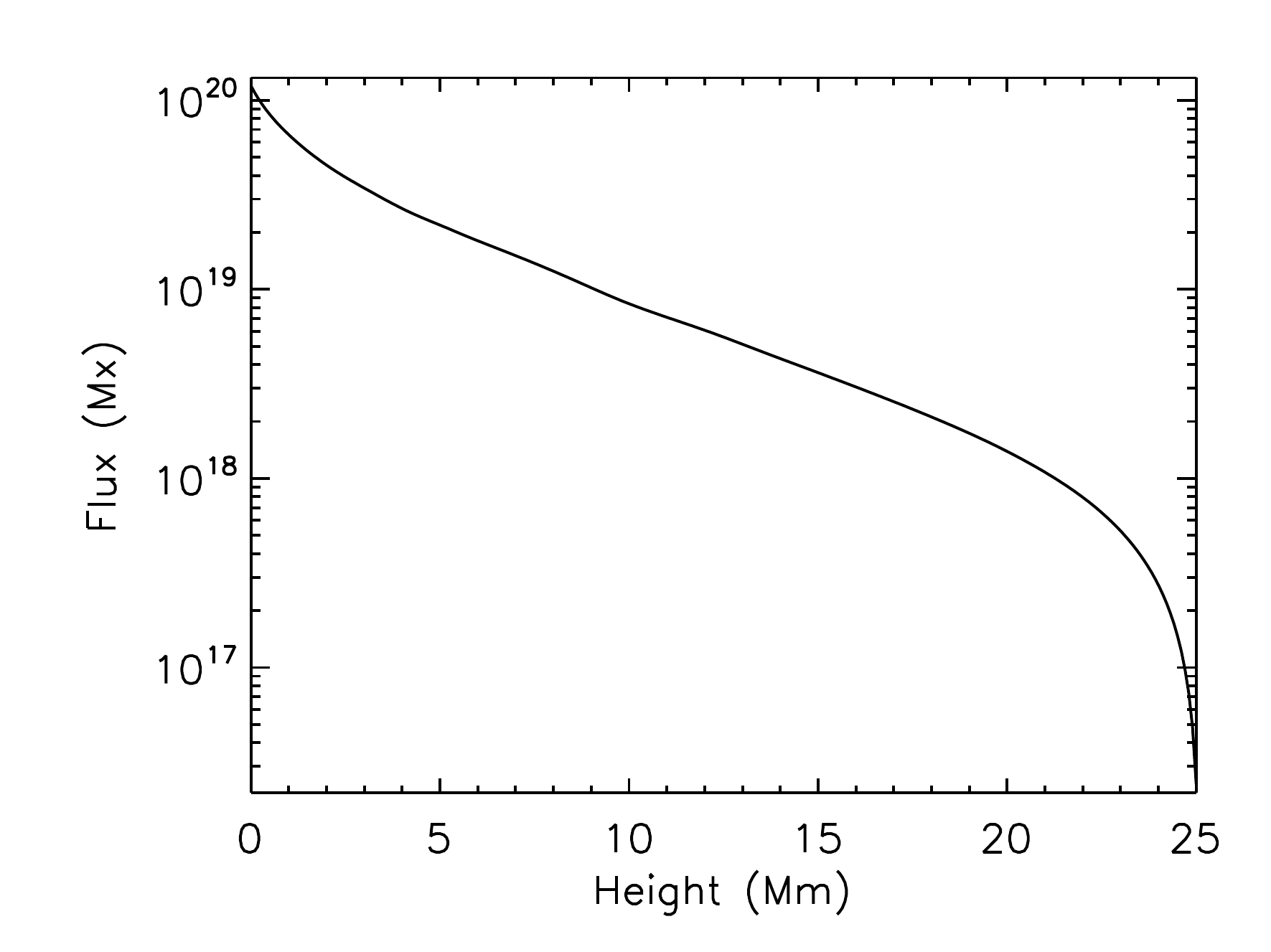}
              }
     \vspace{-0.39\textwidth}   
     \centerline{ \bf     
      \hspace{-0.08 \textwidth}  \color{black}{(e)}
      \hspace{0.49\textwidth}  \color{black}{(f)}
         \hfill}
     \vspace{0.36\textwidth}    

\caption{(a)$-$(d) Energy dissipated per unit area as a function of time for the 3D simulations with no overlying field (green), 1 G (black), 3 G (blue) and 10 G (red) overlying field. (a) Rate of
energy dissipation due to magnetofriction, $\frac{1}{S}\int_V Q_\textrm{\tiny frc} \, dV$,
(b) rate of energy dissipation due to hyperdiffusion, $\frac{1}{S}\int_V Q_\textrm{\tiny
hd} \, dV$, and (c) total rate of energy dissipation, $E_\textrm{\tiny q}(t)$. (d)
Cumulative energy dissipated as a function of time, $E_\textrm{\tiny d}(t)$. (e)
Rate of energy dissipation (integrated over $x$ and $y$) as a function of height
for the 3 G simulation. Six curves are plotted at $50$ min intervals from $t=120.17-124.33$ hr, the remaining six curves are plotted at 8 hr intervals from $t=128-168$ hr. Colours span from blue to red with increasing time. (f) Total absolute flux through surfaces of $z=constant$, as a function of height, at $t=168$ hr.}\label{fig:q1}
   \end{figure}

In addition to the free magnetic energy stored within the system, we consider
energy that is being continually dissipated due to the relaxation processes
within the model applied. This is described by the energy dissipation term,
\begin{equation}\label{eqn:q}
Q = Q_\textrm{\tiny frc} + Q_\textrm{\tiny hd} = \frac{B^2}{4\pi}\nu|\mathbf{v}|^2 + \frac{B^2}{4\pi}\eta_4 |\nabla\alpha|^2.
\end{equation}
The first term, $Q_\textrm{\tiny frc}$, represents energy dissipation due to
magnetofriction, which is released as the coronal magnetic field relaxes towards
a force-free state \cite{yang1986}.
The second term, $Q_\textrm{\tiny hd}$, represents energy dissipation due to
hyperdiffusion \cite{boozer1986}. It has previously been interpreted as the rate at which magnetic energy is converted into heat during the relaxation of the magnetic field \cite{vanballegooijen2008}. For the derivation of $Q$ and a full description of the terms
$Q_\textrm{\tiny frc}$ and $Q_\textrm{\tiny hd}$, see Paper II.

\begin{table}
\begin{center}
\begin{tabular}{cccccccc}
\hline
\vspace{-0.4 cm} \\
 & $B_0$ & 0 G &  1 G & 3 G & 10 G \\
\hline
$\frac{1}{S}\int_V Q_\textrm{\tiny frc} \, dV$  & Mean & 2.66 & 2.67 & 2.74 & 3.25 \\
($\times 10^{4}$ ergs cm$^{-2}$ s$^{-1}$)       & Max  & 4.71 & 4.78 & 4.91 & 5.60 \\
\\
$\frac{1}{S}\int_V Q_\textrm{\tiny hd} \, dV$   & Mean & 0.44 & 0.45 & 0.53 & 0.95 \\
($\times 10^{4}$ ergs cm$^{-2}$ s$^{-1}$)       & Max  & 0.58 & 0.62 & 0.77 & 1.47 \\
\\
$E_\textrm{\tiny q}(t)$ & Mean & 3.10 & 3.12 & 3.27 & 4.20 \\
($\times 10^{4}$ ergs cm$^{-2}$ s$^{-1}$)        & Max  & 5.26 & 5.30 & 5.49 & 6.87 \\
\\
Typical $B_h$ (G) & & 6.5 & 6.6 & 7.2 & 12.1 \\
$P_1/S$ ($\times 10^{4}$ ergs cm$^{-2}$ s$^{-1}$) & & 5.96 & 6.05 & 6.60 & 11.09 \\
\\
$S \, E_\textrm{\tiny d}(t_\textrm{\tiny max})$ ($\times 10^{29}$ ergs) & & 1.35 & 1.37 & 1.43 & 1.83 \\
$E_\textrm{\tiny d}(t_\textrm{\tiny max})$ ($\times 10^{9}$ ergs cm$^{-2}$) & & 5.43 & 5.47 & 5.73 & 7.31 \\
\hline
\end{tabular}
\caption{Mean and maximum values of $Q_\textrm{\tiny frc}$, $Q_\textrm{\tiny
hd}$ and $Q$ integrated over the volume ($E_\textrm{\tiny q}(t)$), and cumulative energy dissipated for
each simulation. Typical horizontal magnetic field, $B_h$, and Poynting flux order of magnitude estimate, $P_1/S$, for each simulation.}\label{tab:q}
\end{center}
\end{table}

Figure~\ref{fig:q1}(a) shows a plot of the rate of energy dissipation per unit area, due to
magnetofriction, $\frac{1}{S}\int_V Q_\textrm{\tiny frc} \, dV$, as a function of time, for
the no overlying field (green), 1 G (black), 3 G (blue) and 10 G (red) simulations. $S=2.5\times 10^{19}$ cm$^2$ is the area of the photospheric boundary surface.  It can be
seen that $Q_\textrm{\tiny frc}$ is not strongly dependent on the overlying field
strength. A stronger overlying field leads to slightly higher $Q_\textrm{\tiny
frc}$, but the variation of values of $Q_\textrm{\tiny frc}$ within each curve
is much larger than the variation of values between the curves for different
overlying field strengths. Table~\ref{tab:q} shows the mean and maximum values
of $\frac{1}{S}\int_V Q_\textrm{\tiny frc} \, dV$ for each overlying field strength simulation. The difference in
mean values between the 0 G and 10 G cases is just $0.59\times 10^{4}$ ergs cm$^{-2}$
s$^{-1}$. However, from the plot in Figure~\ref{fig:q1}(a), each curve has a
variation of around $2.9-3.3\times 10^{4}$ ergs cm$^{-2}$ s$^{-1}$ between its maximum and
minimum. Therefore, the energy dissipation due to magnetofriction is
predominantly dependent upon the evolution of the photospheric magnetic field
driving change within the coronal field.

Figure~\ref{fig:q1}(b) shows the rate of energy dissipation per unit area due to
hyperdiffusion, $\frac{1}{S}\int_V Q_\textrm{\tiny hd} \, dV$, as a function of
time, where lines are coloured as in Figure~\ref{fig:q1}(a). Very little
difference can be seen between the curves for the no overlying field and 1 G cases. The 3 G
case results in slightly higher $Q_\textrm{\tiny hd}$, while the 10 G case
results in significantly higher $Q_\textrm{\tiny hd}$. Also, for the 10 G case there is a larger variation in
the values of $Q_\textrm{\tiny hd}$ than in the no overlying field or 1 G cases. Thus, $Q_\textrm{\tiny hd}$ is clearly dependent on the strength of the overlying field. While this is
the case, the general shape of all of the curves are the same, implying that
like $Q_\textrm{\tiny frc}$, $Q_\textrm{\tiny hd}$ is also predominantly dependent on the
evolution of the photospheric magnetic field driving changes within the coronal
volume.
The mean and maximum values for $\frac{1}{S}\int_V Q_\textrm{\tiny hd} \, dV$ for each strength of
overlying field are given in Table~\ref{tab:q}. For each case, the mean values
for $Q_\textrm{\tiny hd}$ are $3.4-6.0$ times smaller than the mean values for
$Q_\textrm{\tiny frc}$, and the maximum values for $Q_\textrm{\tiny hd}$ are
$3.8-8.1$ times smaller than the maximum values for $Q_\textrm{\tiny frc}$.

Figure~\ref{fig:q1}(c) shows a plot of the total rate of energy dissipation per unit area,
$E_\textrm{\tiny q}(t)=\frac{1}{S}\int_V Q \, dV$, as a function of time, with lines coloured as in
Figure~\ref{fig:q1}(a). Since throughout each simulation, $Q_\textrm{\tiny frc}$
is larger than $Q_\textrm{\tiny  hd}$, the curves for $Q$ follow the same trend
as those for $Q_\textrm{\tiny  frc}$. A stronger overlying field leads to
slightly higher $Q$, but the variation of $Q$ within each curve ($3.1-3.8\times
10^{4}$ ergs cm$^{-2}$ s$^{-1}$) is larger than the variation between the simulations
with different overlying field strengths ($1.1\times 10^{4}$ ergs cm$^{-2}$ s$^{-1}$
difference between the mean values for the no overlying field and 10 G cases). Therefore, the
overall energy dissipation is determined mainly by the photospheric evolution of
the magnetic field. The mean and maximum values of $E_\textrm{\tiny q}(t)$ for each simulation are
given in Table~\ref{tab:q}. Figure~\ref{fig:q1}(d) shows the cumulative energy
dissipated, per unit area, as a function of time, $E_\textrm{\tiny d}(t)$, for each strength of
overlying field, obtained by integrating $Q$ over both the volume and time:
\begin{equation}
 E_\textrm{\tiny d}(t)=\frac{1}{S}\int_0^t \bigg[\int_V Q dV \bigg] dt.
\end{equation}
We see that a stronger overlying field leads to a greater cumulative amount of
energy dissipated. The slopes of these curves are given by the mean values of $E_\textrm{\tiny q}(t)$ in Table~\ref{tab:q}, and can be related to the overlying field strength ($B_0$) by a quadratic:
\begin{equation}
 \overline{ E_\textrm{\tiny q}(t)} = 75.6 B_0^2 + 339.8 B_0 + 3.1\times 10^4.
\end{equation}
The values for the total energy dissipated by the end of each
simulation ($t=t_\textrm{\tiny max}$) are also given in Table~\ref{tab:q}.

The rate of energy dissipation per unit area (ergs cm$^{-2}$
s$^{-1}$) in each simulation is on average $3.1-4.2\times 10^4$ ergs cm$^{-2}$ s$^{-1}$. These values are too low to explain the
radiative losses of the quiet Sun corona, being only $31-42\%$ of the
$10^5$ ergs cm$^{-2}$ s$^{-1}$ calculated by \inlinecite{withbroe1977} and $6.3-8.6\%$ of the
$4.9\times 10^5$ ergs cm$^{-2}$ s$^{-1}$ calculated by \inlinecite{habbal1991}.
Due to the fact that the values are too low, it is of interest to compare the energy dissipated to the Poynting flux through the photospheric boundary, as clearly we cannot dissipate more energy than has been injected. Considering the plots of free magnetic energy as a function of time (Figure~\ref{fig:free1}(a)), they are roughly steady, at least for the $1$ G, $3$ G and no overlying field cases. This implies that the energy dissipated ($E_\textrm{\tiny q}(t)$) should roughly balance the Poynting flux injected. The Poynting flux through the photospheric boundary is given by
 \begin{equation}\label{eqn:poynt}
  P = -\frac{c}{4\pi} \int_S (\BE\times\BB)\cdot d\mathbf{S},
 \end{equation}
 where $d\mathbf{S}=dS \mathbf{\hat{n}}$ and $S$ is the area of the photospheric boundary surface with unit normal vector $\mathbf{\hat{n}}$. From Ohm's law,
 \begin{equation}
  \BE = - \frac{1}{c} \Bv\times\BB,
 \end{equation}
 so Equation~\ref{eqn:poynt} simplifies to
 \begin{equation}\label{eqn:poynt2}
  P = \frac{1}{4\pi}\int_S \bigg[ v_x B_x B_z + v_y B_y B_z - v_z B_x^2 - v_z B_y^2 \bigg] dS.
 \end{equation}
The individual terms in Equation~\ref{eqn:poynt2} may be split into two distinct groups. Those involving $v_x$ and $v_y$ represent energy flow into or out of the domain due to horizontal boundary flows. The terms involving $v_z$ represent energy flow into or out of the domain due to flux emergence or cancellation.
 
As computed in \inlinecite{mackay2011}, an order of magnitude estimate of the Poynting flux due to horizontal boundary flows is:
 \begin{equation}\label{eqn:P}
  P_1 \approx \frac{v_h B_z B_h S}{4\pi},
 \end{equation}
 where $B_h$ and $B_z$ are typical values for horizontal and vertical magnetic field at the photosphere and $v_h$ is a typical value for horizontal velocities at the photosphere.

In all simulations, the mean vertical field strength is $B_z=4.8$ G over a photospheric area of $S=2.5\times 10^{19}$ cm$^2$ and the mean value of our supergranular velocity profile at the photosphere is $v_h=0.24$ km s$^{-1}$. The only other parameter is the typical value for the horizontal field strength, $B_h$, which varies from one simulation to the next due to the overlying field. The mean values of $B_h$ for each simulation are listed in Table~\ref{tab:q}, along with the mean energy dissipation, $E_\textrm{\tiny q}(t)$, and the value of $P_1$ calculated for each simulation. The values in Table~\ref{tab:q} show that our order of magnitude estimate for energy injected due to horizontal flows is within a factor of $2-3$ of the energy dissipated for each simulation, so are in good agreement.

While we are able to estimate the Poynting flux resulting from horizontal flows, we cannot calculate the Poynting flux from vertical flows as, due to our special boundary treatment (Section~\ref{sec:lower}), there is no $v_z$ defined on the photospheric boundary of our model.
The component of the Poynting flux due to $v_z$ would likely lead to both injection and removal of energy respectively during emergence and cancellation events. As the flux domain is in a steady state, with the rates of flux emergence and cancellation roughly identical (see Paper I), we may assume that these processes lead to no net injection of energy. In addition, \inlinecite{parnell2012} have estimated, using observational data, that the Poynting flux injected due to the emergence of magnetic flux is significantly smaller than that injected due to the horizontal motions of existing magnetic flux.

Assuming equal rates of energy input and loss due to emergence and cancellation, this calculation shows that the energy injected does in fact match the energy dissipated within an order of magnitude. This implies that the reason why the energy dissipated within our simulations is not high enough to account for coronal radiative losses is in part due to the fact that not enough energy has been injected in the first place. Therefore, in future simulations, increasing $v_h$, $B_h$ or $B_z$ may lead to an increase in energy dissipation.

We note that other possible reasons for the energy dissipation rate being too low are that this is not a true physical dissipation, and many simplifications have been made for our model. For example, our model contains no plasma and we do not resolve wave motions. The relative simplicity of the synthetic magnetograms compared to observed magnetograms may also be a factor. While the values calculated for $E_\textrm{\tiny q}(t)$ are too low to explain coronal heating, it is of interest for us to consider the location and structure of the energy dissipation term, $Q$.

Figures~\ref{fig:q1}(a)$-$(d) consider the volume integrated rate of energy
dissipation over the entire 3D simulation for each strength of overlying field.
We now consider where $Q$ is spatially located within individual frames of the 3
G simulation. Although results are presented here only for the 3 G simulation,
similar results are found for other strengths of the overlying field.
Figure~\ref{fig:q1}(e) shows the rate of energy dissipation (integrated
in $x$ and $y$) as a function of height. This is
computed with units of ergs cm$^{-2}$ s$^{-1}$ as follows:
\begin{displaymath}
 E_\textrm{\small q}(z)=\frac{L_z}{S} \int_{y_{\textrm{\tiny min}}}^{y_{\textrm{\tiny
max}}}\int_{x_{\textrm{\tiny min}}}^{x_{\textrm{\tiny max}}} Q(x,y,z) dx dy.
\end{displaymath}
Six curves are plotted at intervals of 50 min from $t=120.17-124.33$ hr, the other six are plotted at intervals of 8 hr from $t=128-168$ hr. Colours span from blue to red with increasing time. The curve at the earliest time ($t=120.17$ hr) is lowest, as the coronal magnetic field is still close to potential. The height of the curves increases with increasing time until a near steady rate of dissipation is reached, where very little difference is seen between the curves. 
At each time, the greatest rate of energy dissipation is
found low down, near the photosphere. This is not surprising when we consider
the equation describing $Q$ (Equation~\ref{eqn:q}). From this equation we see that $Q_\textrm{\tiny frc}$ and $Q_\textrm{\tiny hd}$ are both proportional to magnetic field strength, $|\BB|$, $Q_\textrm{\tiny frc}$ is also proportional to velocity, $|\Bv|$, and $Q_\textrm{\tiny hd}$ to gradients in $\alpha$.
Figure~\ref{fig:q1}(f) shows a plot of the total absolute flux through surfaces of constant $z$, as a function of height. This is shown at $t=168$ hr, but similar curves are seen throughout the simulation. This is computed as:
\begin{equation}
 \phi(z) =  \int_{y_{\textrm{\tiny min}}}^{y_{\textrm{\tiny
max}}}\int_{x_{\textrm{\tiny min}}}^{x_{\textrm{\tiny max}}} |B_z(x,y,z)| dx dy.
\end{equation}
The flux is greatest low down, near the magnetic sources, then drops of rapidly with increasing height. This indicates that most connections between magnetic features close low down.
From Figure~\ref{fig:q1}(e), the rate of energy dissipation also rapidly drops with increasing height, having decreased by more than an order of magnitude by $z=2$ Mm. Therefore, the energy dissipation term $Q$
has its largest effect close to the photosphere.
In contrast to energy dissipated (which from Equation~\ref{eqn:q} is always $>0$), localised values of free magnetic energy density can be either positive or negative. In Figure~\ref{fig:free1}(e) and (f) black patches can be seen low down, indicating locations of negative free energy density, most likely occurring due to the cancellation of magnetic features on the photosphere removing energy. Higher up, the free energy density tends to be only positive (white patches). The combined effect of positive and negative regions low down results in the peak in total free energy density as a function of height occurring higher up, just below 1 Mm (Figure~\ref{fig:free1}(b)), where the free energy density becomes solely positive.
The movie, \textcolor{blue}{magnet48b\_q\_ht.mpg}, shows the rate
of energy dissipation as a function of height for the first $8.3$ hr of the 3 G simulation. At
the start of the movie, at greater heights, the rate of energy dissipation
gradually increases until the curve becomes more or less steady. One can see
occasional kinks in the curve at low $z$, which then propagate upward.

  \begin{figure}[!h]

   \centerline{\hspace*{0.07\textwidth}
              \includegraphics[width=0.49\textwidth,clip=]{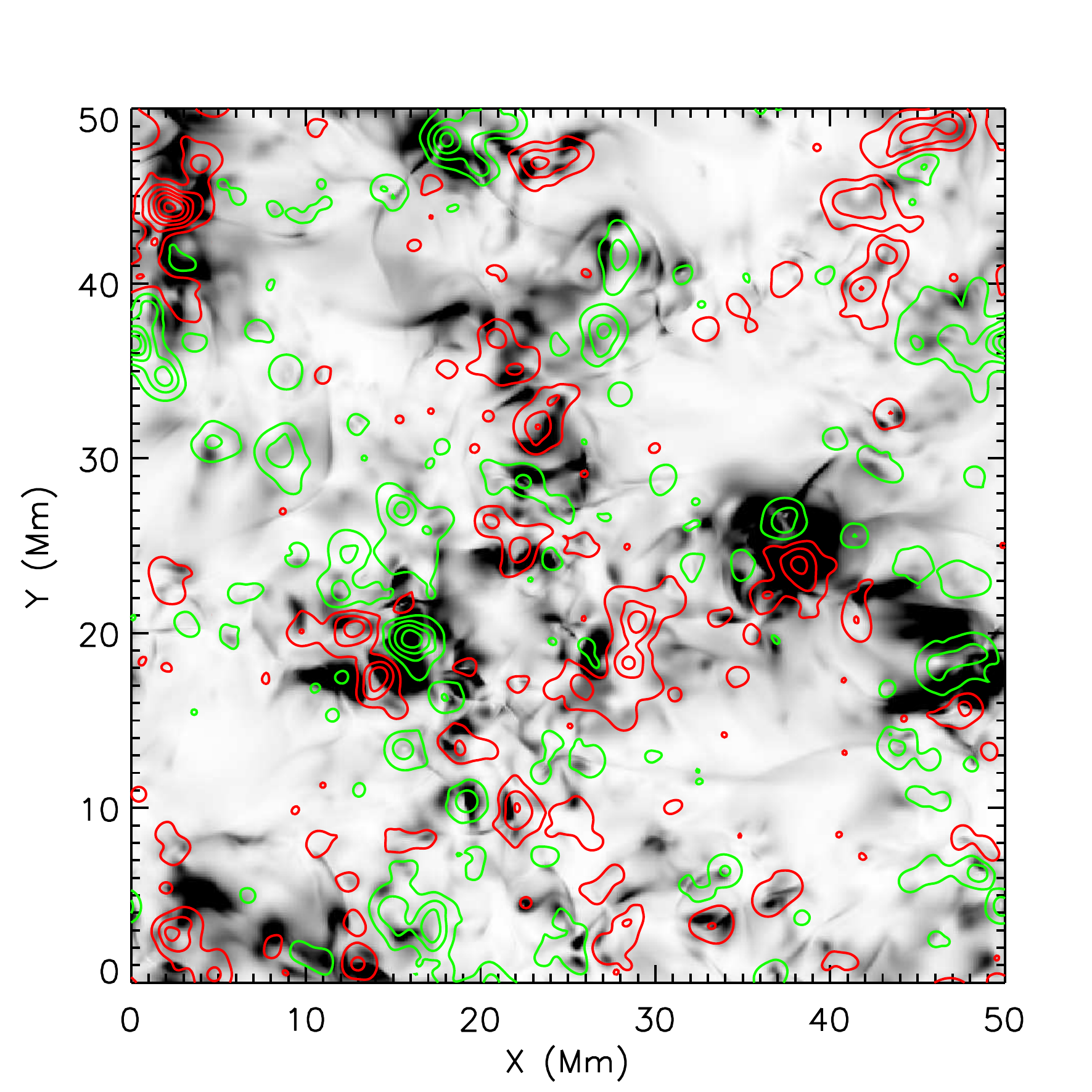}
               \hspace*{0.03\textwidth}
              \includegraphics[width=0.49\textwidth,clip=]{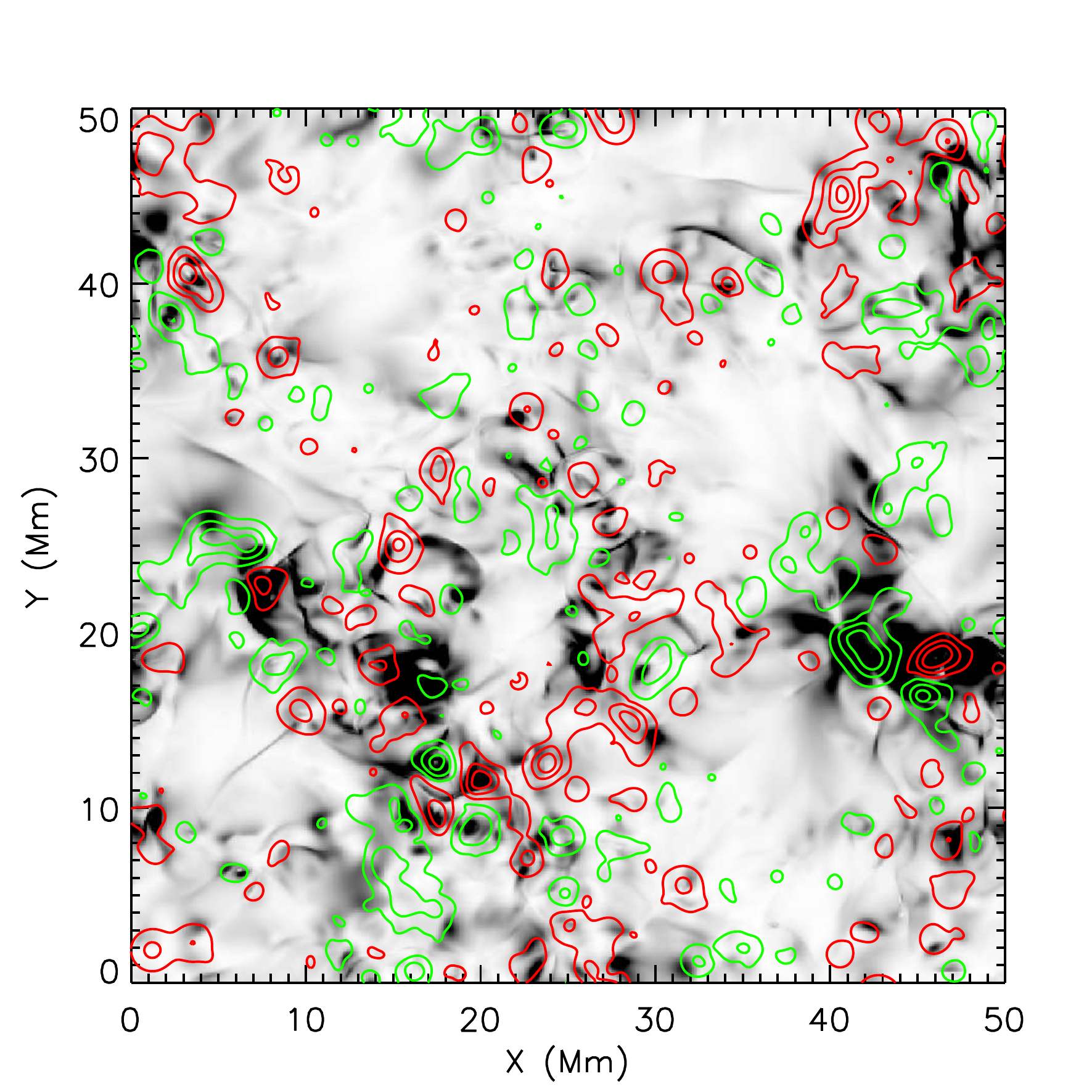}
              }
     \vspace{-0.45\textwidth}   
     \centerline{ \bf     
      \hspace{-0.03 \textwidth}  \color{black}{(a)}
      \hspace{0.46\textwidth}  \color{black}{(b)}
         \hfill}
     \vspace{0.42\textwidth}    
   \centerline{\hspace*{0.07\textwidth}
              \includegraphics[width=0.49\textwidth,clip=]{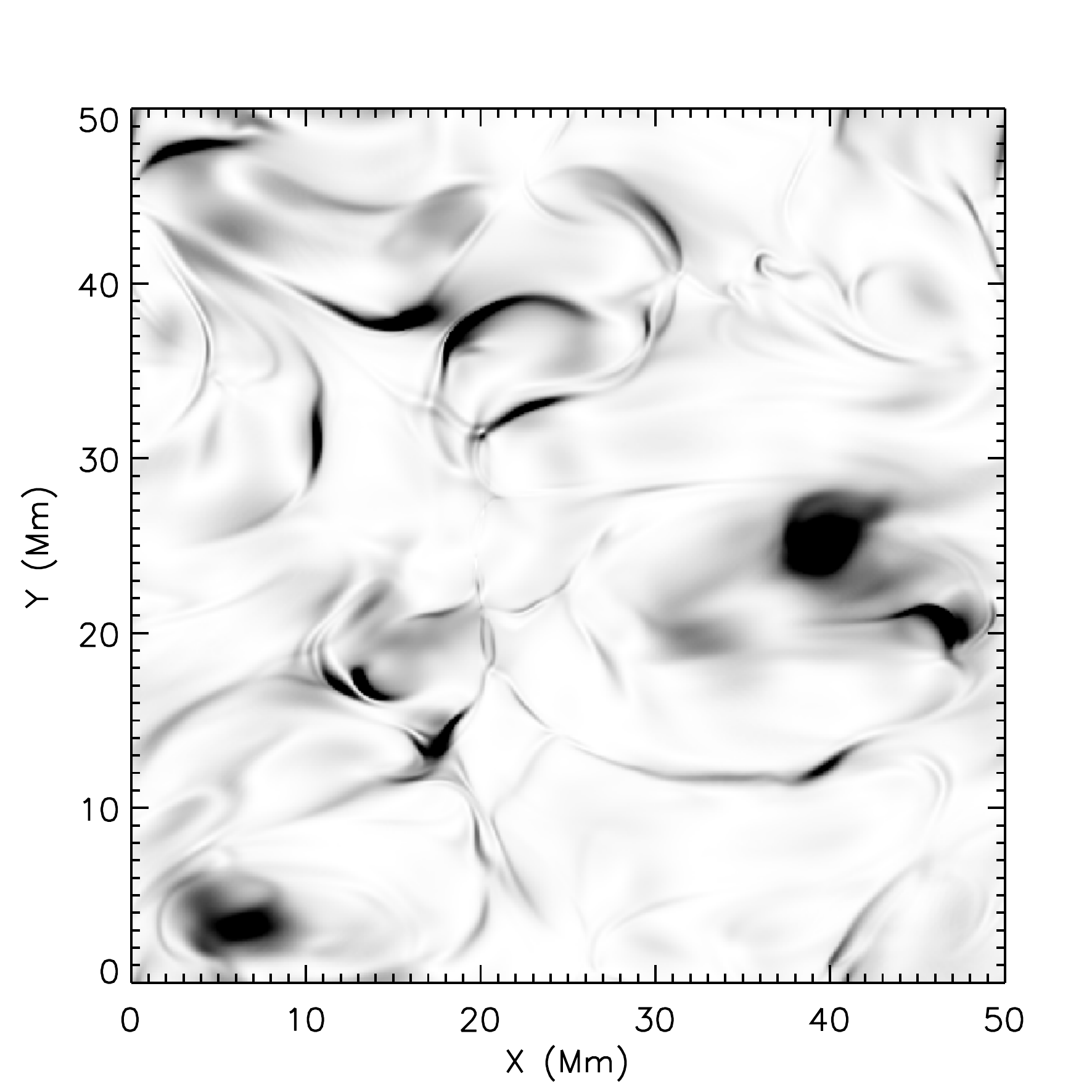}
               \hspace*{0.03\textwidth}
              \includegraphics[width=0.49\textwidth,clip=]{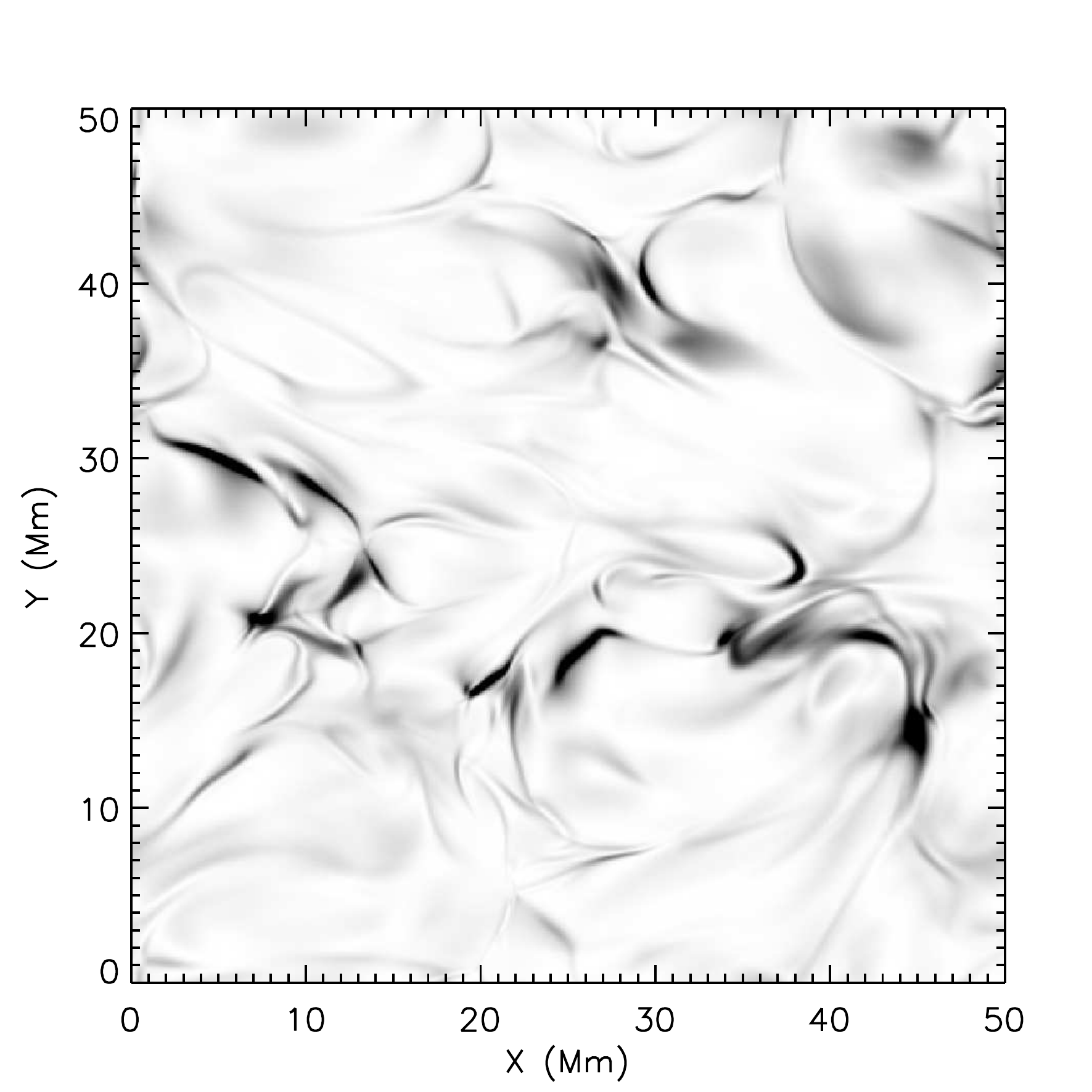}
              }
     \vspace{-0.45\textwidth}   
     \centerline{ \bf     
      \hspace{-0.03 \textwidth}  \color{black}{(c)}
      \hspace{0.46\textwidth}  \color{black}{(d)}
         \hfill}
     \vspace{0.42\textwidth}    
   \centerline{\hspace*{0.03\textwidth}
              \includegraphics[width=0.49\textwidth,clip=]{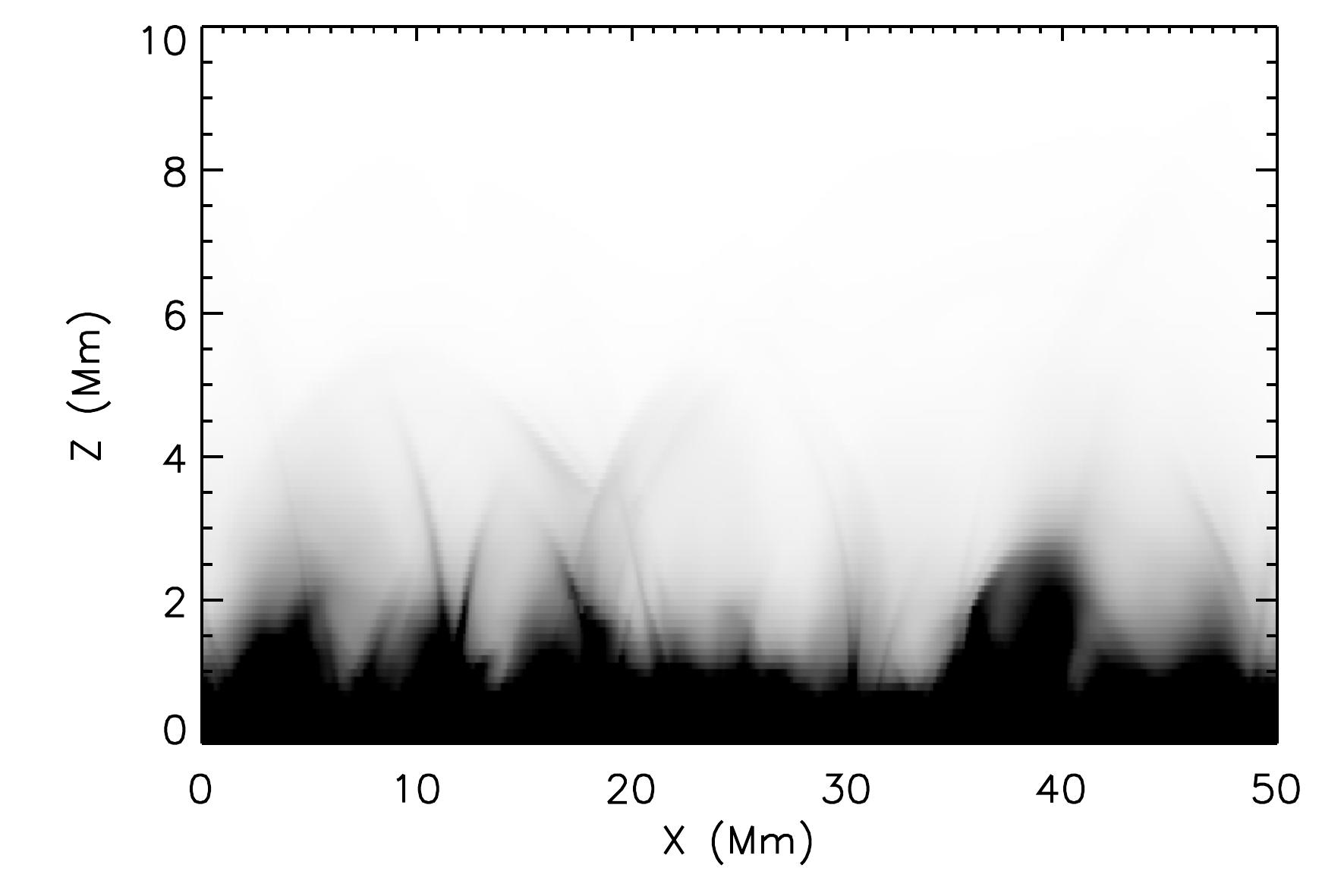}
               \hspace*{0.03\textwidth}
              \includegraphics[width=0.49\textwidth,clip=]{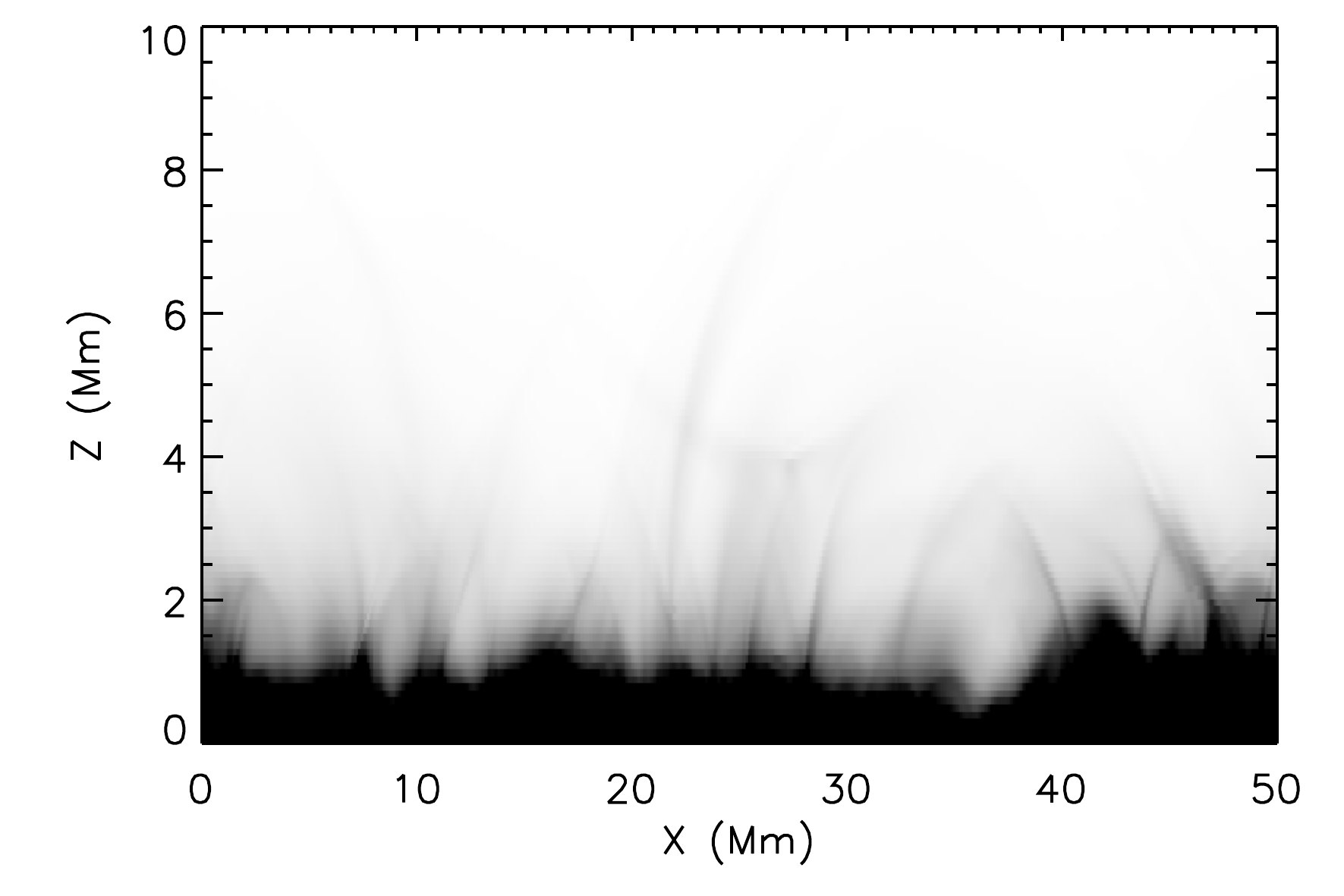}
              }
     \vspace{-0.34\textwidth}   
     \centerline{ \bf     
      \hspace{-0.03 \textwidth}  \color{black}{(e)}
      \hspace{0.46\textwidth}  \color{black}{(f)}
         \hfill}
     \vspace{0.31\textwidth}    

\caption{All images are for the 3 G simulation. Darker regions correspond to higher values. (a) and (b) Rate of energy
dissipation, $Q$, integrated in $z$. The images are shown in the $x-y$ plane,
saturated at $1.5\times 10^5$ ergs cm$^{-2}$ s$^{-1}$. Positive (red) and negative (green) contours of $B_z$ at
$z=0$ Mm are over-plotted at levels of $\pm[7, 13, 27, 53, 106]$ G.
(c) and (d) Rate of energy dissipation, $Q$, saturated at $810$ ergs cm$^{-2}$
s$^{-1}$, shown in the $x-y$ plane at $z=3$ Mm. (e) and (f) Images of $Q$ in the
$x-z$ plane integrated in $y$, saturated at $1.5\times 10^5$ ergs cm$^{-2}$
s$^{-1}$. Images in the left-hand column are shown at $t=128$ hr and in the
right-hand column at $t=168$ hr.}\label{fig:q2}
   \end{figure}

Figures~\ref{fig:q2}(a) and (b) show images of $Q$ integrated in $z$, in the
$x-y$ plane, from the 3 G simulation. This is computed with units of ergs
cm$^{-2}$ s$^{-1}$ as follows:
\begin{displaymath}
 E_\textrm{\small q}(x,y)= \int_{z_{\textrm{\tiny min}}}^{z_{\textrm{\tiny
max}}} Q(x,y,z) dz.
\end{displaymath}
As in Figure~\ref{fig:j}, the colour table has been reversed, so that darker regions correspond to higher values of $E_\textrm{\small q}(x,y)$.
The images are shown at (a) $t=128$ hr and (b) $t=168$ hr, and are saturated at
a level of $1.5\times 10^5$ ergs cm$^{-2}$ s$^{-1}$. As in
Figures~\ref{fig:free1}(c) and (d), contours of $B_z$ at $z=0$ Mm are
over-plotted. One can see that $Q$ is more localised than the free magnetic
energy. Patches of $Q$ tend to lie around the magnetic network, but not
necessarily at the same locations as free magnetic energy seen in
Figures~\ref{fig:free1}(c) and (d). In addition, we do not see far-reaching
bands of $Q$ across the supergranules as we did with the free energy. In Paper
II, it was determined that $Q$ is mainly located at sites of changing magnetic
connectivity and low down, near the magnetic elements. Therefore, it is unsurprising
that in these simulations, $Q$ is seen mainly at the magnetic network, as this
is where large numbers of magnetic elements lie, and the magnetic connectivity is
constantly changing due to the continual interaction of these elements with one
another.

A movie showing $Q$ in the $x-y$ plane for the 3 G simulation is available (\textcolor{blue}{magnet48b\_q\_xy\_bz.mpg}). It shows $Q$ integrated in $z$, and saturated at $1.5\times 10^5$ ergs cm$^{-2}$ s$^{-1}$, with contours of
$B_z$ at $z=0$ Mm over-plotted. The movie shows that the locations and
evolution of $Q$ are very different from those of the positive free magnetic
energy density (\textcolor{blue}{magnet48b\_free\_xy\_bz.mpg}). Indeed, while many of
the patches of positive free magnetic energy density were seen to be long-lived,
regions of $Q$ are seen to be very short-lived, occurring in rapidly evolving
`bursts'. Within the free energy density movie, patches of positive free energy
density are often seen stretched across the supergranular cells, whereas $Q$
tends to be much more localised, occurring predominantly within the magnetic
network where large magnetic elements lie. Several large bursts
of $Q$ can be seen throughout the movie, in regions where many magnetic
elements are emerging and interacting with one another. For readers unable to view the movie, six still images from the movie spaced 1 hr apart are included in Appendix~\ref{sec:app1} (Figure~\ref{fig:app3}). It can be seen that the spatial distribution of $Q$ integrated in the LOS changes significantly from one hour to the next. Note that the evolution
of $Q$ that we see in these movies is dominated by $Q$ low down, since the rate
of energy dissipation rapidly decreases with increasing $z$
(Figures~\ref{fig:q1}(e) and (f)).
We now consider the spatial location of $Q$ as a function of height above the photosphere.

Figures~\ref{fig:q2}(c) and (d) show $x-y$ plane images of $Q$ at $z=3$ Mm in
the 3 G simulation, at $t=128$ hr and $t=168$ hr respectively. $Q$ at height $z=3$ Mm
is given by $Q(x,y,3)L_z$. As in Figures~\ref{fig:q2}(a) and (b), many
patches of $Q$ can be seen located above the boundaries between supergranules.
However, in contrast to Figures~\ref{fig:q2}(a) and (b), within
Figures~\ref{fig:q2}(c) and (d) one can also see much more fine-scale structure
to $Q$ when it is not integrated along the line of sight. Also in contrast to
Figures~\ref{fig:q2}(a) and (b), where $Q$ integrated over $z$ is predominantly
located at the magnetic network, here we see long strands of $Q$ which lie
across the supergranules at $z=3$ Mm. These are localised regions of energy
dissipation that are found mainly at sites of changing magnetic connectivity. A
movie of $Q$ above the photosphere is available
(\textcolor{blue}{magnet48b\_q\_xy\_3\_10.mpg}) showing $Q$ in the $x-y$ plane,
integrated between $z=3$ Mm and $z=10$ Mm:
\begin{displaymath}
 E_\textrm{\small{q,3-10}}(x,y)= \int_{z=3\textrm{ Mm}}^{z=10\textrm{ Mm}}
Q(x,y,z) dz,
\end{displaymath}
in units of ergs cm$^{-2}$ s$^{-1}$.
The movie is saturated at $3\times 10^{3}$ ergs cm$^{-2}$ s$^{-1}$. At this
height, the rate of energy dissipation is much lower than at the photosphere, as
the coronal field is evolving more slowly and is less non-potential. The result of
this is that $Q$ is also less rapidly evolving than it is lower down. In
addition, between these heights, $Q$ is seen to occur anywhere within the $x-y$
plane and not just above the magnetic network. We also see more fine-scale,
further reaching structures. Six images of $Q$ at $z=3$ Mm spaced 1 hr apart are also included in Appendix~\ref{sec:app1} (Figure~\ref{fig:app4}), to illustrate the time-scale of evolution at this height.

Figures~\ref{fig:q2}(e) and (f) show $x-z$ plane images of $Q$ integrated in
$y$, saturated at $1.5\times 10^5$ ergs cm$^{-2}$ s$^{-1}$,
at $t=128$ hr and $t=168$ hr, respectively. In agreement with Figures~\ref{fig:q1}(e)
and (f), the energy dissipation is seen to be greatest low down. Finer-scale
structure can then be seen between $z=2.5$ Mm and $z=5$ Mm. Two movies of $Q$ in
the $x-z$ plane for the 3 G simulation are included with this paper. Similar
evolution is seen in the $y-z$ plane and for other strengths of overlying field.
The first movie, \textcolor{blue}{magnet48b\_q\_xz.mpg}, shows $Q$ integrated in
$y$ and saturated at the same level as in Figures~\ref{fig:q2}(e) and (f). The
second movie, \textcolor{blue}{magnet48b\_q\_xz\_log.mpg}, shows the logarithm
of the first so that the energy dissipation can be seen for a wider range of
values. From this it can be seen that fine-scale structures also exist higher up in
the corona, where the energy dissipation is much weaker. In both movies,
occasional `bursts' can be seen, where a feature will drift upwards and
disappear (or rather, become too faint to be shown at the current level of
saturation). From these movies, we find that the energy dissipated and summed
along the line of sight provides a fine-scale dynamic structure that is in
qualitative agreement with what is observed on the Sun low down.

\section{Discussion and Conclusions} \label{sec:conc}

The aim of this paper was to carry out a preliminary analysis of a set of small-scale,
non-linear force-free field simulations. The simulations were driven by synthetic magnetograms
produced by the model described in Paper I. Four
simulations were run, each driven by the same lower boundary data, three with
different strengths of overlying field: 1 G, 3 G and 10 G, and one with no overlying field. The lower
boundary data consisted of a $48$ hr series of synthetic magnetograms of area
$50\times 50$ Mm$^2$ and of cadence 1 min. The initial coronal magnetic field for each
simulation was potential. This field was then evolved through a
series of quasi-static, non-linear force-free equilibria, via a
magnetofrictional relaxation technique, in response to photospheric boundary
motions. The continuous nature of this coronal evolution technique means that 
current systems are maintained within the corona from one step
to the next and the evolution is smooth. This allows for the build-up and storage of free magnetic energy
$-$ one of the quantities studied within this paper. The presence of free
magnetic energy within our model shows a significant departure from previous
models of the magnetic carpet coronal field, which use potential fields. The
other quantities considered were the energy dissipated and the square of the
electric current density, $j^2$.

Initially, for each simulation, both the free magnetic energy and energy dissipation rate rapidly increase, before levelling off and oscillating about a mean value. The mean free magnetic energy per unit area for each simulation is $6.5-11.7\times
10^{7}$ ergs cm$^{-2}$, whilst the mean energy dissipation rate is $3.1-4.2\times
10^{4}$ ergs cm$^{-2}$ s$^{-1}$, resulting in $5.4-7.3\times 10^{9}$ ergs cm$^{-2}$ ($1.4-1.8\times 10^{29}$ ergs) being
cumulatively dissipated over each $48$ hr simulation. For both the free and
dissipated energies, a stronger overlying field results in higher values,
although the effect is more significant for the free energy.
It is also clear that the evolution of both the free and dissipated energies is
highly dependent upon the evolution of the photospheric magnetic field.

While there are similarities between the evolution of the two types of energy
integrated over the volume, they are seen to be less alike when we consider
their location within each simulation. The bulk of the free magnetic energy is
located above the photosphere, between $z=0.5-0.8$ Mm. This is stored along
closed connections between magnetic elements. Regions of positive free energy
density can be seen both in the magnetic network and across supergranular cells;
such regions may also be long-lived. In contrast, the largest amount of energy
dissipation is found low down, near the magnetic sources, and values decrease
rapidly with increasing height. Regions of increased energy dissipation are seen
predominantly in the magnetic network, although weaker, fine-scale strands are
also seen above the photosphere at sites of changing magnetic connectivity. Also
unlike the free magnetic energy density, the large regions of energy dissipation
seen in the $x-y$ plane are much more rapidly evolving (compare movies
\textcolor{blue}{magnet48b\_free\_xy\_bz.mpg} and
\textcolor{blue}{magnet48b\_q\_xy\_bz.mpg}).

The amount of free magnetic energy built up and stored in each simulation ($1.1-2.1\times 10^{27}$ ergs) is
sufficient to account for such small-scale transient phenomena as nanoflares
($\sim 10^{24}$ ergs, Parker, \citeyear{parker1988}) and X-ray bright points
($10^{22}-10^{24}$ ergs s$^{-1}$, Habbal and Withbroe, \citeyear{habbal1981};
Longcope, \citeyear{longcope1998}). The energy dissipation rate is not high enough to be
able to explain the heating of the quiet corona, providing a contribution of
around $31-42\%$ to the required
$10^5$ ergs cm$^{-2}$ s$^{-1}$ of \inlinecite{withbroe1977} or $6.3-8.6\%$ to the required
$4.9\times 10^5$ ergs cm$^{-2}$ s$^{-1}$ of \inlinecite{habbal1991}. In fact, the discrepancy may be even larger since \inlinecite{habbal1991} did not include the energy losses due to thermal conduction. However, the location and structure of regions of energy dissipation within our model are at least in qualitative agreement with what is
observed on small scales on the Sun. The lower rate of energy dissipation may in part be due to the simplified magnetofrictional model that has been applied, but also due to an insufficient rate of energy being injected in the first place. An order of magnitude estimate of the Poynting flux injected showed that this was within a factor or $2-3$ of the mean energy dissipated in each simulation. This indicates that increasing the magnitude of the horizontal velocities or the horizontal or vertical magnetic field components at the photosphere, hence increasing the Poynting flux, could result in higher energy dissipation. Another reason for our Poynting flux (and hence energy dissipated) being lower than observed values is that in the present model, our magnetic features are treated as `rigid' bodies. If the magnetic field of an actual feature were instead contained within many intense kilogauss flux tubes, the movement of features would be more fluid (as is observed) as the intense flux tubes have high local velocities. This would lead to a greater injection of Poynting flux.
It should also be noted that our model does not include a chromosphere. \inlinecite{habbal1991} determined chromospheric radiative losses to be $3.2\times 10^5$ ergs cm$^{-2}$ s$^{-1}$, while the value determined by \inlinecite{withbroe1977} is an order of magnitude higher at $4\times 10^6$ ergs cm$^{-2}$ s$^{-1}$. A more complex model that included a chromosphere with the effects of, for example, Alfv\'{e}n waves and turbulence would likely result in a much higher rate of energy dissipation ({\it e.g.} \inlinecite{vanballegooijen2011}).

\begin{table}
\begin{center}
\begin{tabular}{cccc}
\hline
\vspace{-0.4 cm} \\
Quantity & Location & Variation &  Effect of O/L  \\
         &          & in height &  field strength  \\
\hline
 Free   & Magnetic network and   & Mostly stored $z=0.5-0.8$ Mm & Significant      \\
 Energy & across s/g cells.      & then rapid decrease          & increase in      \\
        &                        & with increasing $z$.         & total with       \\
        &                        &                              & increasing $B_0$ \\
        &                        &                              &                  \\
 $Q$    & Magnetic network and   & Rapid decrease with          & Slight           \\
        & sites of changing      & increasing $z$               & increase in      \\
        & magnetic connectivity. &                              & total with       \\
        &                        &                              & increasing $B_0$ \\
        &                        &                              &                  \\
$j^2$   & Magnetic network and   &          &     \\
        & non-potential regions. &   $-$    & $-$ \\
        & Follows field lines.   &          &     \\
\hline
\end{tabular}
\caption{Summary of results for free magnetic energy; energy dissipated, $Q$,
and current density, $j^2$.}\label{tab:summary}
\end{center}
\end{table}

Locations of increased $j^2$ are found to be co-located with regions of positive
free magnetic energy density, as both are dependent on the magnetic field being
non-potential. Visually, $j^2$ is seen to follow the shape of the magnetic field
where $\alpha$ is non-zero. Table~\ref{tab:summary} gives a summary of results
for each of the quantities studied in this paper, indicating their locations and
the effect of increasing the strength of the overlying field.

There are several avenues for future work using the non-linear force-free
coronal modelling technique described here. The simulations presented in this
paper will be studied in more detail. For example, a more in depth study of the
connectivity of the magnetic field will be conducted, similar to the study of
\inlinecite{close2003}, who analysed potential coronal fields. It would also be
of interest to investigate in detail the effect of each of the photospheric flux
evolution processes on the evolution of the coronal field and on each of the energy
quantities discussed in this paper. We already know the flux, location and time
of occurrence of each of the processes within the synthetic magnetogram series.
Further 2D simulations will also be run to produce synthetic magnetograms that include an evolving supergranular flow profile and magnetic features on smaller scales. The added complexity of evolving photospheric flows and smaller magnetic features will likely result in a larger build up of free magnetic energy and greater energy dissipation, when the coronal field evolution is simulated.

The rate of energy dissipation per unit area produced by our present simulations was
found to be too low to explain quiet Sun radiative losses. This is unsurprising,
however, as many simplifications have been made for our coronal model and it is not a true physical energy dissipation $-$ we do not have currently have an energy equation. For example, the present simulations do not include the effects of small-scale magnetic braiding and MHD waves, which are believed to be important for heating the chromosphere and corona in active regions \cite{vanballegooijen2011,asgari2012}. In future
we intend to extend the present magnetofrictional model to contain plasma by
including the pressure and density terms in the equations of
magnetohydrodynamics. We will then be able to follow the corresponding plasma
processes associated with energy dissipation and give an estimate of the
resultant plasma heating.

The magnetofrictional technique will also be applied to real magnetogram data,
such as from Hinode/SOT or SDO/HMI. A study will be carried out to compare
regions of interest within a simulated non-linear force-free coronal field
driven by observed magnetograms to events occurring in corresponding coronal
images ({\it e.g.} using various wavelengths of SDO/AIA). What is clear is that within
this single, relatively `simple' simulation, a wide range of processes and
dynamics is occurring. A careful and detailed analysis of these and other
non-linear force-free field simulations of the magnetic carpet coronal field
will be carried out in future.

\begin{acks}
KAM and DHM gratefully acknowledge the support of the Leverhulme Trust and the
STFC. DHM would like to thank the Royal Society for their support through the
Research Grant Scheme. DHM and CEP acknowledge support from the EU under FP7.
\end{acks}

\appendix

\section{Images From Movies}\label{sec:app1}

This appendix contains four figures, each with six images spaced 1 hr apart from $t=144$ hr to $t=149$ hr, taken from some of the $x-y$ plane movies included with this paper.

  \begin{figure}

   \centerline{\hspace*{0.07\textwidth}
              \includegraphics[width=0.49\textwidth,clip=]{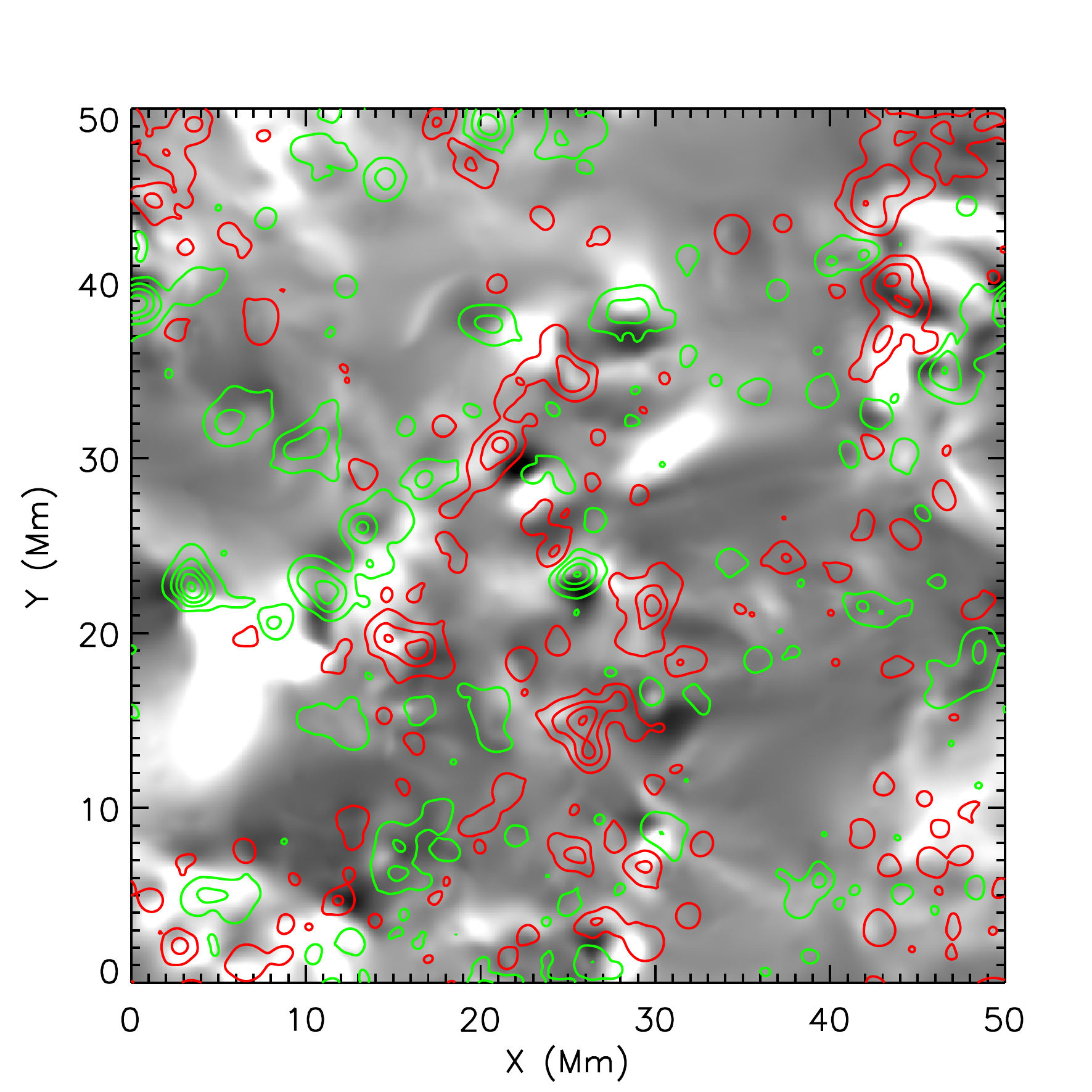}
               \hspace*{0.03\textwidth}
              \includegraphics[width=0.49\textwidth,clip=]{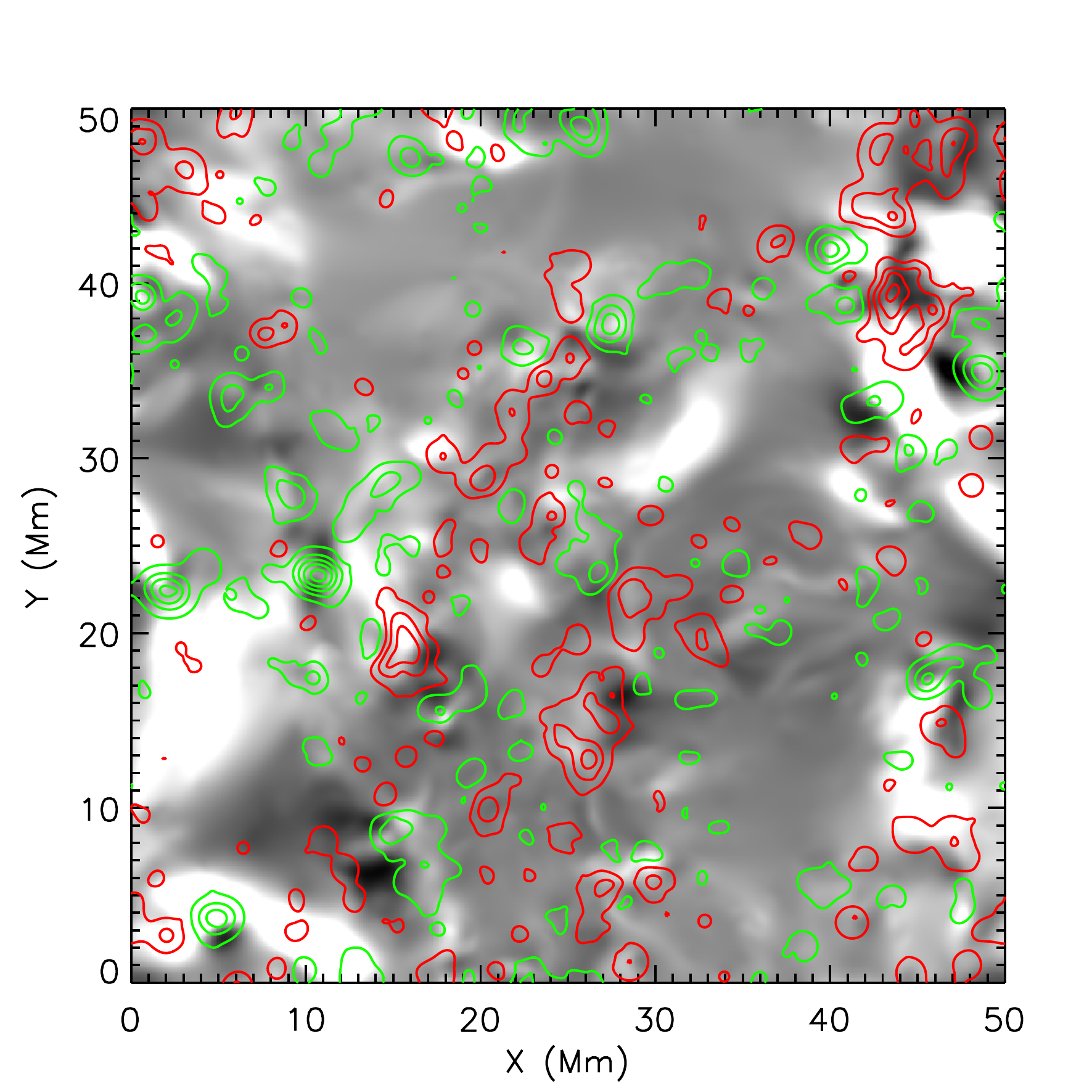}
              }
     \vspace{-0.45\textwidth}   
     \centerline{ \bf     
      \hspace{-0.03 \textwidth}  \color{black}{(a)}
      \hspace{0.46\textwidth}  \color{black}{(b)}
         \hfill}
     \vspace{0.4\textwidth}    

   \centerline{\hspace*{0.07\textwidth}
              \includegraphics[width=0.49\textwidth,clip=]{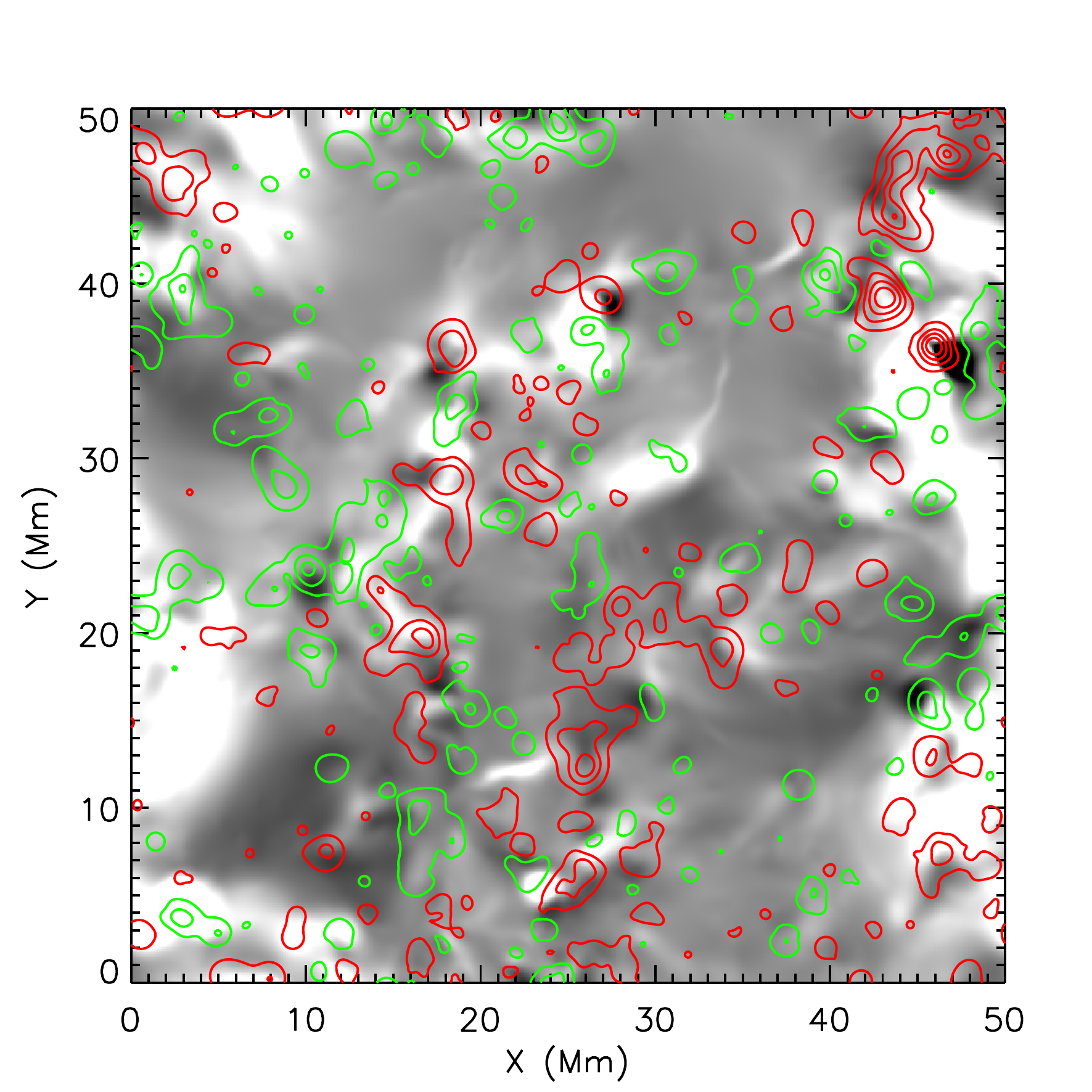}
               \hspace*{0.03\textwidth}
              \includegraphics[width=0.49\textwidth,clip=]{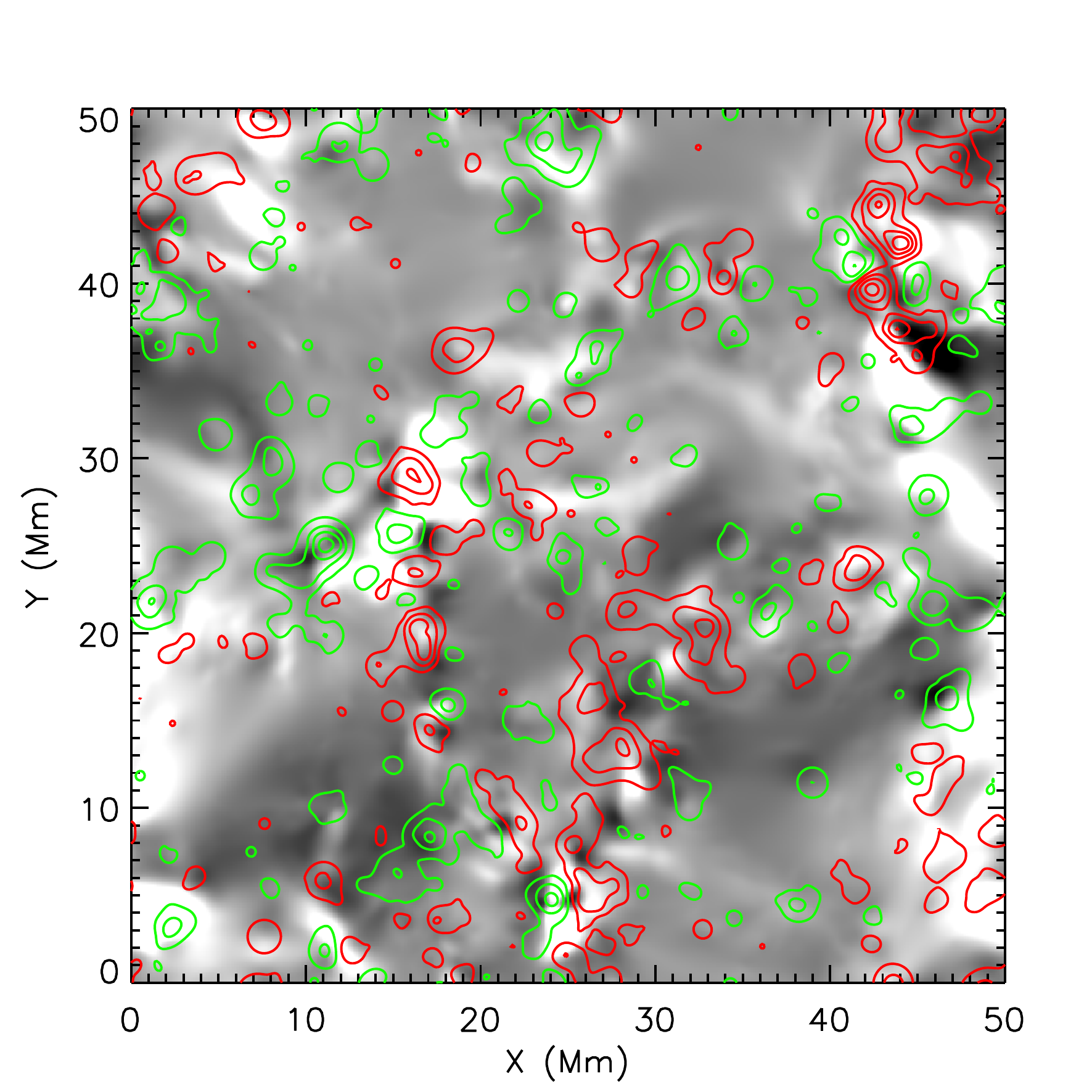}
              }
     \vspace{-0.45\textwidth}   
     \centerline{ \bf     
      \hspace{-0.03 \textwidth}  \color{black}{(c)}
      \hspace{0.46\textwidth}  \color{black}{(d)}
         \hfill}
     \vspace{0.4\textwidth}    
   \centerline{\hspace*{0.07\textwidth}
              \includegraphics[width=0.49\textwidth,clip=]{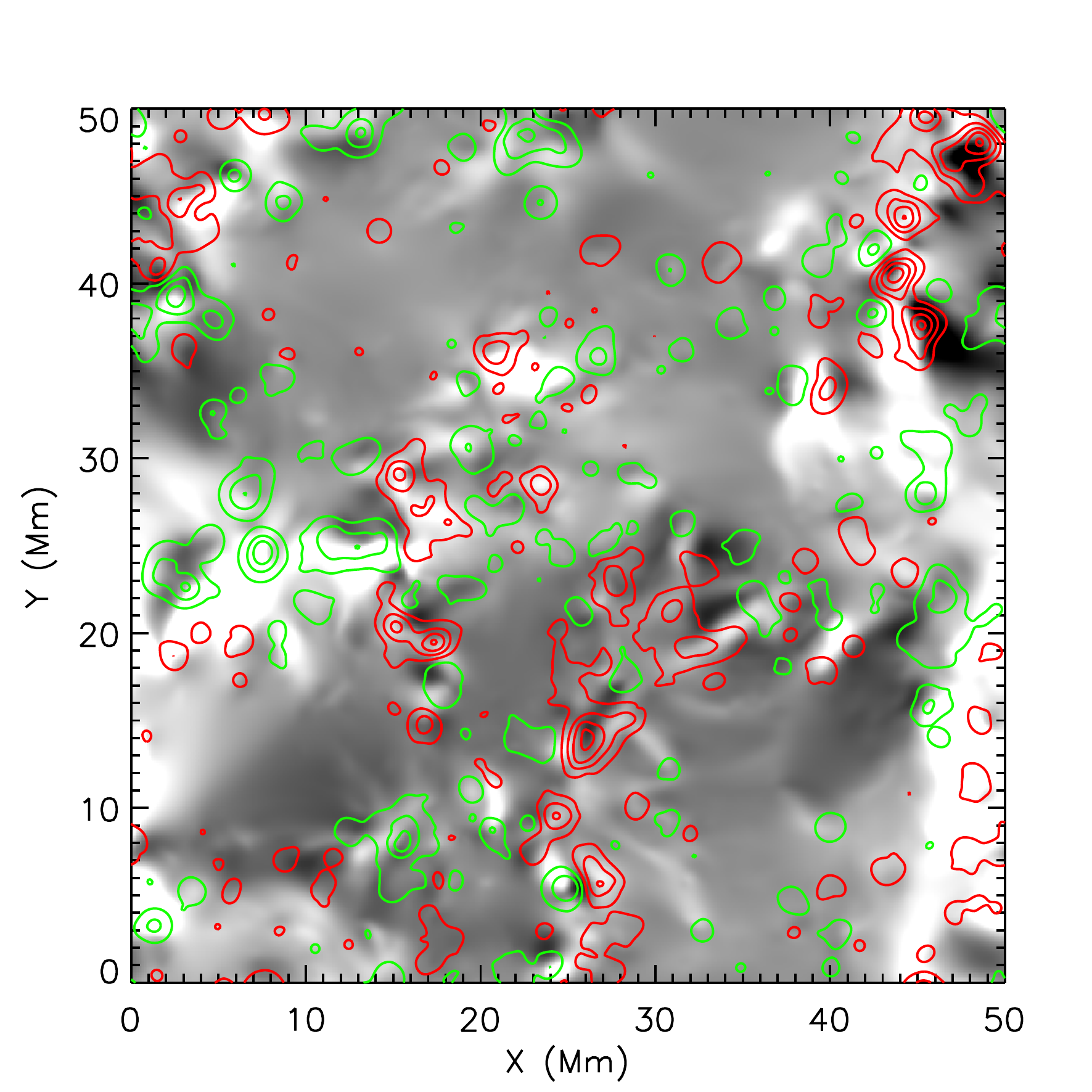}
               \hspace*{0.03\textwidth}
              \includegraphics[width=0.49\textwidth,clip=]{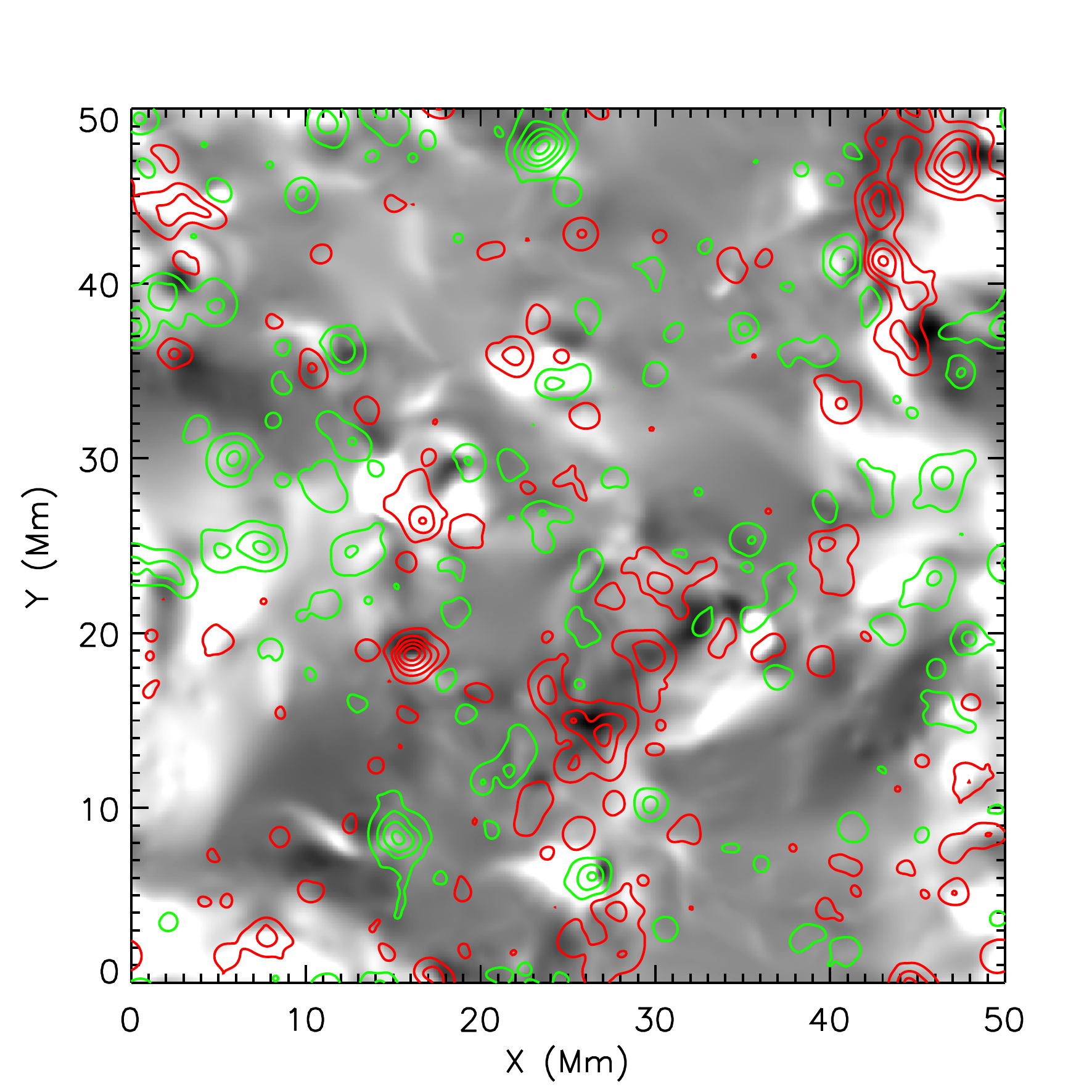}
              }
     \vspace{-0.45\textwidth}   
     \centerline{ \bf     
      \hspace{-0.03 \textwidth}  \color{black}{(e)}
      \hspace{0.46\textwidth}  \color{black}{(f)}
         \hfill}
     \vspace{0.4\textwidth}    

\caption{Free magnetic energy density integrated in the
line-of-sight, for the 3 G overlying field simulation. The images are shown in the $x-y$ plane
saturated at $\pm 1.9\times 10^{22}$ ergs.
Contours of $B_z$ at $z=0$ Mm are over-plotted where red contours represent
positive magnetic field and green contours represent negative, at levels of
$\pm[7, 13, 27, 53, 106]$ G. The images are shown at (a) $t=144$ hr, (b) $t=145$ hr, (c) $t=146$ hr, (d) $t=147$ hr, (e) $t=148$ hr and (f) $t=149$ hr.}\label{fig:app1}
   \end{figure}

  \begin{figure}

   \centerline{\hspace*{0.07\textwidth}
              \includegraphics[width=0.49\textwidth,clip=]{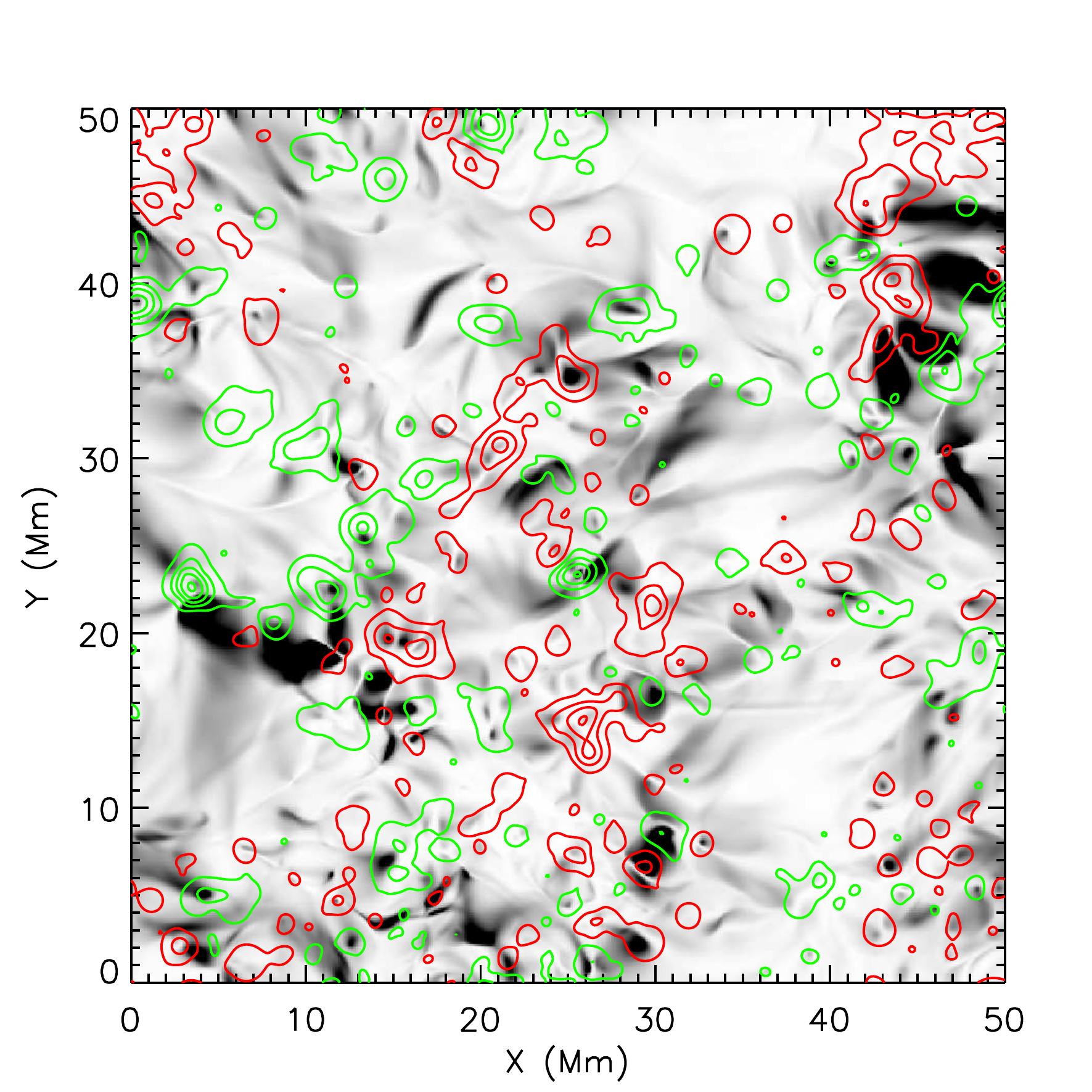}
               \hspace*{0.03\textwidth}
              \includegraphics[width=0.49\textwidth,clip=]{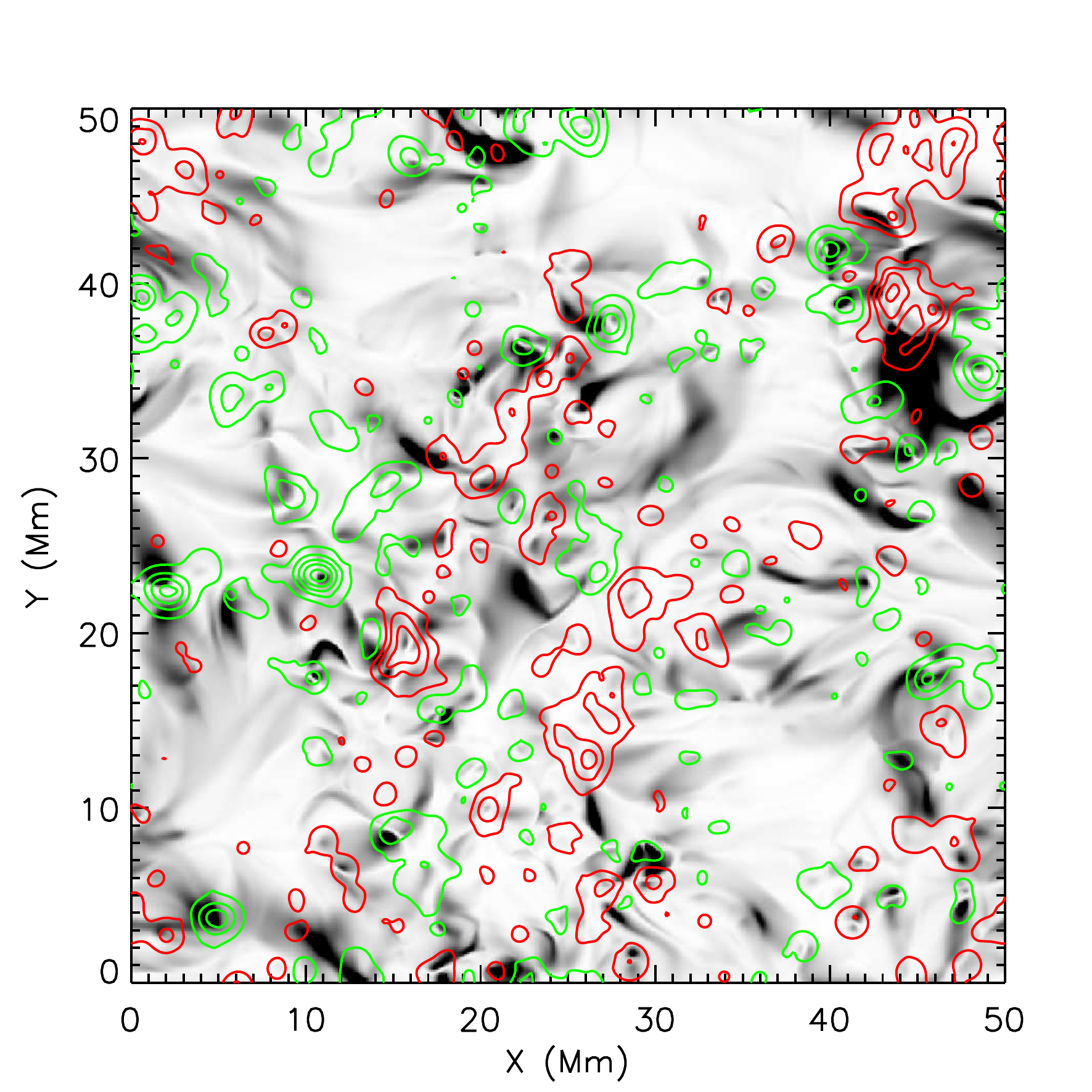}
              }
     \vspace{-0.45\textwidth}   
     \centerline{ \bf     
      \hspace{-0.03 \textwidth}  \color{black}{(a)}
      \hspace{0.46\textwidth}  \color{black}{(b)}
         \hfill}
     \vspace{0.4\textwidth}    

   \centerline{\hspace*{0.07\textwidth}
              \includegraphics[width=0.49\textwidth,clip=]{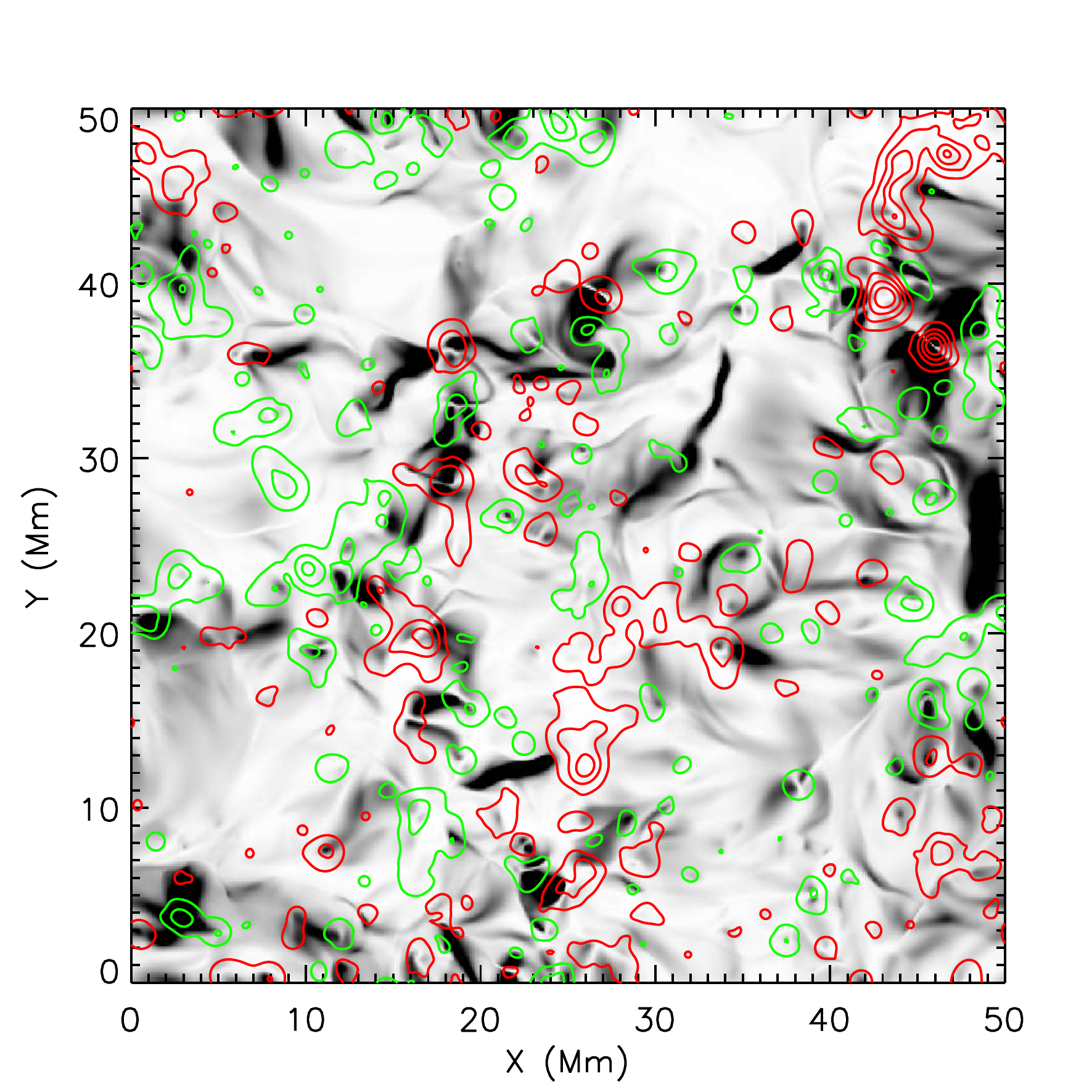}
               \hspace*{0.03\textwidth}
              \includegraphics[width=0.49\textwidth,clip=]{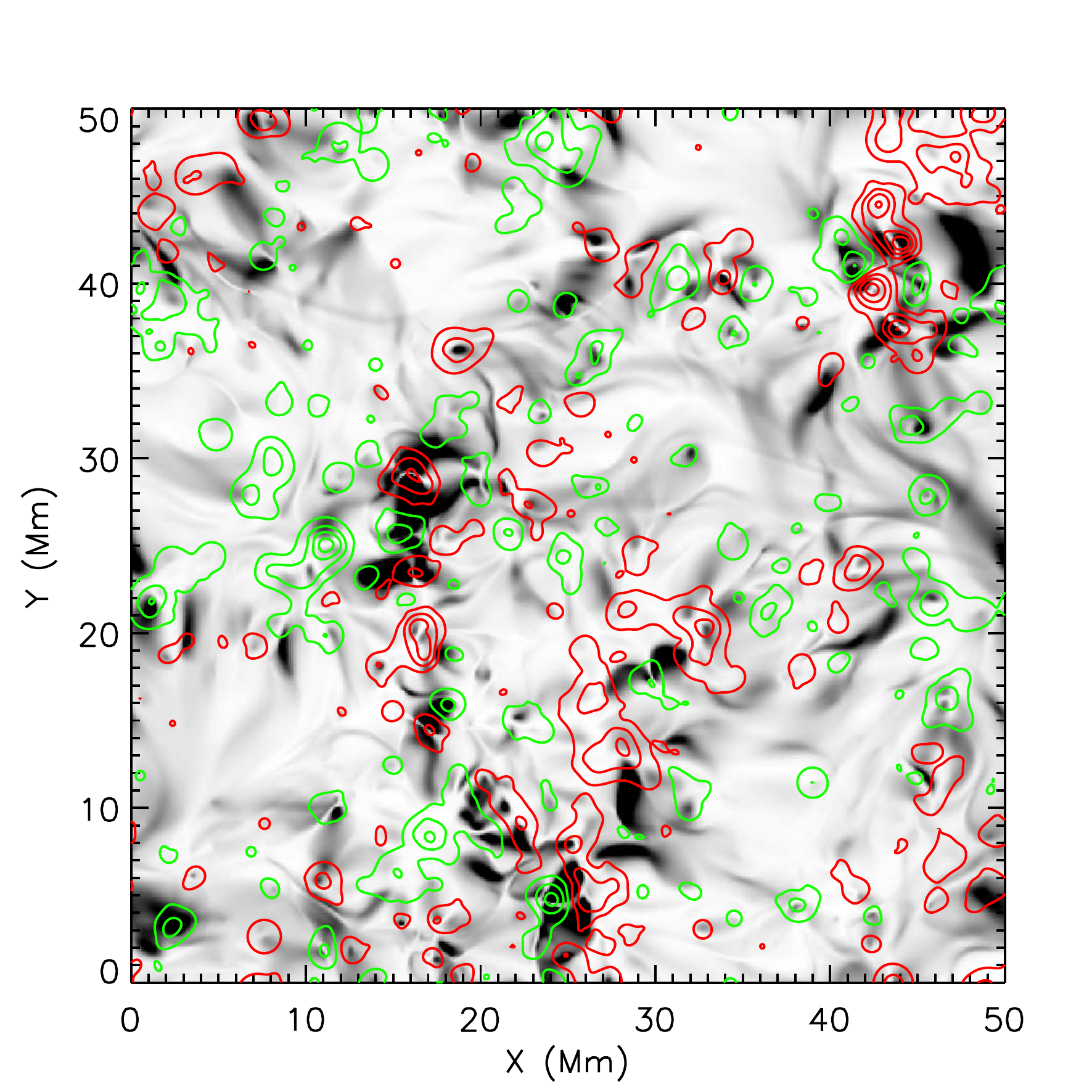}
              }
     \vspace{-0.45\textwidth}   
     \centerline{ \bf     
      \hspace{-0.03 \textwidth}  \color{black}{(c)}
      \hspace{0.46\textwidth}  \color{black}{(d)}
         \hfill}
     \vspace{0.4\textwidth}    
   \centerline{\hspace*{0.07\textwidth}
              \includegraphics[width=0.49\textwidth,clip=]{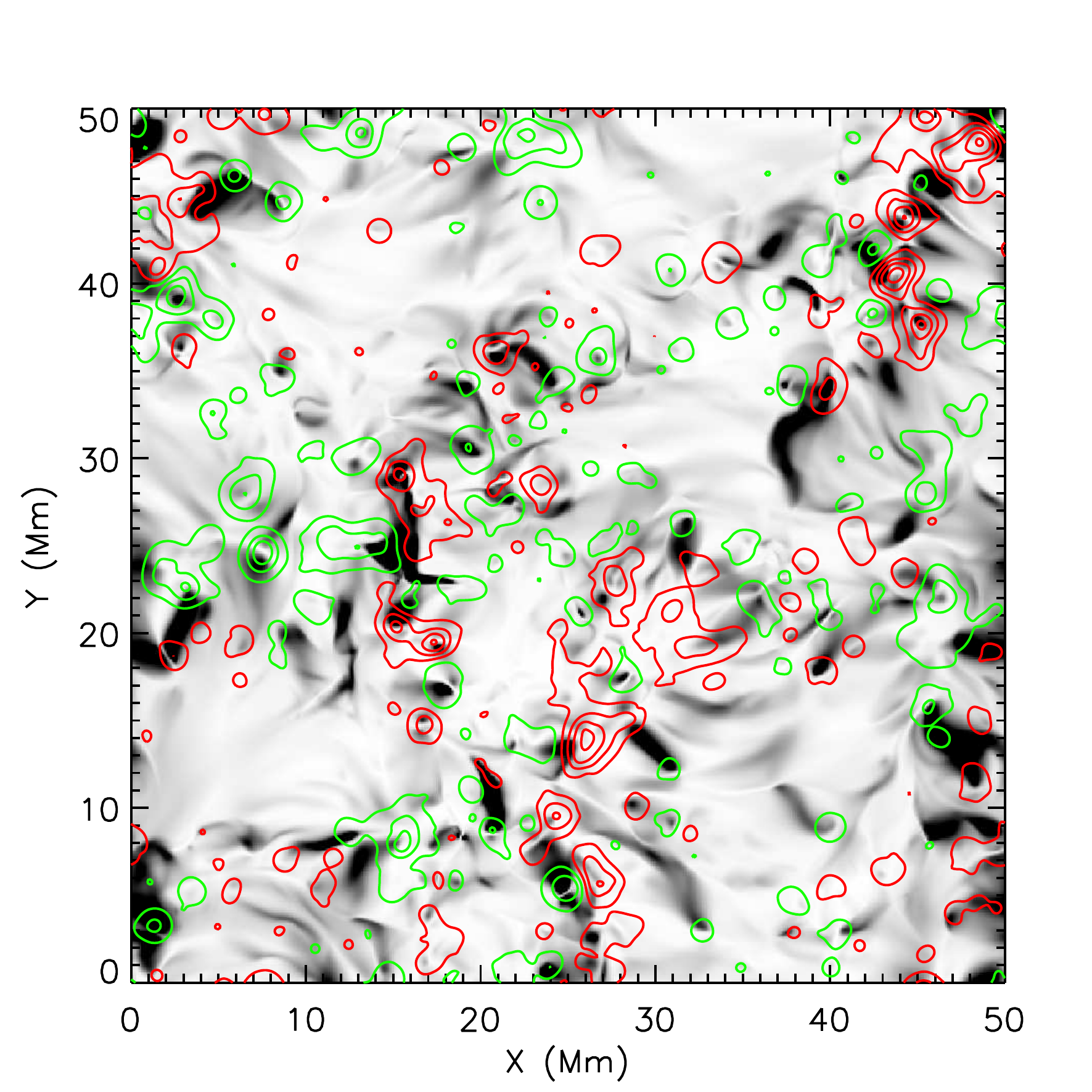}
               \hspace*{0.03\textwidth}
              \includegraphics[width=0.49\textwidth,clip=]{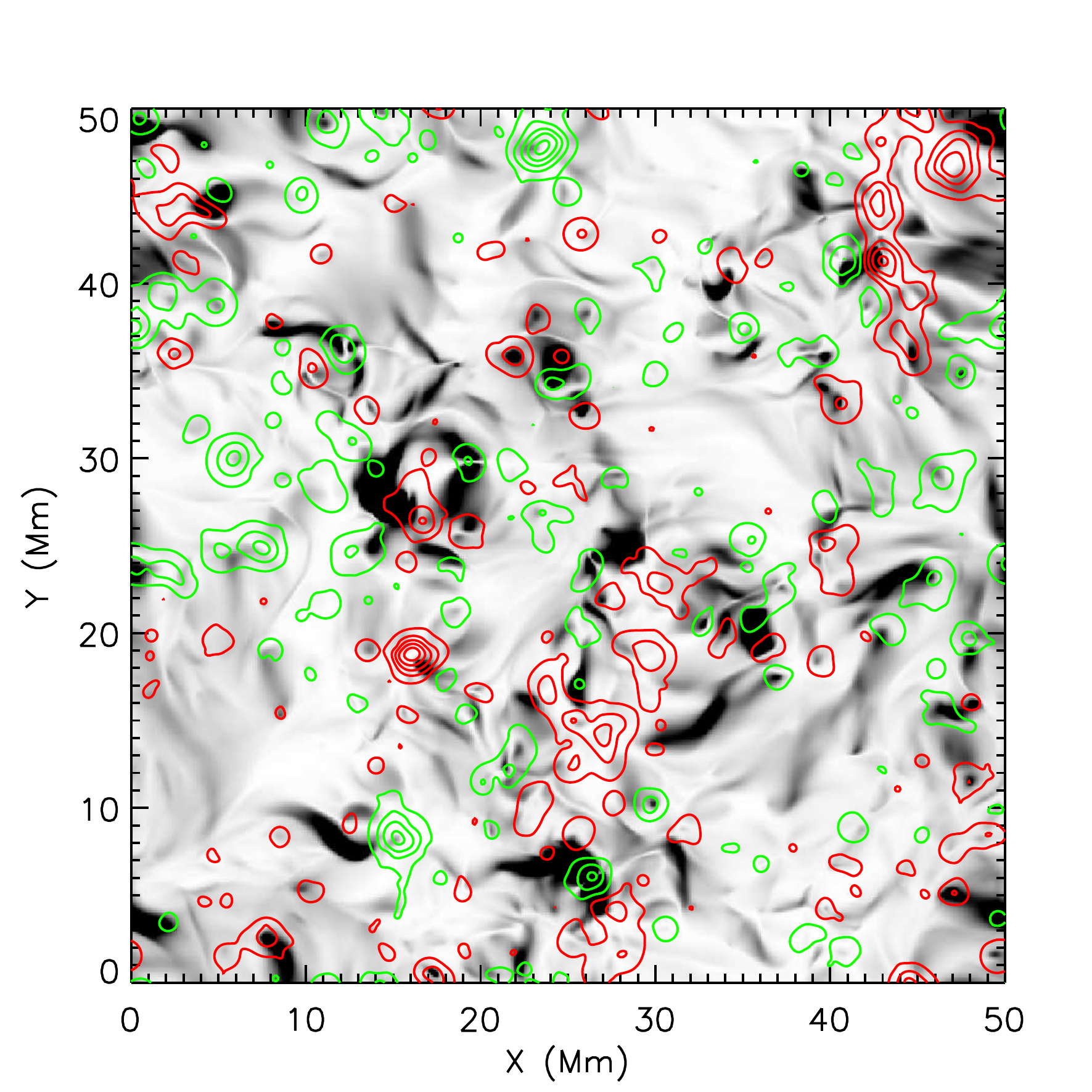}
              }
     \vspace{-0.45\textwidth}   
     \centerline{ \bf     
      \hspace{-0.03 \textwidth}  \color{black}{(e)}
      \hspace{0.46\textwidth}  \color{black}{(f)}
         \hfill}
     \vspace{0.4\textwidth}    

\caption{Normalised $j^2$ integrated in $z$, shown in the $x-y$ plane, for the 3 G overlying field simulation. Darker regions correspond to higher values.
Contours of $B_z$ at $z=0$ Mm are over-plotted where red contours represent
positive magnetic field and green contours represent negative, at levels of
$\pm[7, 13, 27, 53, 106]$ G. The images are shown at (a) $t=144$ hr, (b) $t=145$ hr, (c) $t=146$ hr, (d) $t=147$ hr, (e) $t=148$ hr and (f) $t=149$ hr.}\label{fig:app2}
   \end{figure}

  \begin{figure}

   \centerline{\hspace*{0.07\textwidth}
              \includegraphics[width=0.49\textwidth,clip=]{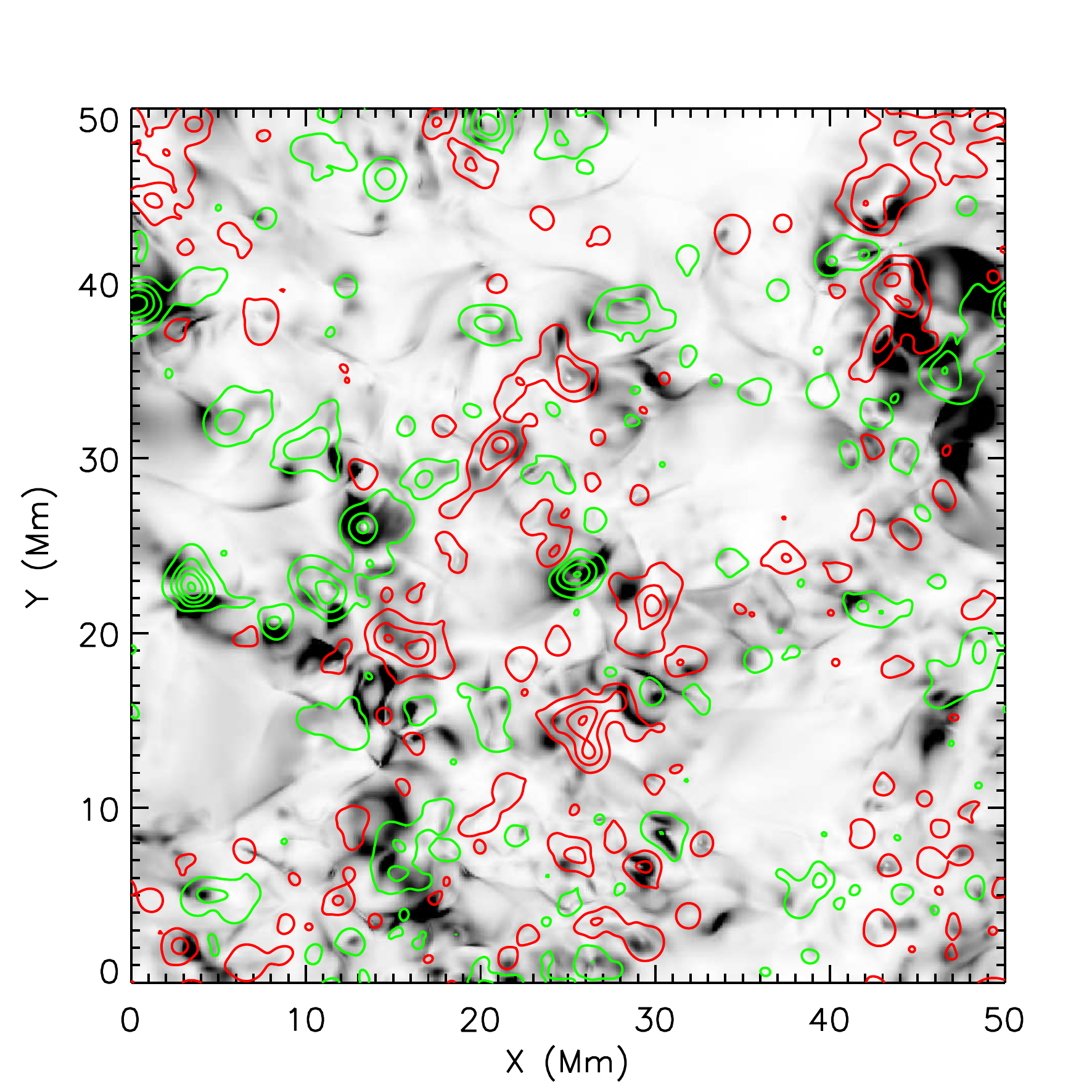}
               \hspace*{0.03\textwidth}
              \includegraphics[width=0.49\textwidth,clip=]{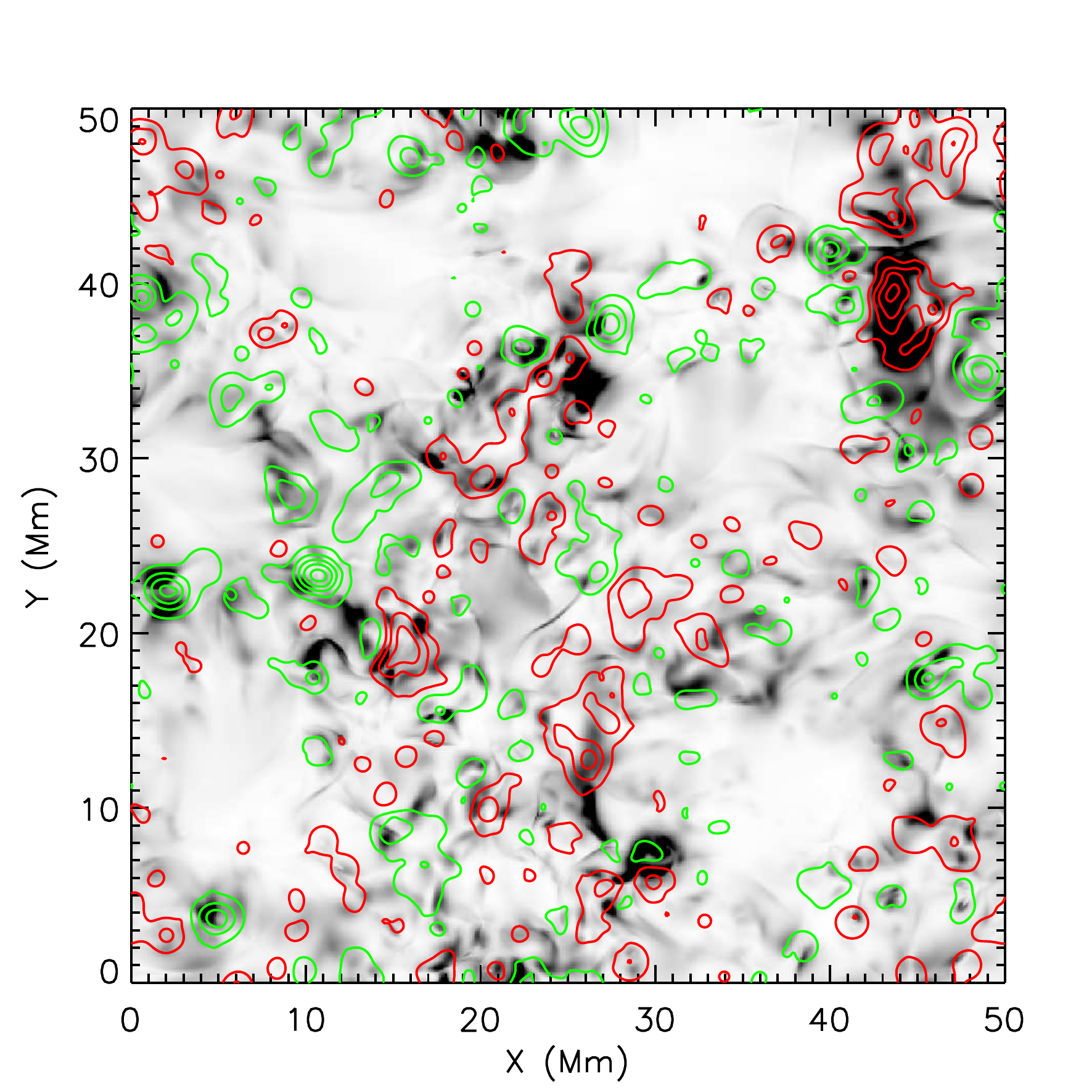}
              }
     \vspace{-0.45\textwidth}   
     \centerline{ \bf     
      \hspace{-0.03 \textwidth}  \color{black}{(a)}
      \hspace{0.46\textwidth}  \color{black}{(b)}
         \hfill}
     \vspace{0.4\textwidth}    

   \centerline{\hspace*{0.07\textwidth}
              \includegraphics[width=0.49\textwidth,clip=]{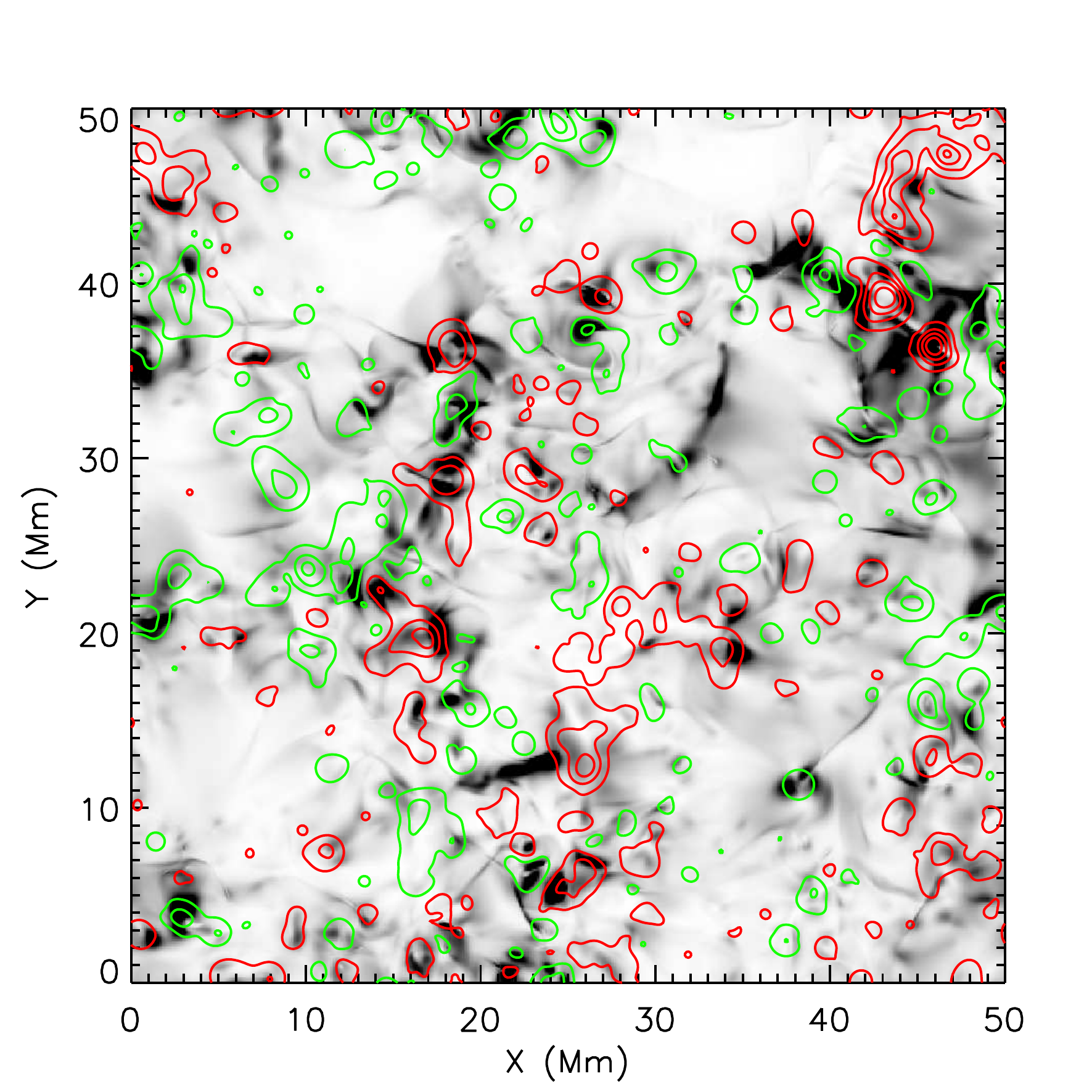}
               \hspace*{0.03\textwidth}
              \includegraphics[width=0.49\textwidth,clip=]{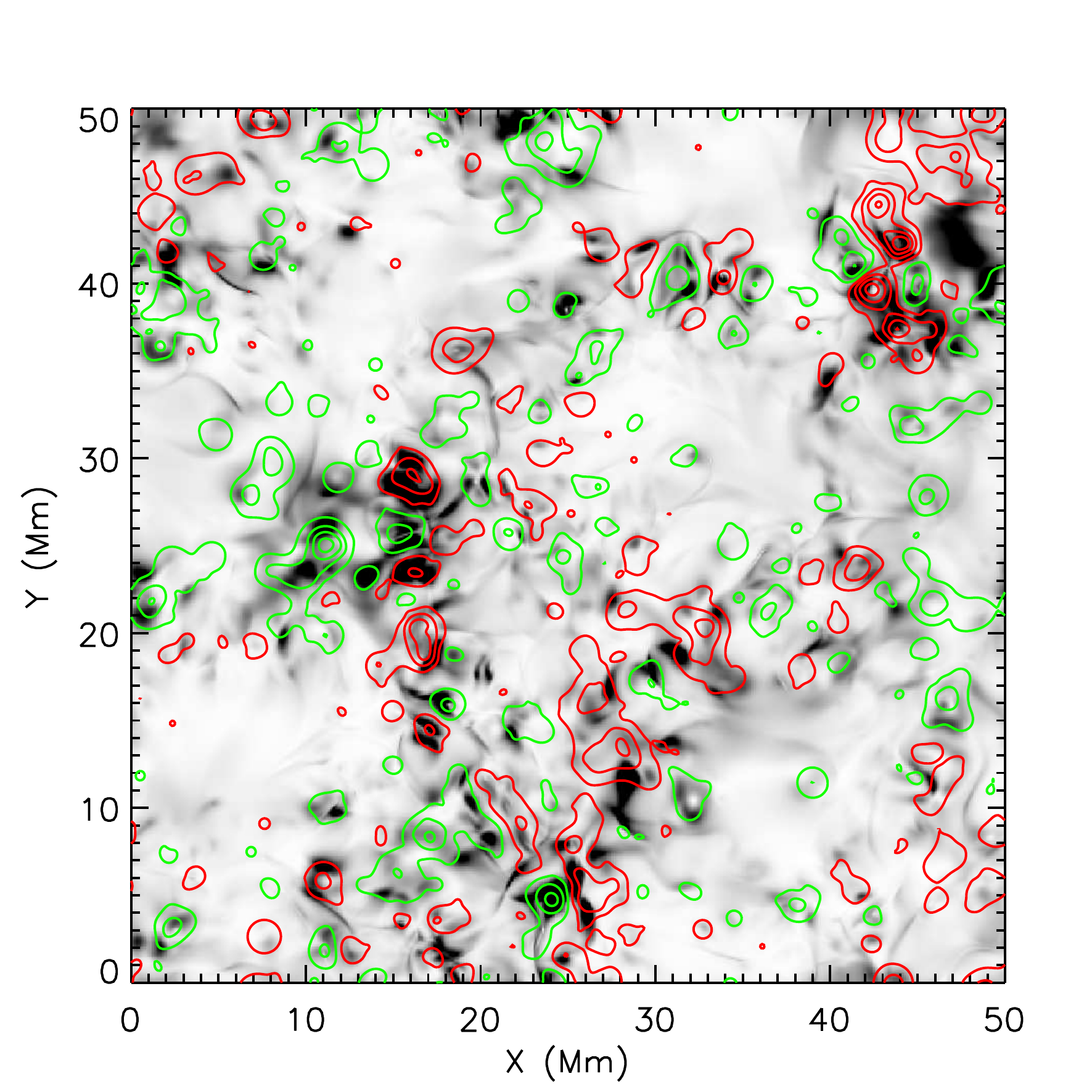}
              }
     \vspace{-0.45\textwidth}   
     \centerline{ \bf     
      \hspace{-0.03 \textwidth}  \color{black}{(c)}
      \hspace{0.46\textwidth}  \color{black}{(d)}
         \hfill}
     \vspace{0.4\textwidth}    
   \centerline{\hspace*{0.07\textwidth}
              \includegraphics[width=0.49\textwidth,clip=]{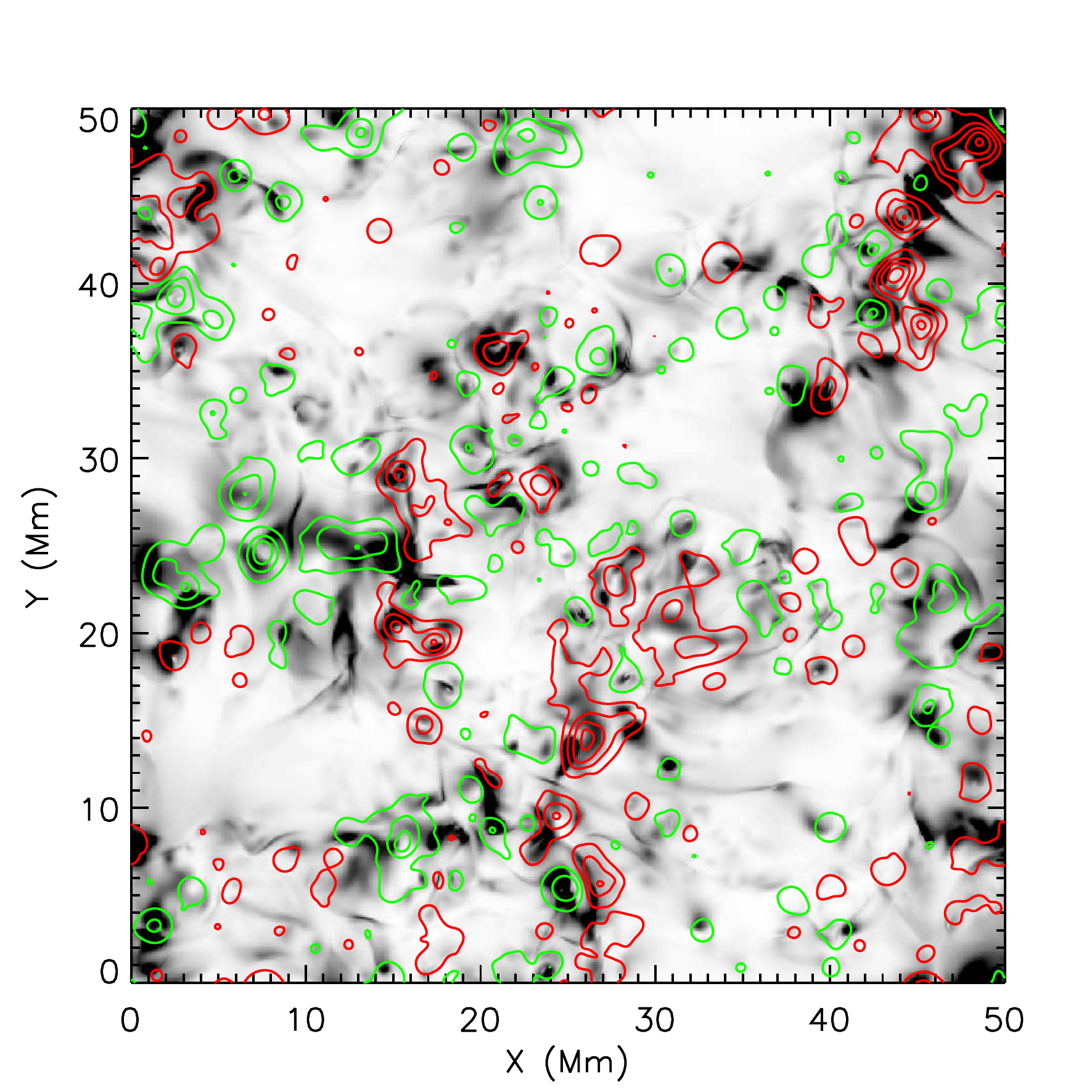}
               \hspace*{0.03\textwidth}
              \includegraphics[width=0.49\textwidth,clip=]{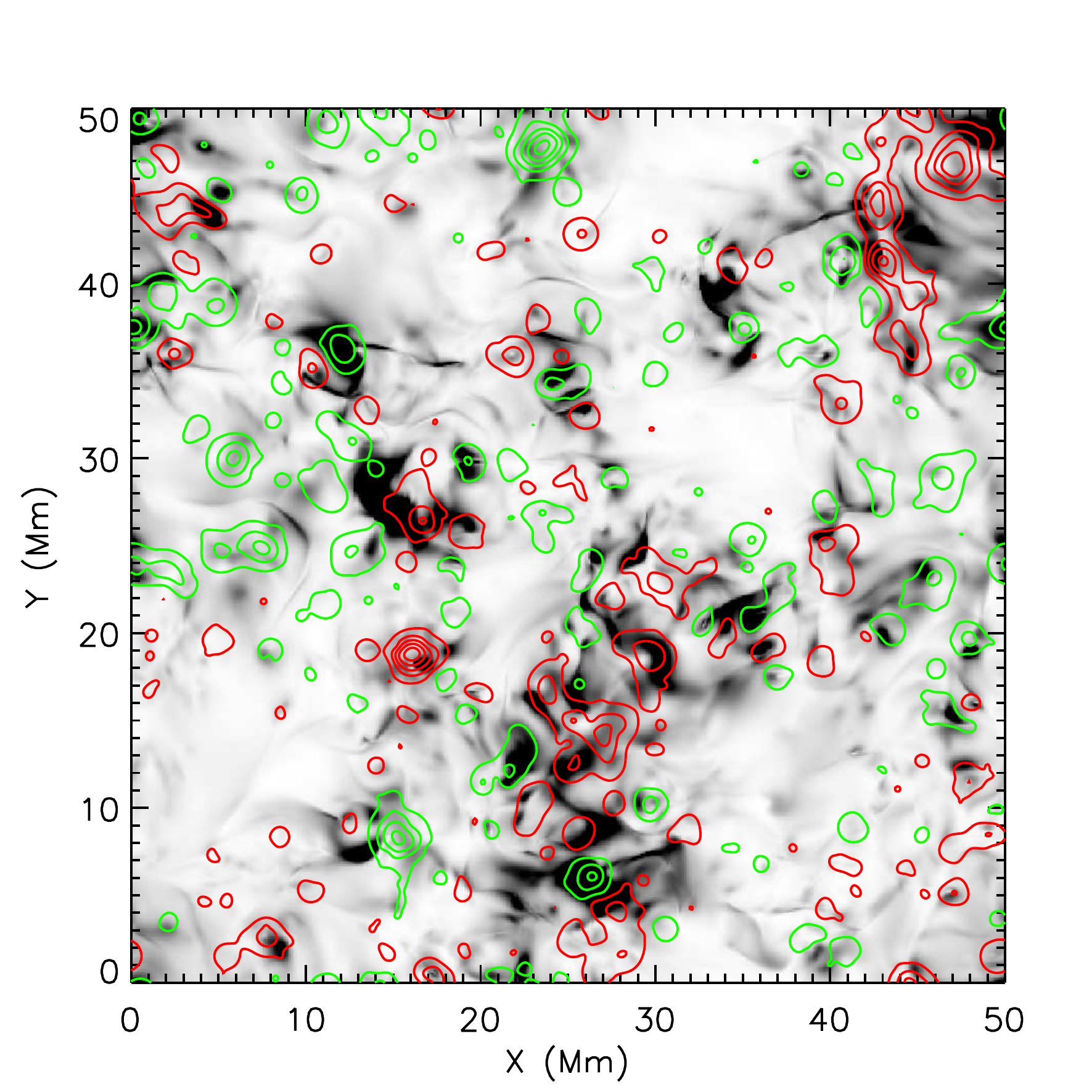}
              }
     \vspace{-0.45\textwidth}   
     \centerline{ \bf     
      \hspace{-0.03 \textwidth}  \color{black}{(e)}
      \hspace{0.46\textwidth}  \color{black}{(f)}
         \hfill}
     \vspace{0.4\textwidth}    

\caption{Rate of energy dissipation, $Q$, integrated in the
line-of-sight, for the 3 G overlying field simulation. Darker regions correspond to higher values. The images are shown in the $x-y$ plane
saturated at $1.5\times 10^{5}$ ergs cm$^{-2}$ s$^{-1}$.
Contours of $B_z$ at $z=0$ Mm are over-plotted where red contours represent
positive magnetic field and green contours represent negative, at levels of
$\pm[7, 13, 27, 53, 106]$ G. The images are shown at (a) $t=144$ hr, (b) $t=145$ hr, (c) $t=146$ hr, (d) $t=147$ hr, (e) $t=148$ hr and (f) $t=149$ hr.}\label{fig:app3}
   \end{figure}

  \begin{figure}

   \centerline{\hspace*{0.07\textwidth}
              \includegraphics[width=0.49\textwidth,clip=]{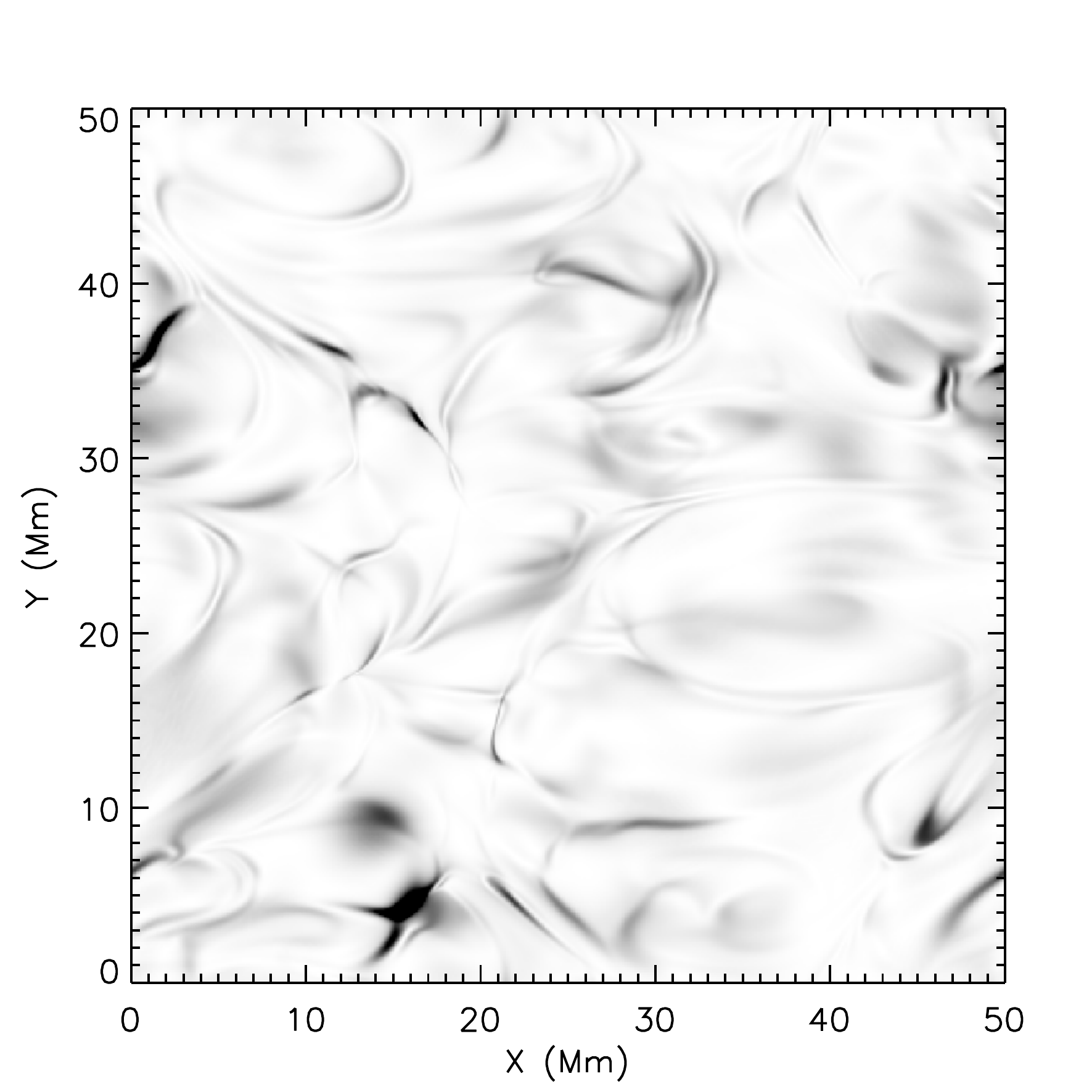}
               \hspace*{0.03\textwidth}
              \includegraphics[width=0.49\textwidth,clip=]{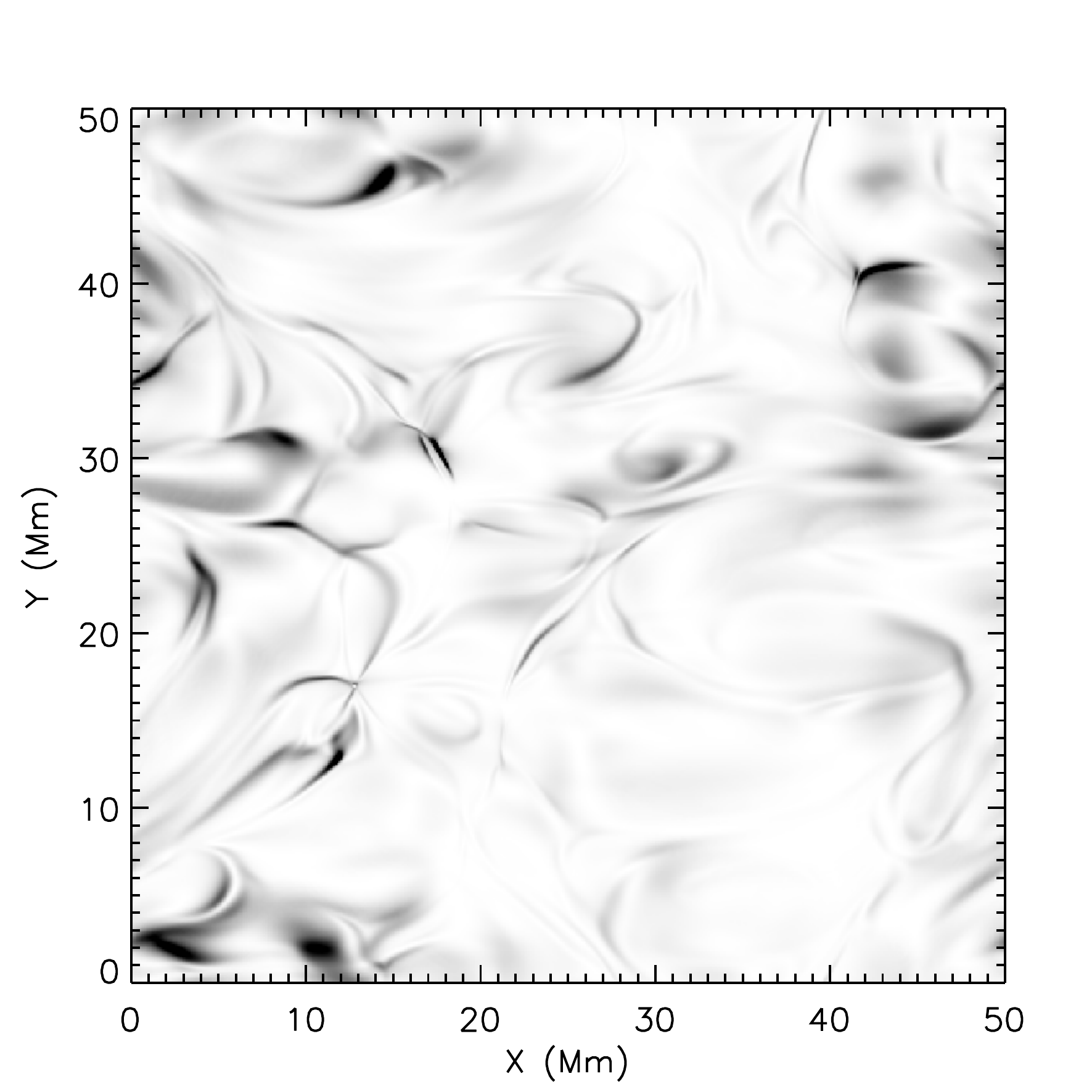}
              }
     \vspace{-0.45\textwidth}   
     \centerline{ \bf     
      \hspace{-0.03 \textwidth}  \color{black}{(a)}
      \hspace{0.46\textwidth}  \color{black}{(b)}
         \hfill}
     \vspace{0.4\textwidth}    

   \centerline{\hspace*{0.07\textwidth}
              \includegraphics[width=0.49\textwidth,clip=]{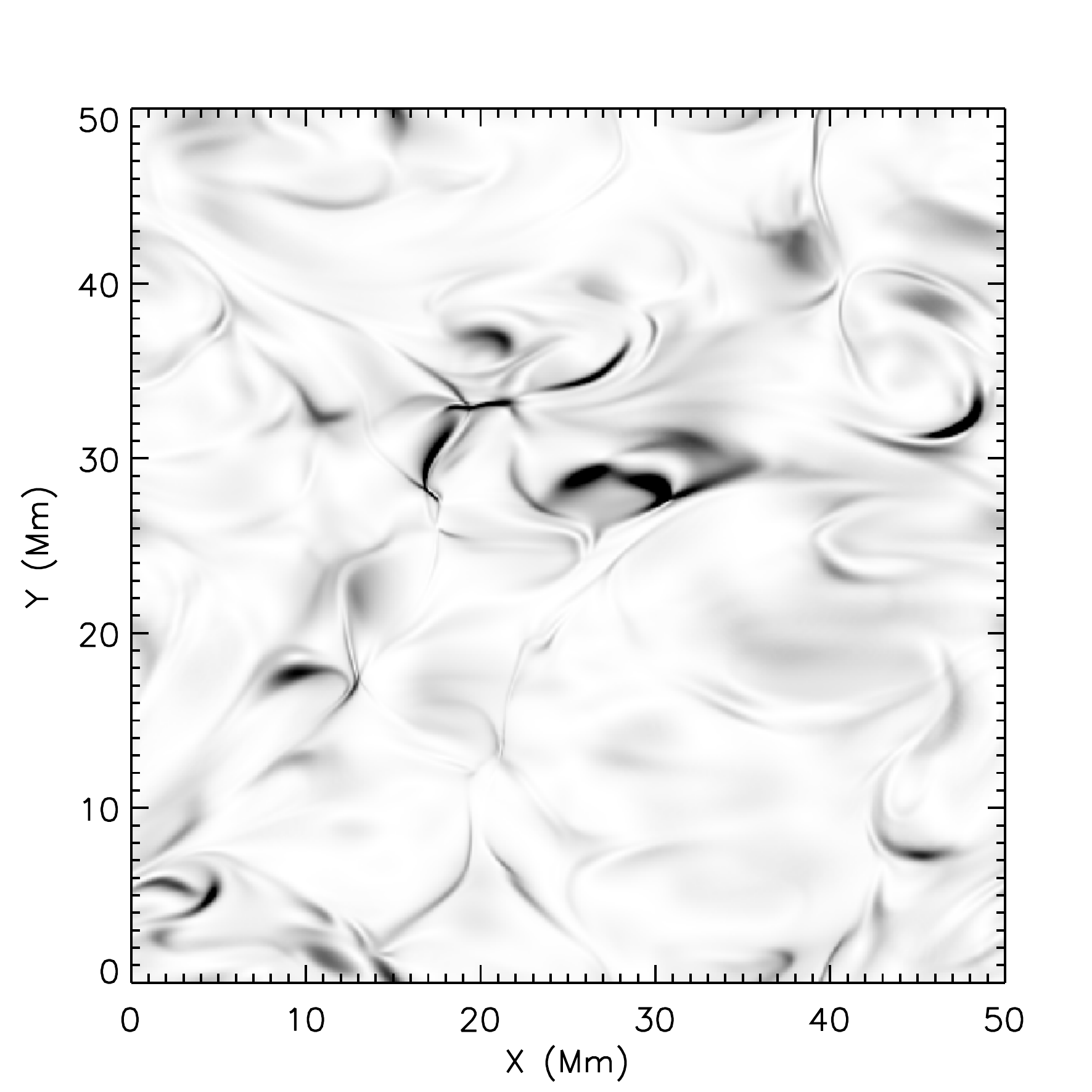}
               \hspace*{0.03\textwidth}
              \includegraphics[width=0.49\textwidth,clip=]{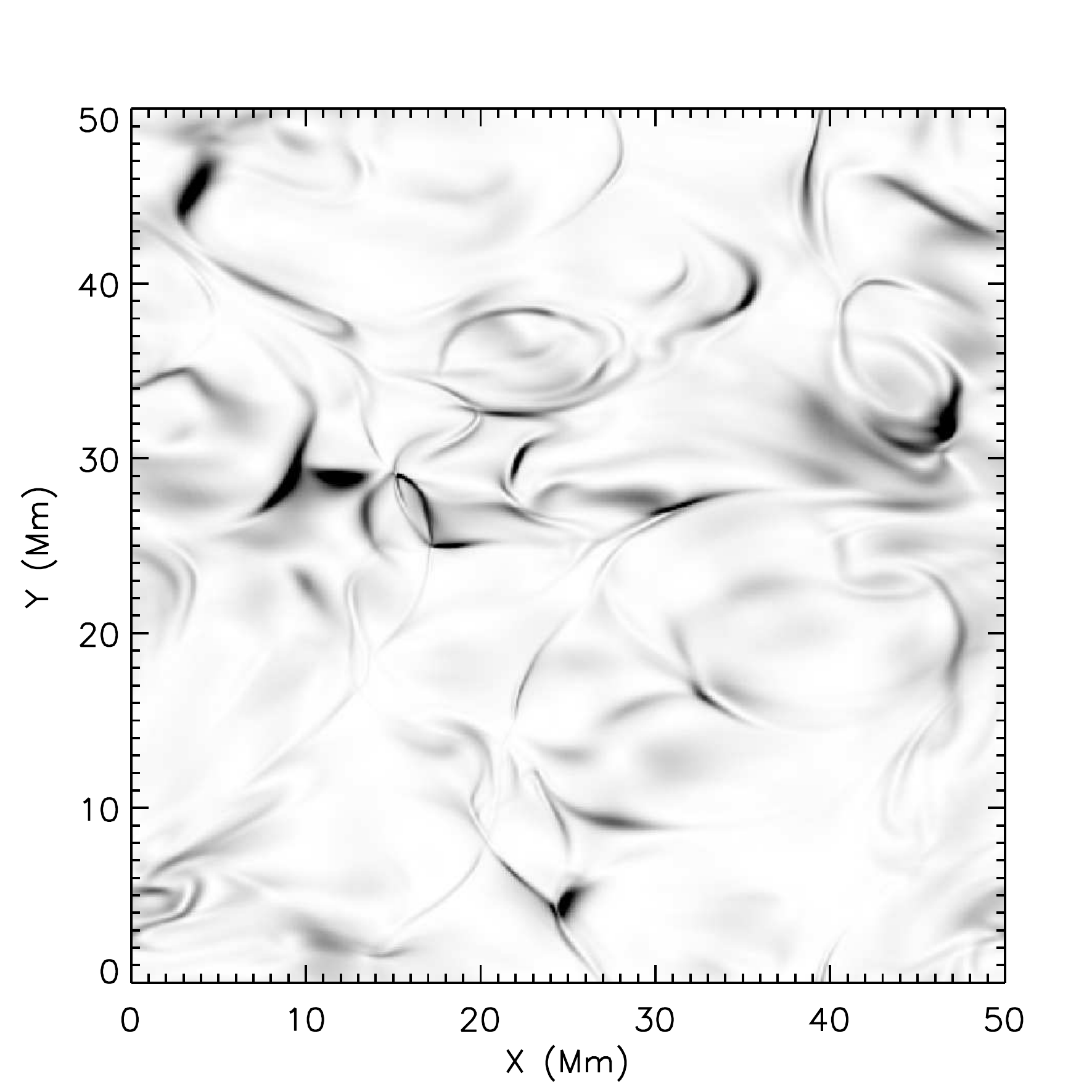}
              }
     \vspace{-0.45\textwidth}   
     \centerline{ \bf     
      \hspace{-0.03 \textwidth}  \color{black}{(c)}
      \hspace{0.46\textwidth}  \color{black}{(d)}
         \hfill}
     \vspace{0.4\textwidth}    
   \centerline{\hspace*{0.07\textwidth}
              \includegraphics[width=0.49\textwidth,clip=]{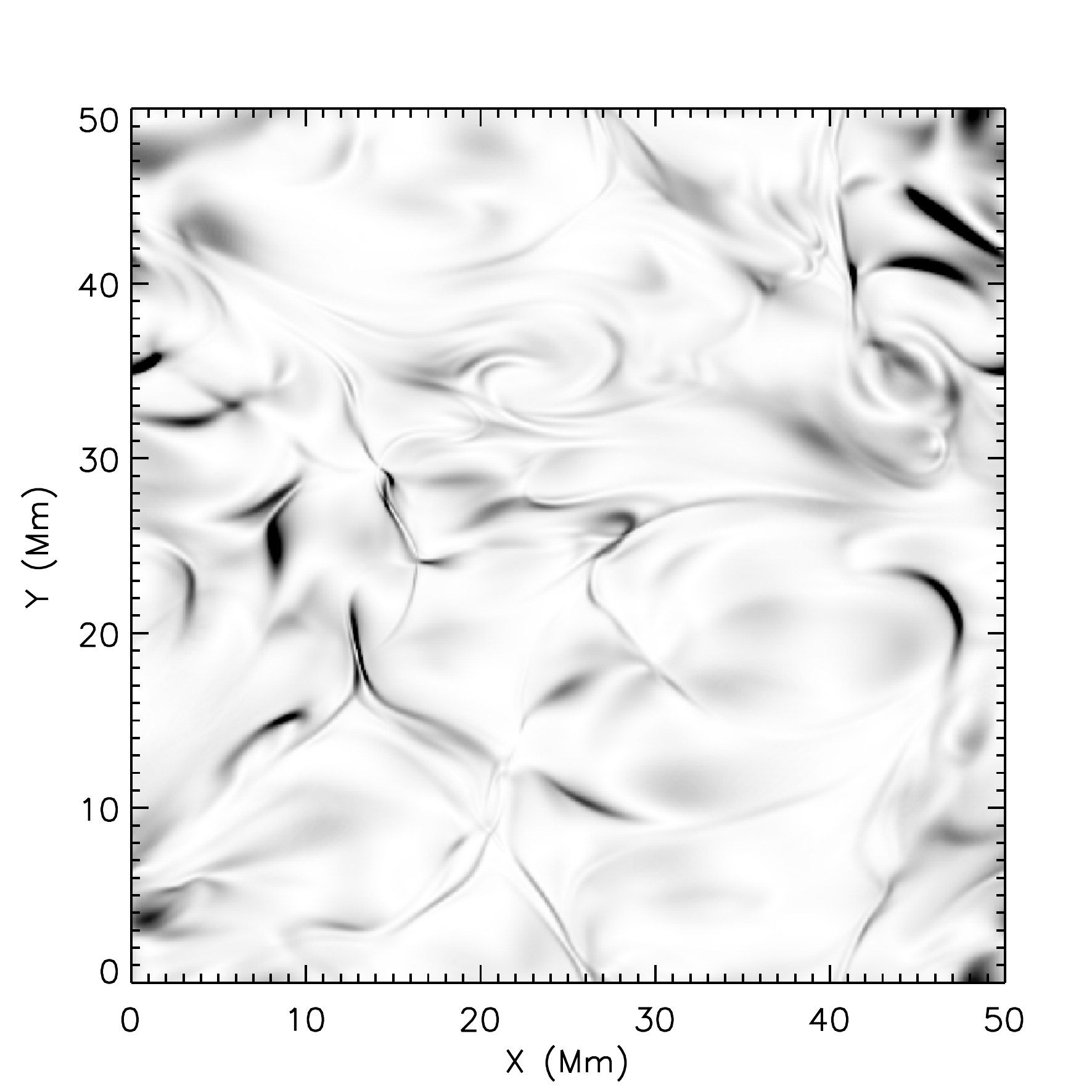}
               \hspace*{0.03\textwidth}
              \includegraphics[width=0.49\textwidth,clip=]{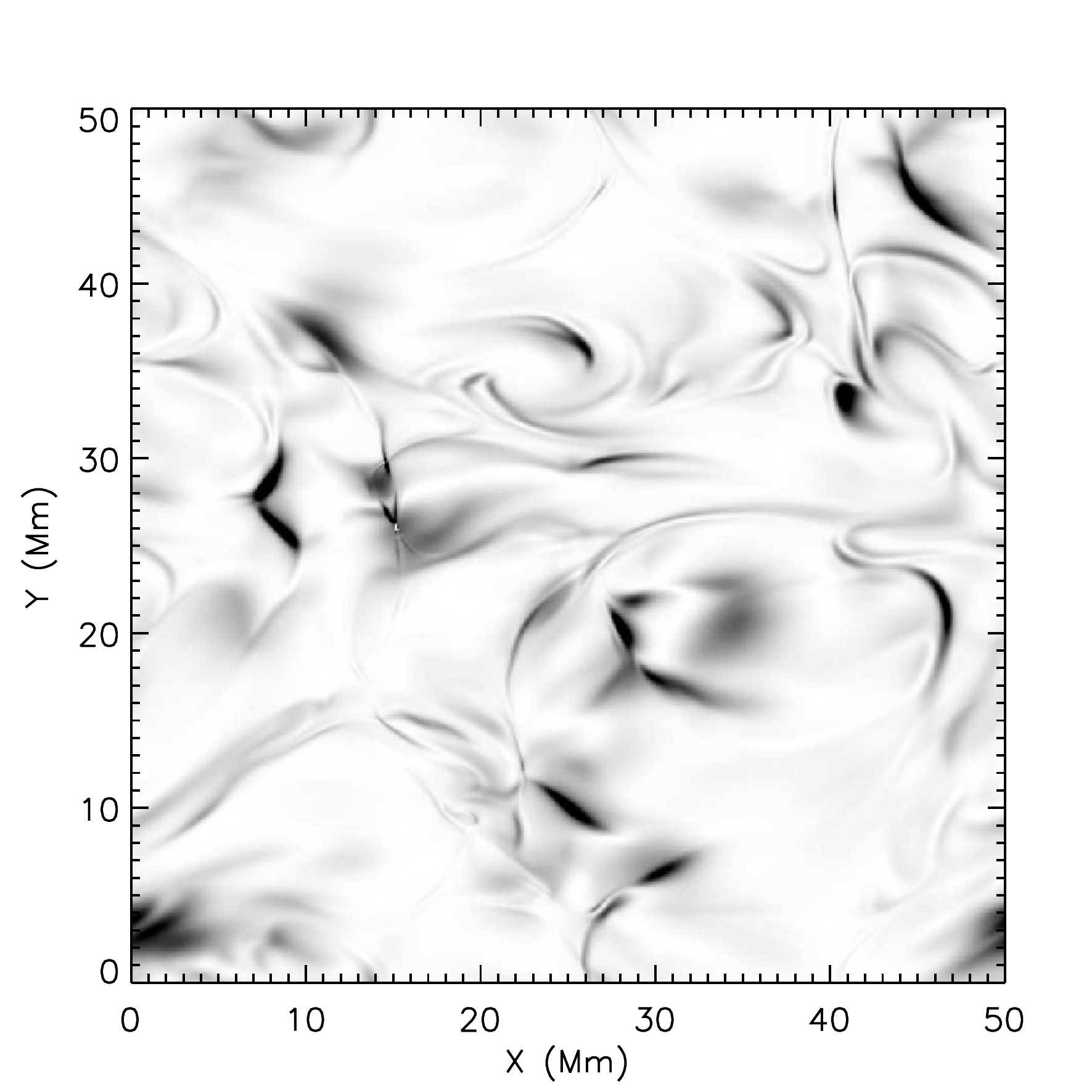}
              }
     \vspace{-0.45\textwidth}   
     \centerline{ \bf     
      \hspace{-0.03 \textwidth}  \color{black}{(e)}
      \hspace{0.46\textwidth}  \color{black}{(f)}
         \hfill}
     \vspace{0.4\textwidth}    

\caption{Rate of energy dissipation, $Q$, for the 3 G overlying field simulation. The images are shown in the $x-y$ plane at $z=3$ Mm,
saturated at $810$ ergs cm$^{-2}$ s$^{-1}$. Darker regions correspond to higher values. They are shown at (a) $t=144$ hr, (b) $t=145$ hr, (c) $t=146$ hr, (d) $t=147$ hr, (e) $t=148$ hr and (f) $t=149$ hr.}\label{fig:app4}
   \end{figure}

\end{article} 

\end{document}